\documentclass{xx-ijmpa}

\newcommand{\ipb}{\ensuremath{\mathrm{~pb^{-1}}}}
\newcommand{\inb}{\ensuremath{\mathrm{~nb^{-1}}}}

\newcommand{\GeV}{\ensuremath{\mathrm{~Ge\kern -0.1em V}}}
\newcommand{\GeVc}{\ensuremath{\mathrm{~Ge\kern -0.1em V\!/}c}}
\newcommand{\MeVc}{\ensuremath{\mathrm{~Me\kern -0.1em V\!/}c}}
\newcommand{\GeVcc}{\ensuremath{\mathrm{~Ge\kern -0.1em V\!/}c^2}}
\newcommand{\MeVcc}{\ensuremath{\mathrm{~Me\kern -0.1em V\!/}c^2}}
\newcommand{\MeV}{\ensuremath{\mathrm{~Me\kern -0.1em V}}}
\newcommand{\TeV}{\ensuremath{\mathrm{~Te\kern -0.1em V}}}

\newcommand{\MeVNS}{\ensuremath{\mathrm{Me\kern -0.1em V}}}
\newcommand{\MeVccNS}{\ensuremath{\mathrm{Me\kern -0.1em V\!/}c^2}}

\begin{document}

\markboth{G. Bauer}
{The $X(3872)$ Meson and ``Exotic'' Spectroscopy at CDF~II   }

\catchline{}{}{}{}{}
%
%%%%%%%%%%%%%%%%%%%%%%%%%%%%%%%%%%%%%%%%%%%%%%%%%%%%%%%%%%%%%%%%%%%%

\title{
The $X(3872)$ Meson and\\  ``Exotic'' Spectroscopy at CDF~II\\
         ---Or NOT?---
%\footnote{For the title, try not to use more than 
%three lines. Typeset the title in 10 pt Times Roman, uppercase and 
%boldface.}
}

\author{\footnotesize G. Bauer
%\footnote{
%Typeset names in 8 pt Times Roman, uppercase. Use the footnote to 
%indicate the present or permanent address of the author.}
}

\address{Laboratory of Nuclear Science, Massachusetts Institute of Technology, \\
77 Massachusetts Avenue, Cambridge, MA 02139, USA
%\footnote{State completely without abbreviations, the 
%affiliation and mailing address, including country and e-mail address. 
%Typeset in 8 pt Times Italic.
%}
% \\
% bauerg@fnal.gov
}

\maketitle

\pub{Received (24 May 2005)}{Revised (Day Month Year)}

\begin{abstract}
A spate of remarkable new hadrons reported in 2003 may lead to  unequivocal 
proof of  states beyond conventional $q\bar{q}$ and $qqq$ structure.
Claimed baryonic states $\Theta^+$, $\Phi$, and $\Theta^0_c$
would consist of five quarks, and new $D^+_{sJ}$-states and/or  $X(3872)$
might contain four quarks.
I review efforts to search for and study this ``new'' spectroscopy
in $\bar{p}p$-collisions with the CDF~II detector.
Pentaquark searches are negative, and
no evidence for exotic analogs of $D_{sJ}$-states was found.
CDF has confirmed the  $X(3872)$.
My main focus is  the production and decay properties of the $X(3872)$,
and its possible interpretations.

\keywords{X(3872), Charmonium, Pentaquark, Exotic Hadrons}
\end{abstract}

\ccode{PACS Nos.: 13.25.Gv,    14.40.Gx,  13.85.Ni,  12.39.Mk }

\section{2003: Annus Mirabilis?}
	
After decades of relatively mundane additions to the hadron spectrum,
2003 may one day be recounted as the  dawn of a new era in spectroscopy.
This year witnessed reports that may lead to the {\it first unequivocal proof\,} that 
Nature is not limited to simple  $q\bar{q}$ and $qqq$ constructions.
But these claims  are  dogged by controversy,
and  may instead be recalled as an ignominious tale
told to future graduate students.

The idea of unconventional quark structures is quite old.
If one glosses over delicate  distinctions between
2-baryon {\it nuclei} and 
6-quark {\it particles}---and
pardons the anachronism---``exotic'' hadrons {\it pre-date}  the quark model.
Far back in antiquity  Fermi and Yang considered
$N\overline{N}$ bound states as a model of the pion.\cite{FermiYang}
Later the $SU(3)$ symmetry of the Eightfold Way\cite{8Fold}
was used to put the deuteron in a  dibaryon multiplet\cite{Oakes}---with
some evidence for a $\Lambda p$-state.\cite{DahlEtAl}
In the 1964 birth of the quark model Gell-Mann\cite{Gell-Mann}
actually mentions $qq\bar{q}\bar{q}$ and $qqqq\bar{q}$ 
as  mesons and baryons---but only
their lighter $q\bar{q}$ and $qqq$ siblings
were considered relevant at the time.

In the mid-1960s enhancements in  $KN$ scattering\cite{Cool}
pointed to +1 strangeness bar\-y\-on resonances, implying  minimal $qqqq\bar{s}$ content.
These very broad structures  required careful partial wave analysis
to justify them as resonances, called $Z^*$'s.
About the same time
$K\overline{K}$ bound states were suggested
to explain a low mass $I\!=\!1$ enhancement
in $\bar{p}p \rightarrow\! K\overline{K}\pi$.\cite{AstierKK}
And theoretically, duality arguments for baryon-antibaryon scattering
via meson exchanges implied, in quark language,  $qq\bar{q}\bar{q}$
systems.\cite{Rosner}

With the advent of QCD in the early 1970s 
the $q\bar{q}$/$qqq$-pattern was explained  by  $SU(3)_c$.
It was soon realized  that  not only 
were more complex quark structures allowed, but  also new types
exploiting gluons: ``hybrids''  with valence gluons added to quarks,
and ``glueballs'' without any quarks at all.\cite{Jaffe}
It is, however, a {\it dynamical}
issue whether any exotics  are manifest 
in an observationally  meaningful way.
Using a bag model  Jaffe and Johnson not only answered positively,  %affirmatively,
but argued that some known $0^{++}$ mesons ($f_0$, $a_0$\ldots)
were better viewed as  $qq\bar{q}\bar{q}$  than as a $^3P_0$  nonet of $q\bar{q}$.
Later, a $K\overline{K}$ state was invoked 
to explain $\pi\pi\!\rightarrow\! f_0(980) \!\rightarrow\! K\overline{K}$
data.\cite{ABWicklund}
Based on a potential model, both  $f_0(980)$ and $a_0(980)$
made good  $K\overline{K}$ ``molecules''---and 
likely  the only ones.\cite{WeinIsgur}
The $s$-quark mass  seemed to strike the right balance for binding.

Today exotics remain a dynamic topic.\cite{Review}
The  $f_0(980)$ and $a_0(980)$ are still promoted as  $K\overline{K}$-molecules,
and hybrid  and glueball   candidates are bandied about.
For a full list of suspects see
the PDG's {\it Non-$q\bar{q}$ Candidates} review.\cite{PDG04}
Despite decades of progress, 
no exotic meson has been conclusively identified.
Many are claimed as ``probably exotic,'' but proof is difficult.
Candidates are very wide,  and thus hard to study; and 
those with $q\bar{q}$ quantum numbers (``cryptoexotics'')
mix with ordinary mesons and are thus hard to understand.
More mesons
{\it are} known than needed as $q\bar{q}$-states,
hinting of {\it something} exotic.
But resonances can arise  dynamically,
opening another loophole. 
% {\it caveat} for exotic interpretations.
The ultimate smoking gun, 
a state with non-$q\bar{q}$ quantum numbers ({\it e.g.} $1^{-+}$),
has yet to be acclaimed.\cite{ExoticQuatNum}
This messy soup  demands
a painfully detailed understanding of data {\it and} theory
before there is consensus on non-$q\bar{q}$ {\it light} mesons.

For baryons the situation was worse.
After great hope for  $Z^*$ pentaquarks and di\-baryons
in the late 1960s and 70s, 
a grim reality set in  in the early 80s.\cite{BaryonRev}
Claims were either ruled out, or were simply unconvincing.
The PDG became so disillusioned that 
they last listed  $Z^*$'s
in 1986,\cite{PDG86}
and dibaryons in 1988.\cite{PDG88}
In spite of this dismal verdict, 
theoretical and experimental work continued out of the spotlight.%\cite{LastHurrah}

In summary, despite the valiant effort
of experimentalists and theorists for nearly forty years, 
the question of whether Nature elects 
to form systems beyond  $q\bar{q}$ and  $qqq$ 
remains  open.
But events in 2003 were to begin a new chapter in this saga.

\section{The Tevatron and the CDF II Detector}

\begin{figure}[t]
\centerline{
~\begin{minipage}[b]{0.43\textwidth}
\psfig{file=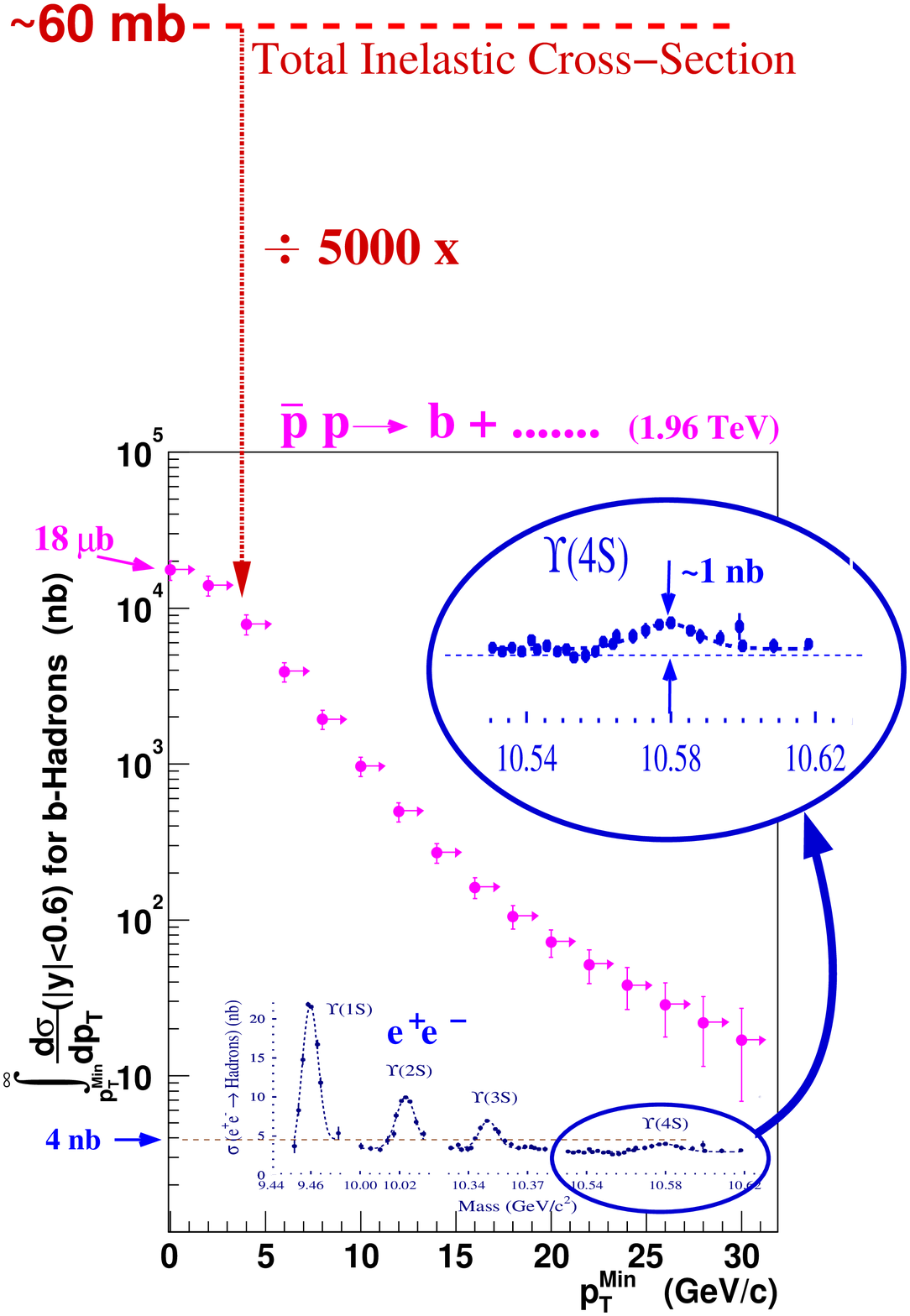,width=1.01\textwidth}
 \end{minipage}
\hfill
\begin{minipage}[b]{0.44\textwidth}
\begin{center}
\psfig{file=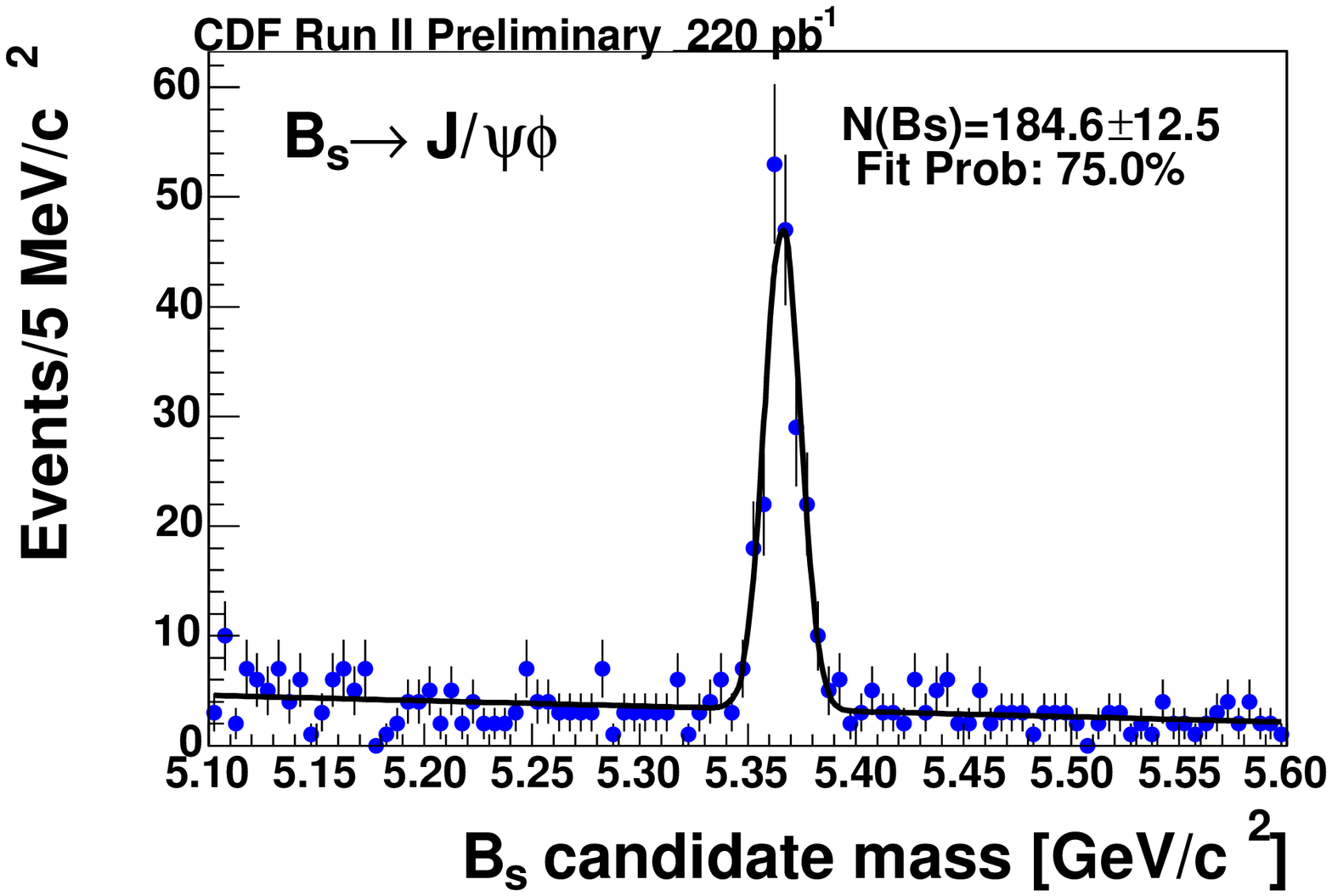,width=0.93\textwidth}\\
\psfig{file=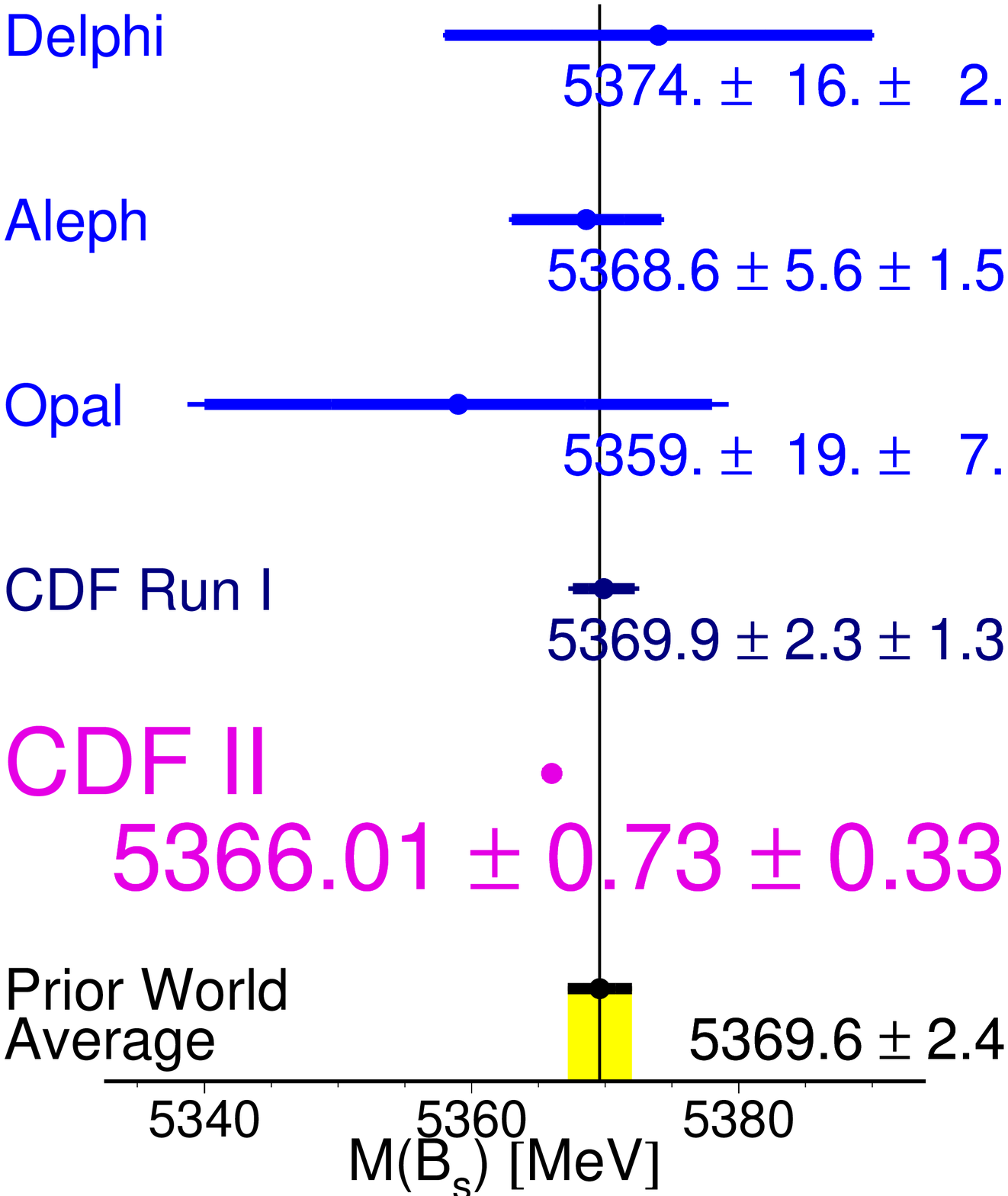,width=0.65\textwidth}
\end{center}
 \end{minipage}
           }
\vspace*{8pt}
\caption{{\bf LEFT:} Comparison of  the $b$-quark cross section
          at the Tevatron,\protect\cite{CDFbXSec} integrated above 
          a minimum $p_T$, $p_{T,\rm min}$,
          to the total inelastic cross section\protect\cite{InElasticX}  on a log-scale.
          Overlayed at the bottom is the $e^+e^-$ cross-section\protect\cite{XePlusEminus}
          on linear scale aligned to match the log-scale at 4 nb, 
         {\it i.e.} at the $\Upsilon(4S)$ where $B$-factories operate.
         {\bf TOP:} The CDF~II  $J/\psi \phi$ mass distribution  
                    ($\sim\! 8\MeVcc$ resolution)
                     used  for a $B^0_s$ mass measurement. 
         {\bf Bottom:} Compilation of world  $B^0_s$ mass measurements.\protect\cite{PDG04,BsMassII} 
\label{Fig:XSection}
}
\end{figure}

CDF~II is a general purpose detector at Fermilab's
$\bar{p}p$ collider\cite{Tevatron} ($\sqrt{s} \!\sim\! 2\TeV$).
Originally designed %\cite{CDFI} 
in the late
1970s for high-$p_T$ physics ($W$, $Z$, top.\ldots),
CDF became~an important venue for bottom/charm physics\cite{CDFI-Paulini}
as luminosities increased
and the detector enhanced.
The Tevatron produces hadrons
with very large cross sections,
as seen in Fig.~\ref{Fig:XSection},
where  $b$-production  is compared to 
$e^+e^-\!\rightarrow\! \Upsilon(4S)\!\rightarrow\! B\overline{B}$.
At the same time, CDF has excellent tracking for  spectroscopy, 
illustrated in  Fig.~\ref{Fig:XSection} by a $B^0_s$-mass 
measurement  to  sub-$\MeVNS$  precision.
The challenge is  to exploit this bounty:
just as $b$-production is very large,
the total inelastic cross section~(Fig.~\ref{Fig:XSection})~is~huge!
One lives or dies at a hadron collider by being able to selectively trigger on events.

\begin{figure}[t]
\centerline{
~\begin{minipage}[b]{0.51\textwidth}
\psfig{file=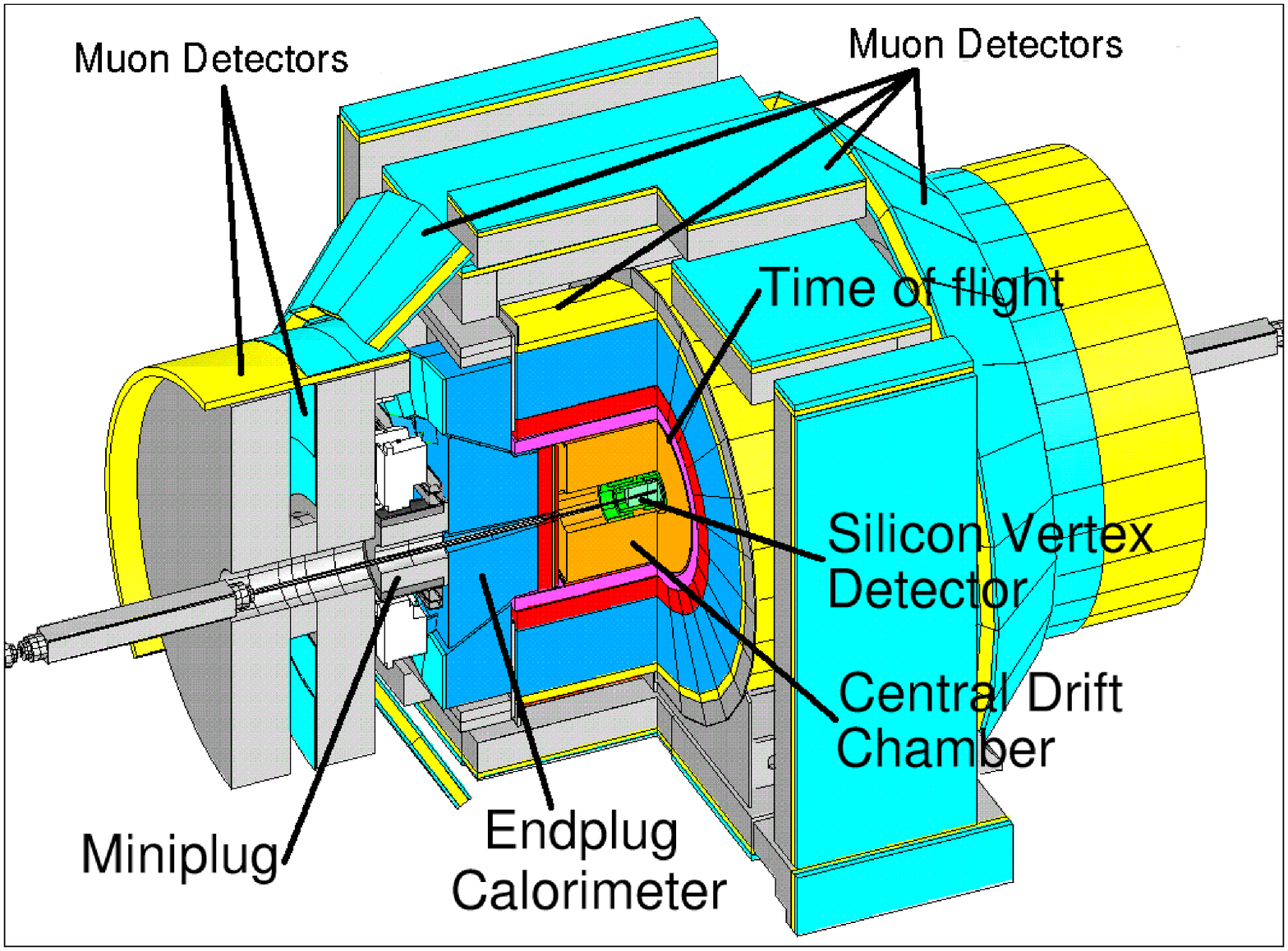,width=0.99\textwidth}
 \end{minipage}
\begin{minipage}[b]{0.45\textwidth}
\psfig{file=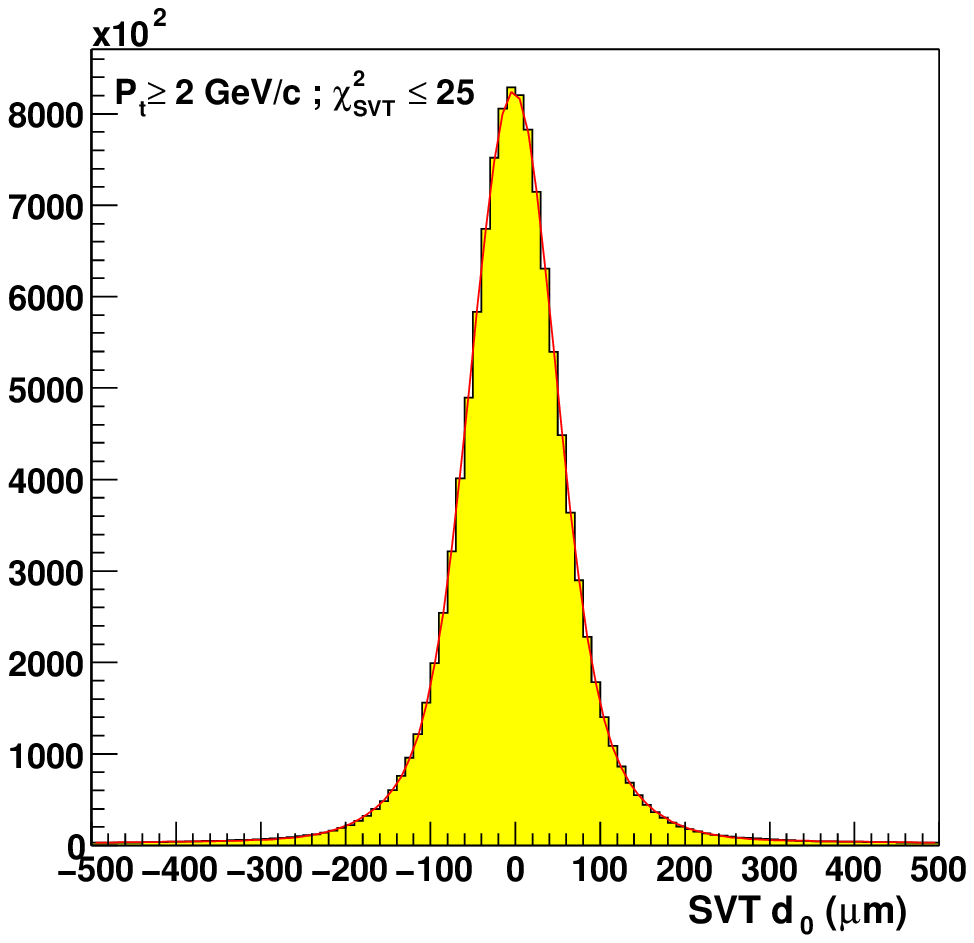,width=0.99\textwidth}
 \end{minipage}
           }
\vspace*{8pt}
\caption{{\bf LEFT:}   CDF~II detector.
         {\bf Right:} Online impact parameter measured by the SVT.
\label{Fig:CDFII}
}
\end{figure}

CDF~II is the product of a major upgrade\cite{CDFUpgradeReport}  for Run~II.
Only a cursory description of the detector,
sketched in Fig.~\ref{Fig:CDFII},
is given here.
The tracking system consists of a Si-strip vertex detector (SVX)\cite{SVX}
comprising 5 layers of double-sided sensors
(axial and stereo coordinates), that
span radii from 2.5-10.6 cm from the beamline.
This is surrounded by the Central Outer Tracker (COT),\cite{COT}
a 3.1 m long  open-cell drift chamber 
spanning radii of 43-132 cm.
Both trackers are immersed in a  1.4~T solen\-oid\-al magnetic field,
enabling measurement of the transverse momenta, $p_T$,
of charged particles.
The SVX is able to resolve the
displacement of  decay vertices ($\vec{x}_{decay}$) of long-lived $c/b$-hadrons 
from the collision point ($\vec{x}_{prim}$), \mbox{and expressed as:}
\begin{equation}
   L_{xy} \equiv {(\vec{x}_{decay}- \vec{x}_{prim})\cdot\vec{p}_T}/{|\vec{p}_T|}.
\label{eq:Lxy}
\end{equation}

Between the COT and solenoid  is a  TOF\cite{TOF} system for particle ID,
supplementing that from $dE/dx$-measurements of the COT.
Outside the solenoid are scintillator-based EM (Pb) 
and then hadronic (Fe) sampling calorimeters,\cite{CDFCal}
with a tower geometry 0.1 wide in pseudorapidity $\eta$, and
$15^\circ$ in azimuth $\phi$ ($5^\circ$ for $|\eta|\!>\!1.2$).
Towers with energy depositions are clustered together in 
$\Delta R \!\equiv\! \sqrt{(\Delta \phi)^2+(\Delta \eta)^2}$
to form ``jets.''
The calorimeter design was  aimed at $W$-physics,
and is not well suited for low-energy $\gamma$-related
spectroscopy.
Beyond the calorimeters are a series of multi-layer muon chambers.\cite{Muons}
The central muon system (CMU) covers $|\eta|\!\leq \! 0.6$,
and additional chambers (CMX) extend the coverage up to  $|\eta|\!\leq\! 1.0$.

The trigger has three Levels.
Important here at L-1 is the track trigger (XFT),\cite{XFT}
which uses COT hits to trigger on tracks above 
a $p_T$-cut, typically 1.5 or 2.0\GeVc. 
At L-1, XFT tracks are matched to 
hits in triggered  $\mu$-chambers.
XFT tracks are also fed to the Si-vertex trigger (SVT)\cite{SVT}
for a L-2 decision on tracks displaced from the collision vertex.
L-3 is a farm of  PC's\cite{CDF-L3} running offline code
using the full event.

Distinctive features of heavy quarks make triggering practical.
Traditionally lepton ($e$, $\mu$) triggers were the backbone 
of  heavy flavor physics at hadron colliders, 
either through semileptonic decays 
or $J/\psi\!\rightarrow\!\mu^+\mu^-$.
Lepton triggers are well established, and we gloss over them
other than to note that the CDF $J/\psi\!\rightarrow\!\mu^+\mu^-$
trigger requires:\cite{CDFbXSec} two opposite-sign XFT tracks with 
$p_T\geq 1.5$ ($2.0$)\GeVc\ which are
matched to CMU (CMX) tracks, and lie in the
mass range from 2.7 to 4.0~\GeVcc.

A dramatic new  capability in Run~II is a displaced track trigger,
thereby keying-in on the long lifetime 
of  weak $c$/$b$  decays.
Originally driven  by $B\!\rightarrow\!\pi\pi$
physics,\cite{SVTMotivat} this trigger
is a tremendous advantage over leptons
for accessing fully reconstructed decays.
For our purposes the ``SVT trigger'' 
is: a L-1 demand for two opposite-sign XFT tracks with $p_T\!\geq\! 2.0\GeVc$,
and scalar sum $p_{T1} \!+\! p_{T2} \!\geq\! 5.5\GeVc$.
At L-2 this seed is used by the SVT to assign $r$-$\phi$ SVX
measurements  and   
find the impact parameter of the tracks, $d_0$,
with respect to the beamline.
An affirmative decision requires that both tracks
have $120\,\mu$m$\leq\! d_0 \!\leq\! 1.0\,$mm,
a transverse opening angle 
of $2^\circ\!\leq\! |\Delta\phi| \!\leq\! 90^\circ$,
and  $L_{xy} \!>\!200\,\mu$m.
The impact parameter distribution is shown in Fig.~\ref{Fig:CDFII}.
The $d_0$-resolution is $50\,\mu$m, which includes $\sim\!30\,\mu$m
from the beam profile.

CDF and the Tevatron are not a universal forum
for spectroscopy, but the strengths brought
to bear nevertheless present important opportunities.
I review  searches for 
possible exotic hadrons in CDF~II data that
were recorded from February 2002 until 
as recently as August 2004.

\section{The Pentaquark Revolution}

\begin{figure}[t]
\centerline{\psfig{file=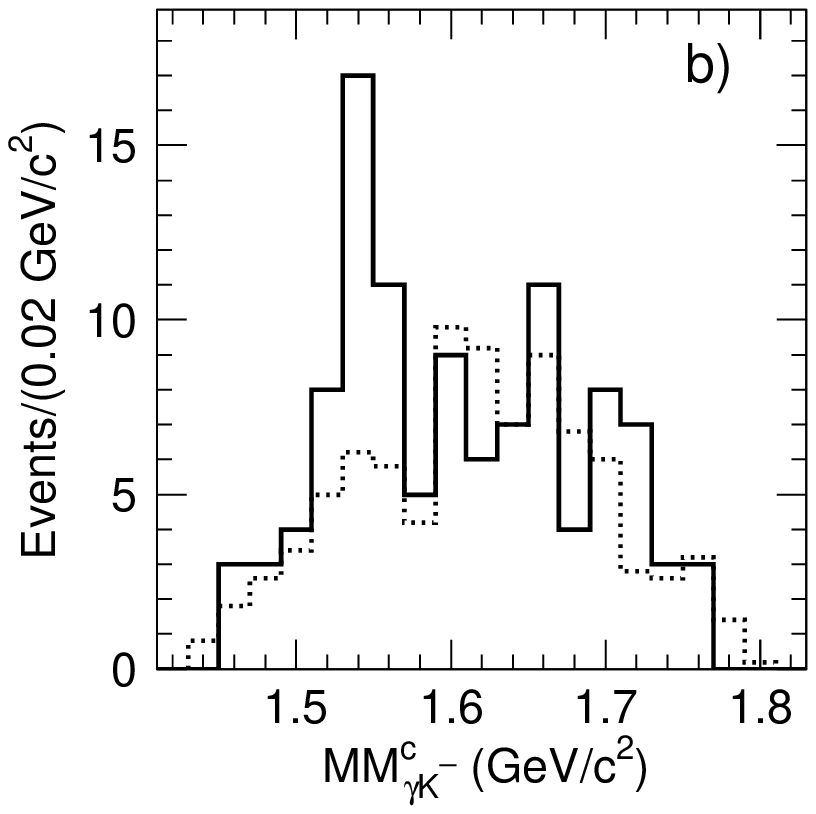,width=2.0in}~~~~\psfig{file=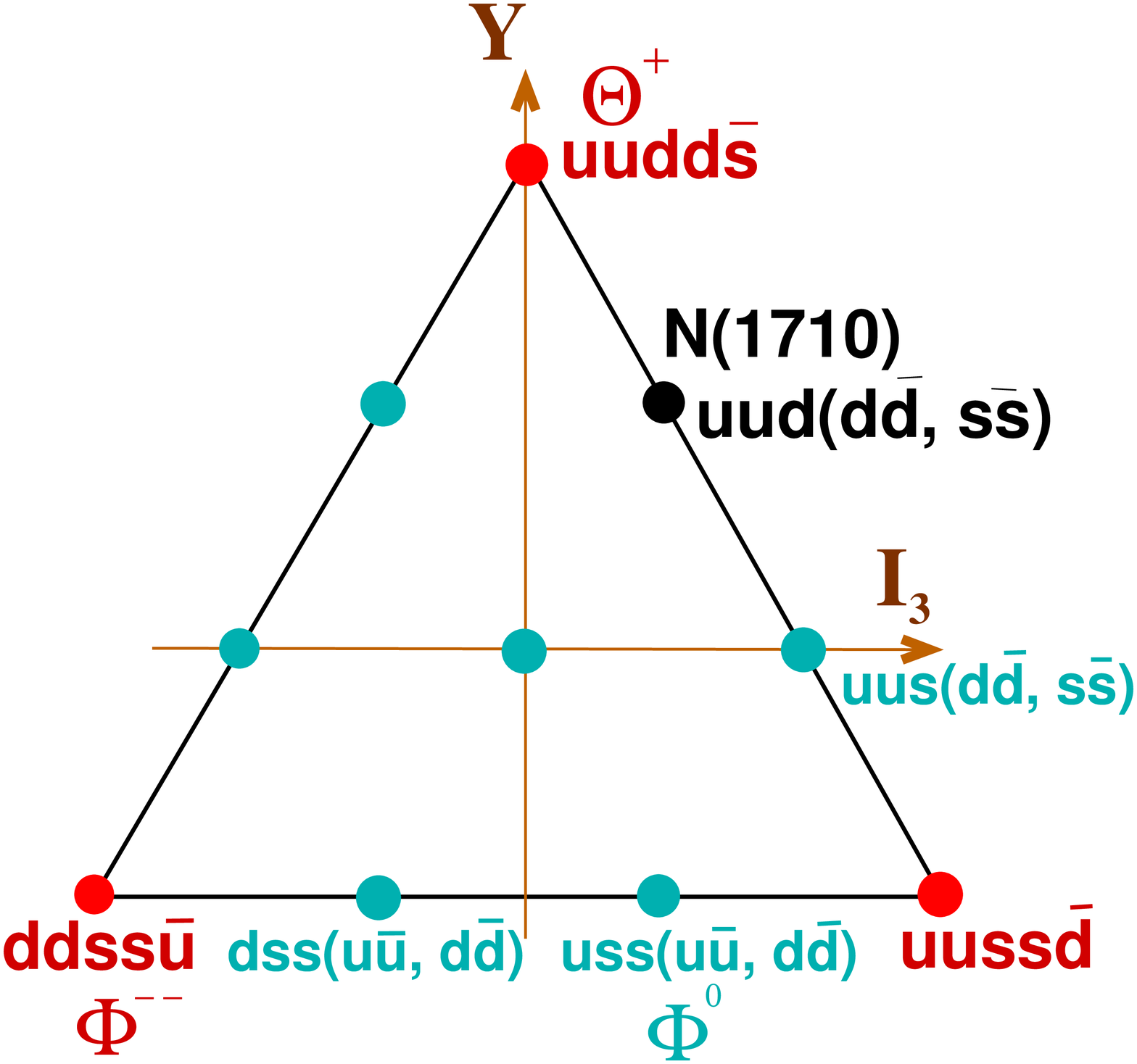,width=1.9in}
           }
\caption{{\bf LEFT:} The `plot that launched a thousand preprints,' the  LEPS  $\Theta^+$ 
         signal  in the $nK^+$ mass 
         [missing mass recoiling against $\gamma K^-$]
         spectra (solid line), and a $pK^+$ control distribution (dotted line).
         [{\it Figure reprinted with permission from T. Nakano {\it et al.},  Phys. Rev. Lett. {\bf 91}, 012002 (2003).
         Copyright 2003 by the American Physical Society.}]
         {\bf RIGHT:} The   baryon anti-decuplet 
         of Diakonov {\it et al.}\protect\cite{Diakonov}
         Note that only the corner states are manifestly exotic.
\label{Fig:ThetaLEPS}
}
\end{figure}

After decades of disappointments, triumph seemed to be at hand in January 2003:
the LEPS Collaboration reported a resonance, 
now called   $\Theta^+$, decaying to $nK^+$ 
at  $1540\!\pm\!10\!\MeVcc$ (Fig.~\ref{Fig:ThetaLEPS}) 
in photoproduction ($E_\gamma\!\sim\!1.5$-$2.4\GeV$)
off of neutrons.\cite{LEPS5Q}
With strangeness $+1$ the   $\Theta^+$ is manifestly exotic for a baryon.
The  minimal quark content is  $uudd\bar{s}$,
like the old $Z$-states, but dramatically narrower: 
$\Gamma_\Theta\!<\!25\MeVcc$.

The LEPS search was prompted by the  1997 predictions
of Diakonov, Petrov, and Polyakov\cite{Diakonov} 
for a  light, $\sim\!1530\MeVcc$,   
and remarkably narrow, $\lesssim\! 15\MeV$, member of an  exotic baryon anti-decuplet 
anchored by the $N(1710)$ resonance (Fig.~\ref{Fig:ThetaLEPS}).
The authors motivated the LEPS and DIANA collaborations to conduct a search.\cite{LASS}
After a couple of years both groups independently isolated a signal,
although DIANA\cite{DIANA} reported some months after LEPS.
DIANA's signal was in the isospin analog $pK^0_S$ 
at  $1539\!\pm\!2\MeVcc$ in  $K^+$Xe data ($p_K\!<\!750\MeVc$).
While  $pK^0_{S} $ has indefinite $s/\bar{s}$ content,
the incident $K^+$ is
strong  evidence for $+1$ strangeness.

\begin{figure}[t]
\begin{minipage}[b]{0.64\textwidth}
\centerline{\psfig{file=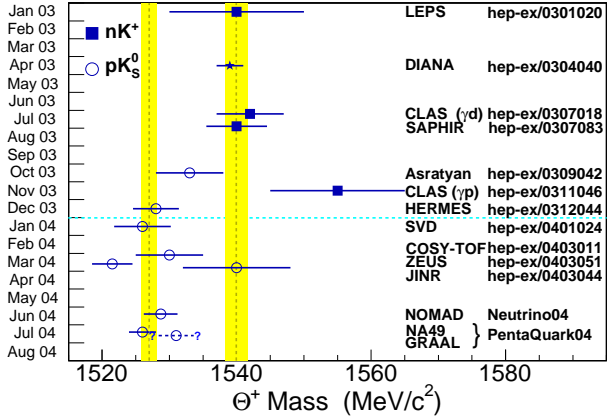,width=\textwidth}
           }
\end{minipage}
\hspace*{-0.15cm}\
\begin{minipage}[b]{0.34\textwidth}
\caption{
  Time-line of   $\Theta^+$ reports. 
   Dates are from the listed {\tt hep-ex} postings for published results, 
   and conference dates for unpublished\protect\cite{NOMAD}$^-$\protect\cite{GRAALtheta} 
   sightings.
   Vertical bands show
   the separate  $pK^0_S$  and $nK^+$  mass averages and error bands.
   GRAAL quoted no error, and is excluded from the average.
   The DIANA result is grouped with  ``$nK^+$'' because
   it is flavor specific even though they observe a $K^0_S$ in the final 
   state. 
\label{Fig:ThetaOthers}
}
\vspace*{20pt}
\end{minipage}
\end{figure}

An avalanche of confirmations ensued (Fig.~\ref{Fig:ThetaOthers}),
although individually results are  only low to moderate significance.
Many are  $pK^0_S $ signals, and thus are evidence
for an exotic baryon only by virtue of their consistency in mass 
with  $nK^+$ observations.

Placing the $\Theta^+$ in an anti-decuplet is not the only option,\cite{ThetaOptions}
but failure to find a  $\Theta^{++}$ partner\cite{CLASThetapp}$^{\rm-}$\cite{BaBarThetapp}
supports $\Theta^+$ as an isosinglet.  
Finding  related states  is key, such as excited states\cite{CLAS2Theta},
but perhaps more telling: other members of the multiplet, {\it e.g.}
the exotic $ddss\bar{u}$  (Fig.~\ref{Fig:ThetaLEPS}).\cite{STARN5}
In the fall of 2003 NA49  ($pp$ at $\sqrt{s}\!=\!17.2\GeV$)
reported -2 strangeness baryons at $1862\!\pm\!2\MeVcc$ in $\Xi^-\pi^-$, 
as well as indications of a partner in  $\Xi^-\pi^+$.\cite{NA49}
The  $\Xi^-\pi^-$ is necessarily exotic and is interpreted as 
$ddss\bar{u}$, the $\Phi^{--}(1860)$ [formerly $\Xi^{--}_{3/2}$]; 
and the other as $udss\bar{d}$, the $\Phi^{0}(1860)$  [or $\Xi^{0}_{3/2}$].
To set the scale of the signal,  2191 charged            % a review says 2664
$\Xi$'s were used to obtain  67.5  $\Phi^{--,0}$  
candidates---quite a plentiful yield of $\sim\!3\%$ 
of  $\Xi$'s---over a background of 76.5.
NA49's observation would be an important first step in filling
in the anti-deculplet, although the chiral model predicted
a heavier mass, around $2070\MeVcc$.\cite{Diakonov}

Pentaquark sightings advanced to the charm sector\cite{OldCharmQ5} 
in March 2004.
At a DESY seminar  H1 reported\cite{H1} a narrow ($\sigma\!\sim\!\!12\!\MeVcc$) structure
at  $3099 \!\pm\! 3  \!\pm\! 5\!\MeVcc$ in  $pD^{*-}$
and interpreted it as the charm analog of the $\Theta^+$,
{\it i.e.}  $uudd\bar{c}$. 
With  $75\ipb$ of Deep Inelastic data ($ep$ collisions), they selected
3400 $D^{*-}$'s                      % from R Stamen's slide from PentaQuark04 at Spring-8 July04
after  $dE/dx$ particle ID, yielding $50.6\!\pm\!11.2$  $\Theta^{0}_c$'s.
Another analysis with 4900  $D^{*-}$'s from photoproduc\-tion
reproduced the signal---albeit with higher backgrounds---for
$43\!\pm\!14$ $\Theta^{0}_c$'s.
At the same seminar, however,
ZEUS reported\cite{ZEUSseminar}  no signal  in $126\ipb$ with almost 
43k inclusive $D^{*-}$'s,        %quoted from  R Stamen's slide from PentaQuark04 at Spring-8 July04
                                 %was original analysis less???...yes, see same in Gladilin DESY slides
or  $\sim\!10$k in DIS data.
ZEUS expects a distinct signal if the  $\Theta_c^0$ is a few tenths of a percent   %$0.1$'s\%
of $D^{*-}$'s,  whereas the raw  H1 yield per  $D^{*-}$  was $\sim\!1\%$.
%%% +++> BUT: formally the ZEUS search had different cuts and sensitive to different part of
%%%      phase space, or so it is claimed by some....a alternate search was done(?), which
%%%      still failed to find Thetac???       (see hep-ex/0502018

%chiral soliton prediction of  $\Theta^0_c$ around 2710\cite{JaffeWilczek}
%%3099 is heavier and goes to D*---maybe  $\Theta^{*0}_c$?

Doubt is not limited to the $\Theta_c^0$.
The $\Phi$ was quickly {challenged by old  WA89  data,}
a high-statistics hyperon experiment.\cite{WA89}
A broader survey concluded that the  $\Phi$ was
``at least partially inconsistent''\cite{FischerWenig} 
with a large amount of earlier $\Xi$ data.
And, despite  many $\Theta^+$ claims, 
skepticism surfaced here too, including the  spectre of kinematic reflec\-tions.\cite{ThetaReflec}
As widely noted,  the   $nK^+$ and  $pK^0_S $
claims do not share a consistent mass (Fig.~\ref{Fig:ThetaOthers}).
Also,  the absence  of  $\Theta^+$ in prior $KN$ data 
limit $\Gamma_\Theta\!\lesssim\! 1\!\MeVcc$,\cite{ThetaWidthLim} 
too narrow to easily explain.\cite{NarrowHypoth}
Then, in early 2004,  null $\Theta^+$ searches started surfacing.

The Tevatron is an important venue for pentaquark
searches by virtue of large hadronic rates
and  access to all  flavors.
Conceivably the Tevatron  might not be conducive to
the manufacture  of complex and fragile quark systems,
but if  so, this too would be interesting.
Preliminary results of  CDF searches
are, so far, all negative.

\subsection{The  $\Theta^+(1540)$  at CDF$\,$\protect\cite{CDFThetaC}}

\begin{figure}[t]
\centerline{\psfig{file=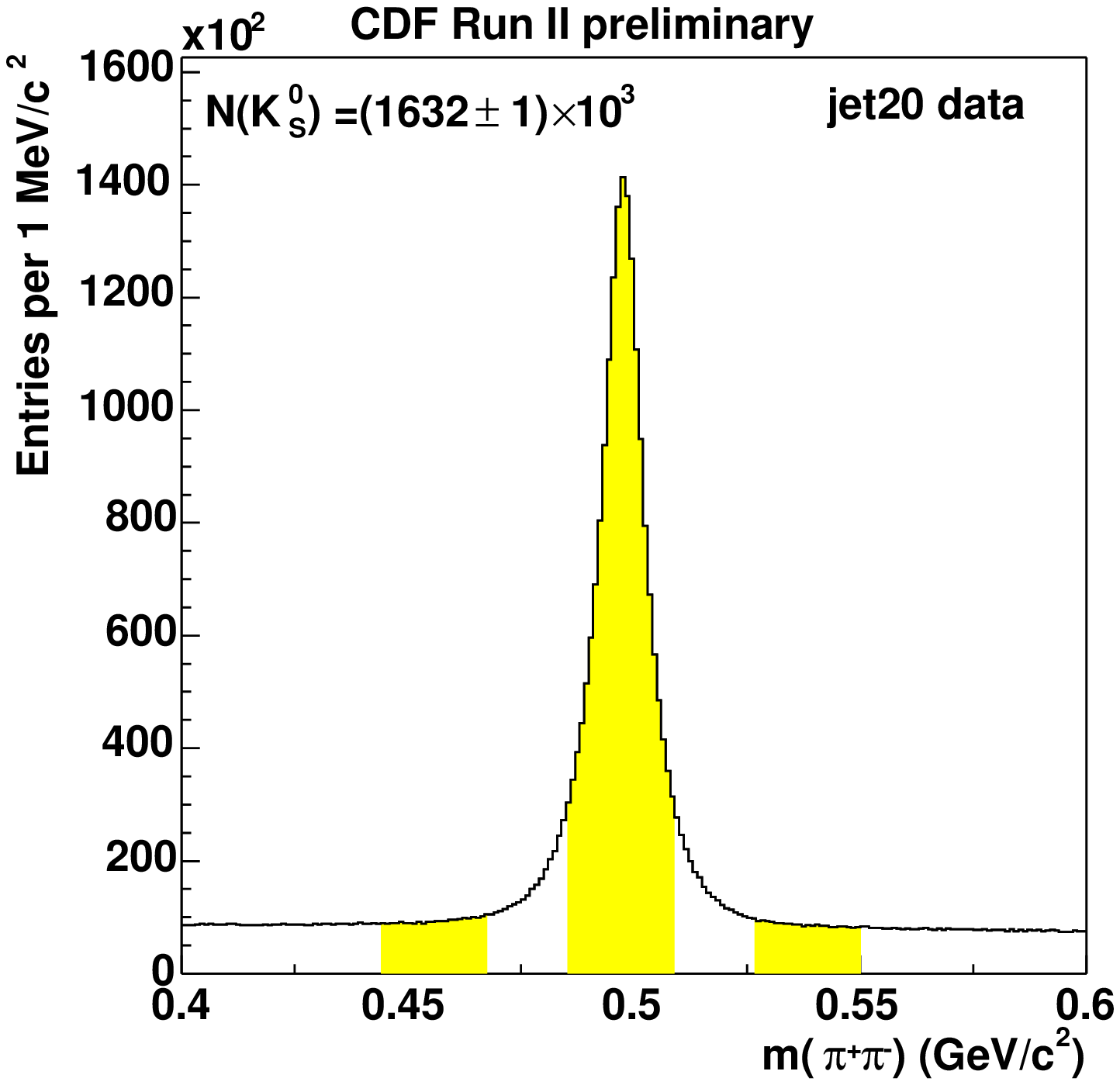,width=1.7in}\psfig{file=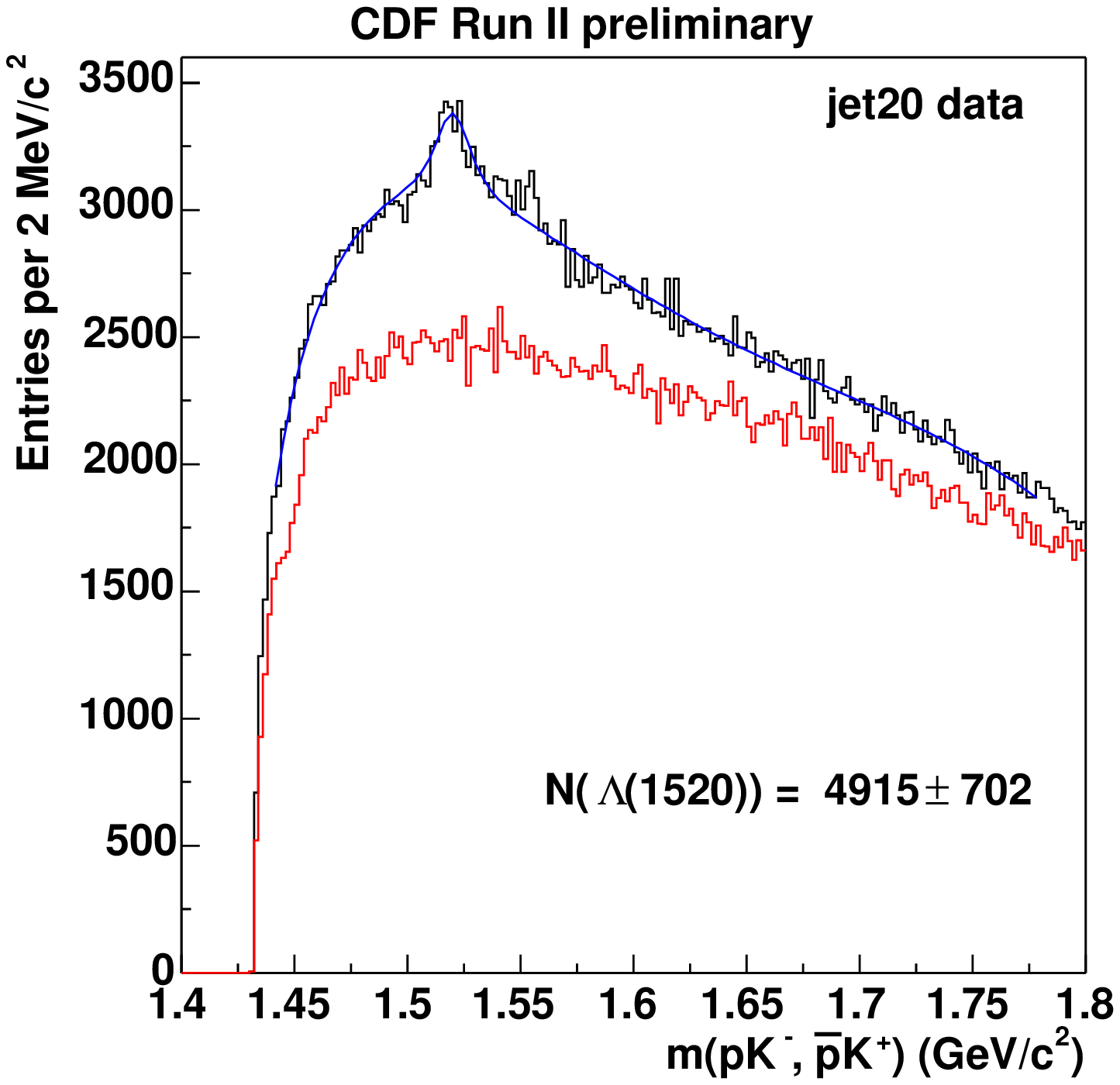,width=1.7in}
\psfig{file=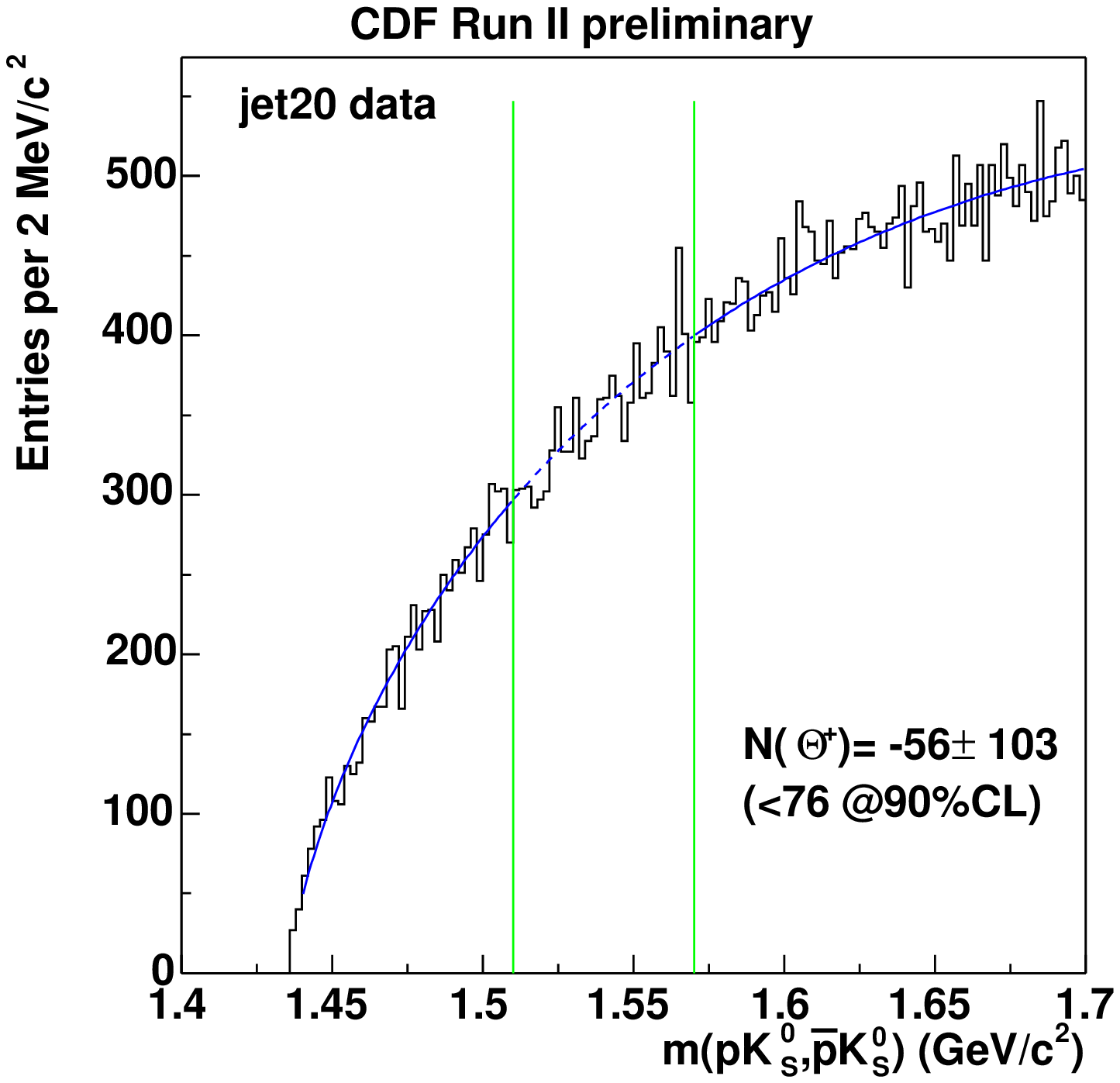,width=1.7in}
           }
\vspace*{8pt}
\caption{{\bf LEFT:}  The Jet-20 $K^0_s$ sample used for a $\Theta^+$ search.
         {\bf Center:} The  $pK^-$ spectrum showing the $\Lambda(1520)$ reference signal
         (upper curve), and same-sign $Kp$  (lower curve).
         {\bf RIGHT:} The $pK^0_s$ mass distribution from the  Jet-20 sample.
         Vertical lines mark the  $\Theta^+$ search window.
\label{Fig:ThetaCDF}
}
\end{figure}

As in many detectors, neutron detection is not viable 
in CDF, and  $\Theta^+(1540) \!\rightarrow\! pK^0_S $ is searched for.
No CDF trigger preferentially selects  these decays.
Because   $\Theta^+$ production 
is not understood, two contrasting types of events are used:
\mbox{soft inelastic}
collisions with minimal trigger requirements, 
a.k.a. ``Min-Bias'' events;
and hard-scatters which produce jets---at 
least one that passes a  20\GeV\ calorimeter jet trigger.
The two samples respectively  consist of 22.2M and 14.2M events,
but as these are very large cross-section triggers the integrated
luminosities are only  $0.37\inb$ and  $0.36\ipb$.
Even so,  a large sample of
$0.67$M and $1.6$M  $K^0_S$'s are available
in these respective samples.
The  $K^0_S$'s from the  Jet-20 sample are shown in Fig.~\ref{Fig:ThetaCDF}.

$\Theta^+$ candidates are formed by adding to $K^0_S$'s
a charged track,  which must be identified 
by TOF  within at least $2\sigma$ of a proton.
This effectively restricts the protons to momenta from 0.5-2.1\GeVc.
The selection, as well as the use of the TOF,
are monitored by reference
signals: $\phi\!\rightarrow\! K^+K^-$,  
$\Lambda(1520)\!\rightarrow\! K^-$p (Fig.~\ref{Fig:ThetaCDF}), and
  $K^{*+}\!${$\rightarrow$}{$ K^0_S\pi^+$}.
The  $pK^0_S$ mass distribution for Jet-20 data
is shown in Fig.~\ref{Fig:ThetaCDF}, the Min-Bias distribution
is similar but with about $1/3$ the statistics.
In both cases no signal is apparent around 1540\MeVcc.
Counting events in the signal region of 1510 to 
1570\MeVcc\  (vertical bars on the plot) and using $K^0_S$ sidebands to subtract
back\-ground, the fitted $\Theta^+$ ``excess'' is $18\pm56$ Jet-20 candidates
and $-56\pm103$ for Min-Bias, or:
not more than 76 (89)  $\Theta^+$ candidates
\mbox{for Jet-20 (Min-Bias) at 90\% CL.}

Incisive comparisons across the diverse $\Theta^+$ reports are problematic
as we lack theoretical bridges to connect them.
The only signal in a environment  analogous to CDF's comes from HERA,
a high-energy $ep$-collider.
There, based on $0.87$M $K^0_S$'s, ZEUS 
sees  $221\!\pm\!48$  $\Theta^+$'s.\cite{ZEUStheta}
In terms of raw  $K^0_S$'s, CDF should have a fair signal.

%-------------------------------------------------------------
\subsection{The $\Phi(1860)$ at CDF$\,$\protect\cite{CDFXi}}

As in the  $\Theta^+$ search, no CDF trigger explicitly
keys on $\Phi(1860)\!\rightarrow\! \Xi\pi$.
Two complementary triggers are  used:
Jet-20 again, and 220\ipb\ of SVT triggers.
Displaced tracks {\it are} produced in $\Xi$ decays, 
but these are {\it too far} away  
for the SVT to trigger.

\begin{figure}[t]
\centerline{\psfig{file=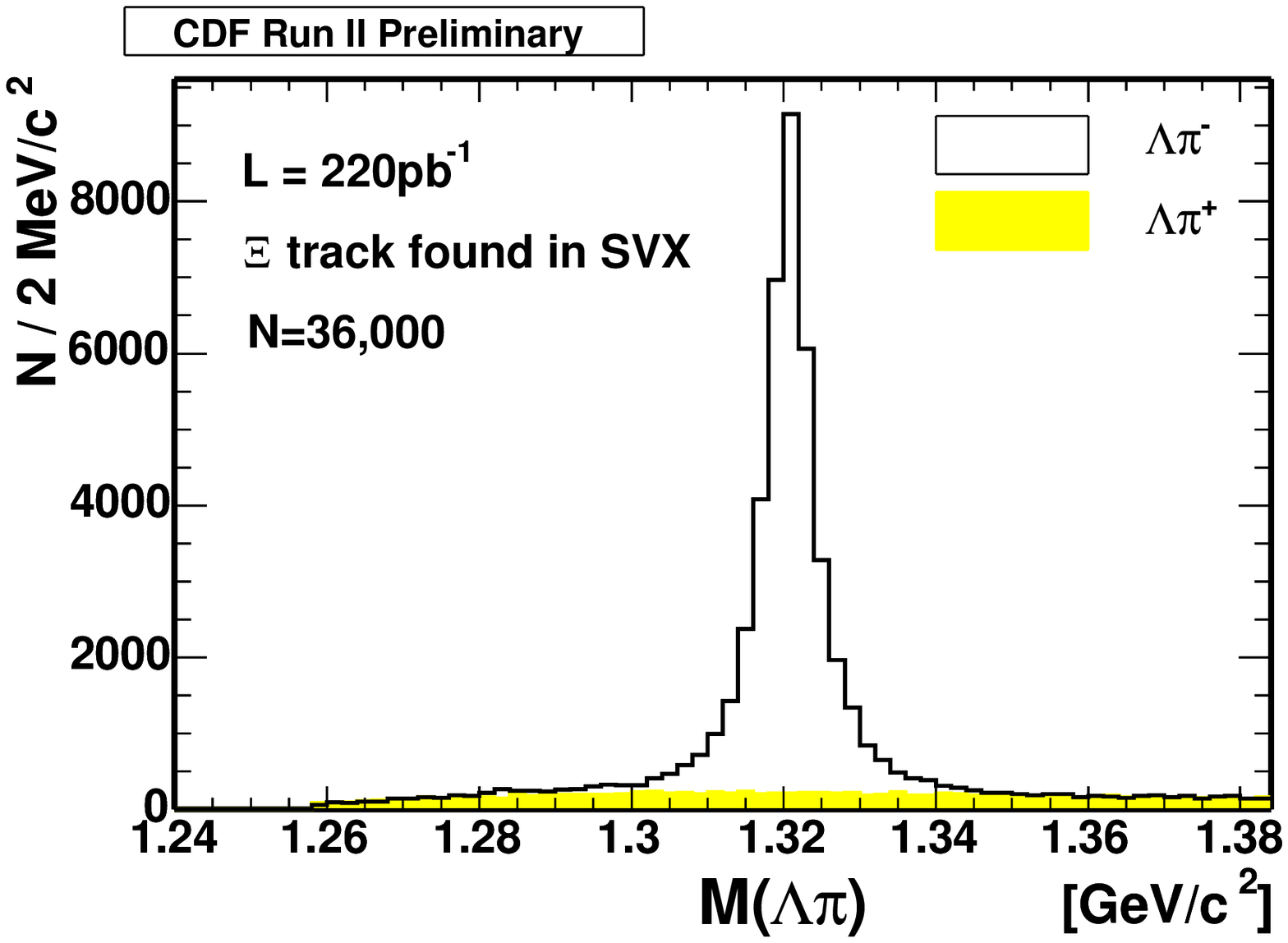,width=2.5in}\psfig{file=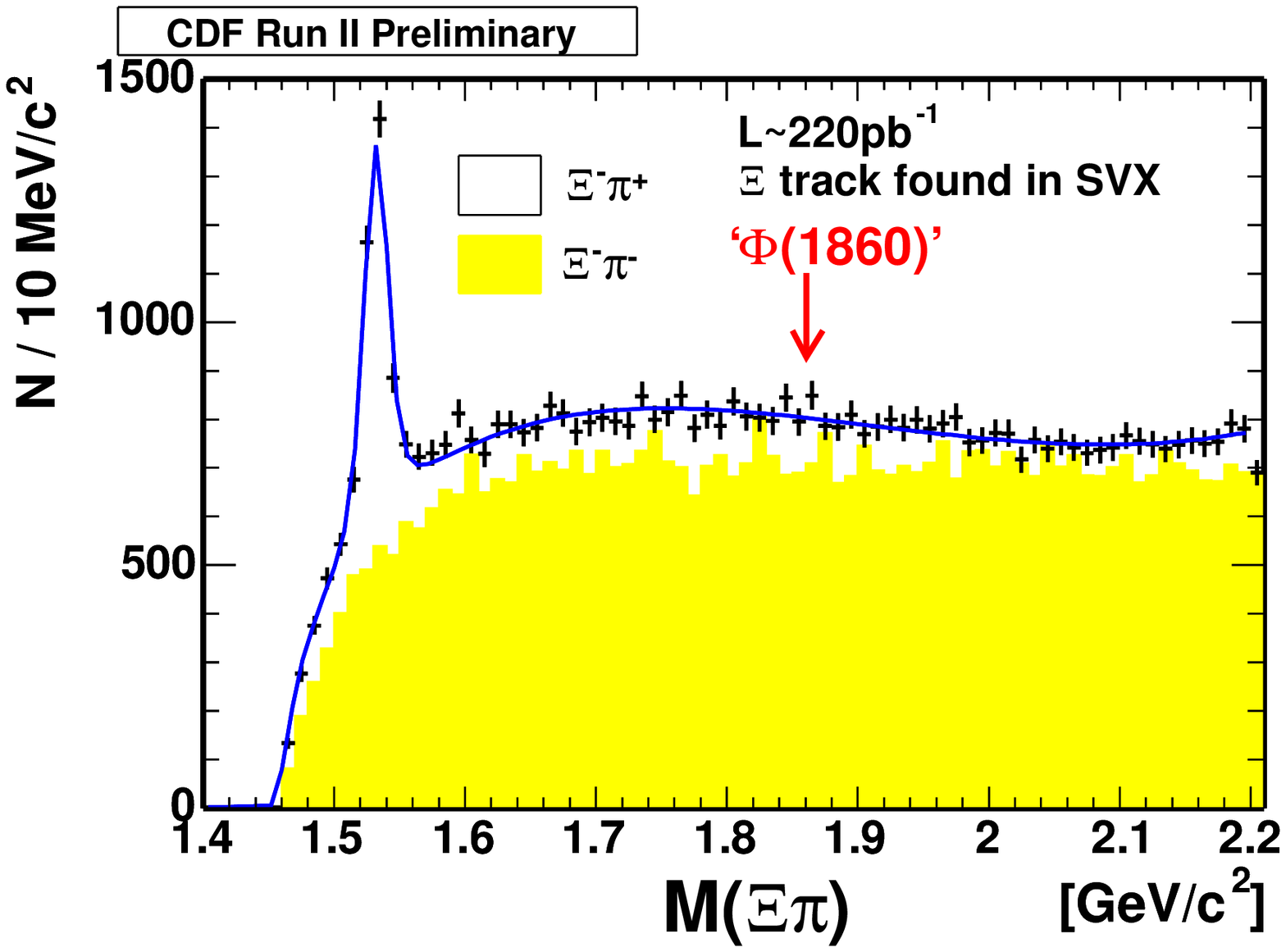,width=2.5in}
           }
\vspace*{8pt}
\caption{{\bf LEFT:}   $\Lambda \pi$ mass spectrum 
                        (and ``wrong-sign''  $\Lambda^0\pi^+$ background) 
                       for candidates where a  $\Xi$ track was found in SVX (SVT trigger sample).
         {\bf RIGHT:}  The $\Xi^-\pi^+$  (points) and  $\Xi^-\pi^-$ (shaded histogram)
                       mass distributions. The arrow marks the  $\Phi(1860)$-mass 
                       reported by NA49.
\label{Fig:XiCDF}
}
\end{figure}

Reconstructing  $\Lambda^0\!\rightarrow\! p\pi^-$ is straightforward.
More subtle is $\Xi^-\!\rightarrow\! \Lambda^0\pi^-$.
The  $\Xi$ is {\it charged},
with  almost half the $\Lambda^0$ lifetime, 
and will  often leave hits in the SVX.
A specialized reconstruction is used whereby displaced pions are
added to $\Lambda^0$'s to form  $\Xi^-$ candidates, and potential  $\Xi^-$ SVX-hits
are sought for a full  $\Xi^-$ track fit.
In the SVT data $\sim\!36$k   $\Xi^-$'s are 
cleanly reconstructed (Fig.~\ref{Fig:XiCDF}), and $\sim\!5$k in Jet-20.

A  $\Phi\!\rightarrow\! \Xi\pi$ search has a good control signal 
in  $\Xi^0(1530)\!\rightarrow\! \Xi^-\pi^+$, of which there 
are  $2,200\!\pm\! 100$ in the  SVT data, and $390\!\pm\!30$ in Jet-20.
The $\Xi^0(1530)$ is prominent in the  $\Xi^-\pi^+$ distribution
of Fig.~\ref{Fig:XiCDF}, but no other structures are seen there,
or, in the $\Xi^-\pi^-$ masses.
The limit on the number of $\Phi$  candidates
is expressed relative to the raw number of observed $\Xi^0(1530)$'s.
Imposing an 1860-resonance fit in the $\Xi^-\pi^-$ SVT data 
yields $-54\!\pm\!47$ candidates, or a 90\% CL limit of 51 $\Phi^{--}(1860)$'s.
This translates into the limit $R^{--}\!\equiv\! N(\Phi^{--})/N(1530)\!<\!0.03$ at  90\% CL.
Similarly,  $R^{0} \!<\! 0.06$, 
or combining both channels  $R^{Tot} \!<\! 0.07$  at  90\%~CL.
The limit on the ratio is not corrected for acceptance,
but this is not expected to be a large effect.
For the Jet-20 samples the limits are   $R^{--}_{20} < 0.07$, $R^0_{20} < 0.06$, 
and  $R^{Tot}_{20} < 0.09$.%  at  90\% CL.

CDF's raw sensitivity compares well with NA49's.
CDF's $\Xi^-$ sample is more than $10\times$ the $\sim\!2000$ $\Xi^-$'s of  NA49.
With a looser selection\cite{NA49More} that is more sensitive to the $\Xi(1530)$,
the NA49  $\Phi$ yield appears  to be $\sim\!50\%$ of $\Xi(1530)$,
well above CDF's $<\!10\%$ limits.
Note   that the $\Xi(1530)/\Xi$ ratio is similar for both experiments.

%==========================================================================================
%==========================================================================================
\subsection{Charm Pentaquarks  at CDF$\,$\protect\cite{CDFThetaC,CDFThetaCMore}}

\begin{figure}[t]
\centerline{\psfig{file=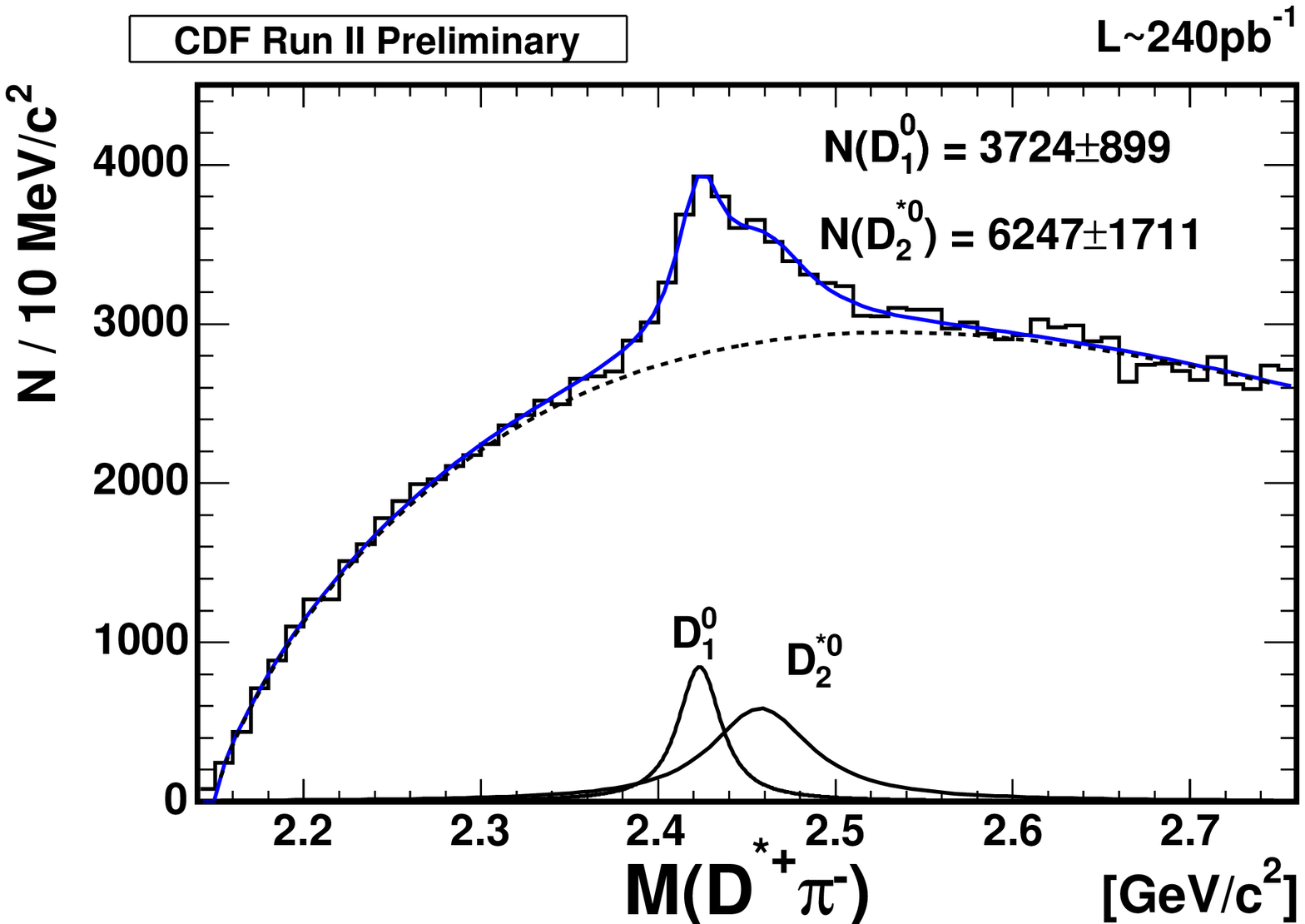,width=2.1in}\psfig{file=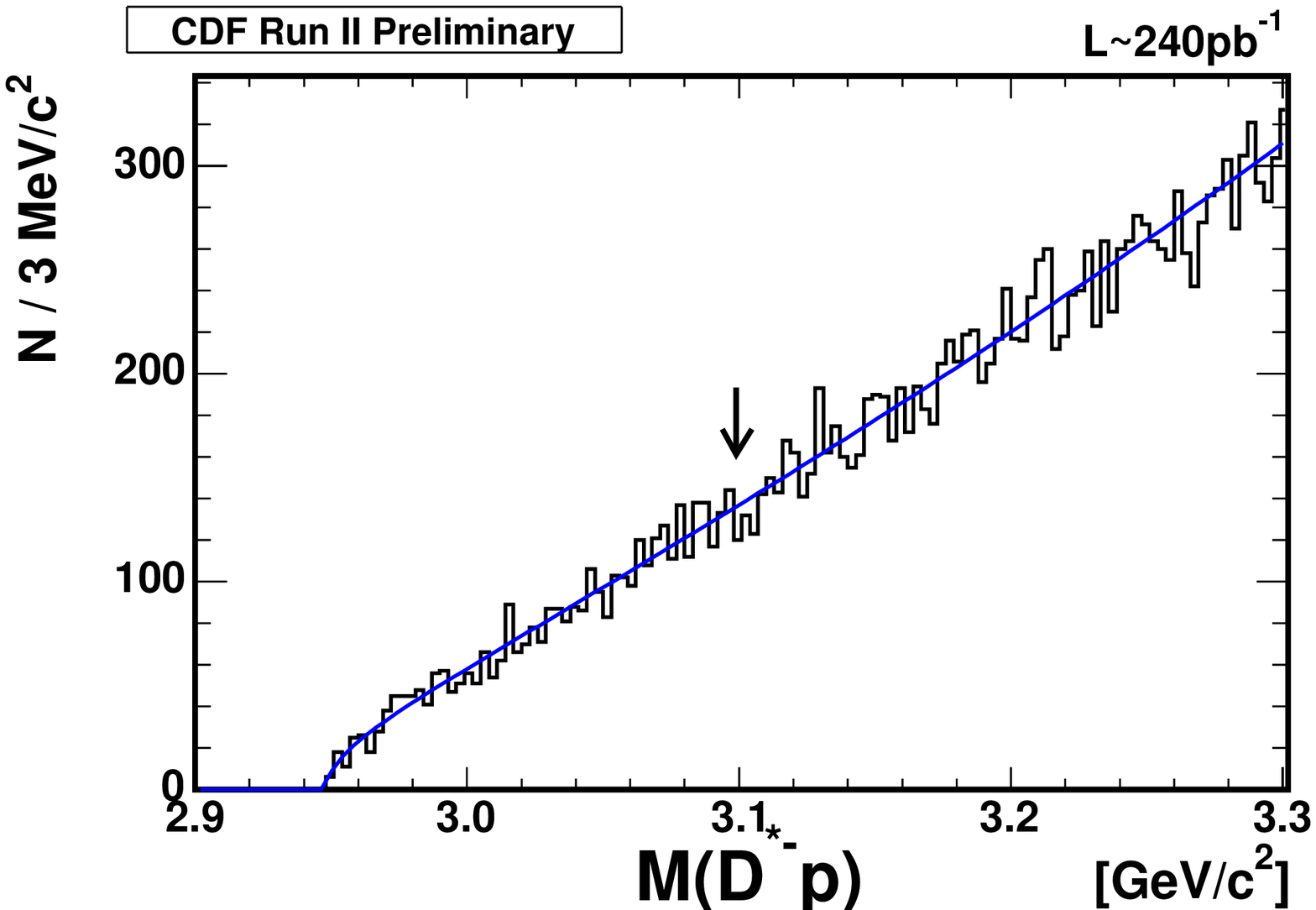,width=2.1in}} 
\centerline{\psfig{file=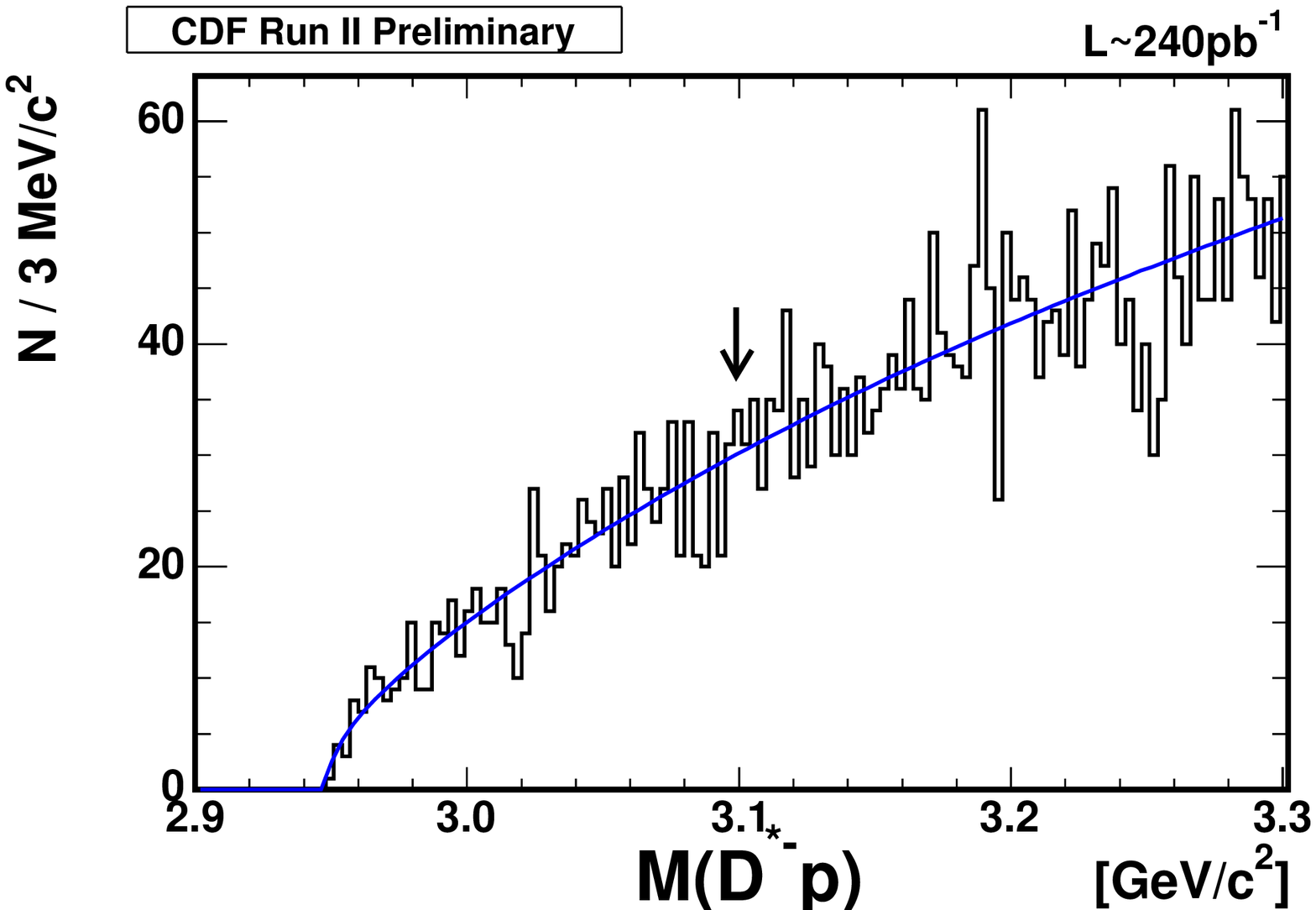,width=2.1in}\psfig{file=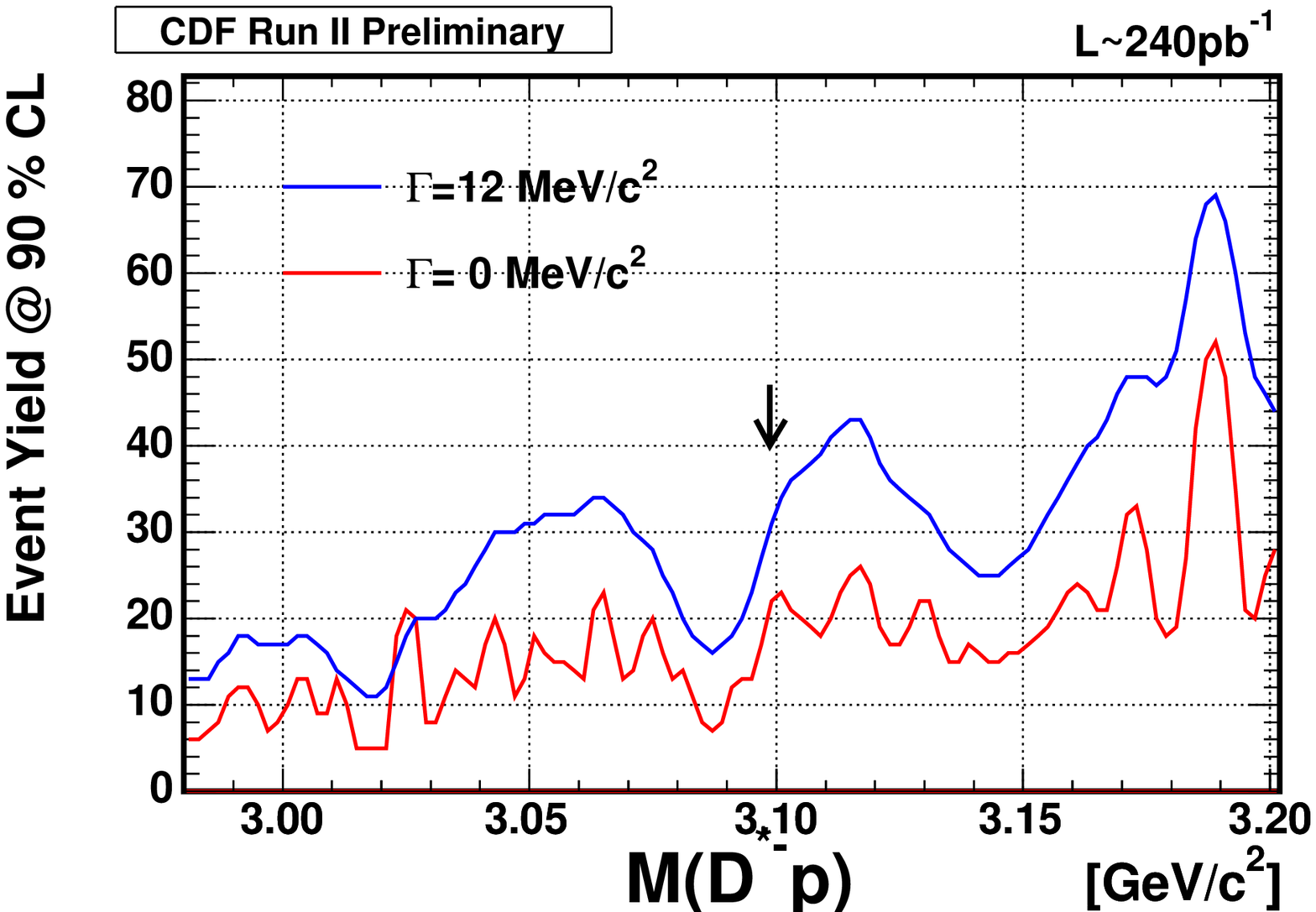,width=2.1in}}
\vspace*{8pt}
\caption{{\bf TOP-LEFT:} prompt $D^{*+}\pi^-$ mass spectrum, where overlapping 
                     $D^0_1(2420)$ and  $D^0_2(2460)$ are clearly visible.
         {\bf TOP-RIGHT:} $pD^{*-}$ masses for the prompt sample
                      (no PID).
         {\bf BOTTOM-LEFT:} $pD^{*-}$ masses for the long-lived  sample
                      (no PID).
         {\bf BOTTOM-RIGHT:} 90\% upper limit as a function of mass
                      in the long-lived sample
                      for two $\Theta_c$ widths.
                      The arrows mark H1's  $\Theta^0_c$  mass.
\label{Fig:ThetacCDF}
} 
\end{figure}

An important distinction for a $\Theta_c^0(3100)\!\rightarrow\! pD^{*-}$ search in CDF,\cite{CDFThetaC}
versus those for $\Theta^+$ and $\Phi$, is that
the SVT trigger is aimed at $D$ decays.
In 240\ipb\ of data CDF has $\sim\!3$M $D^0\!\rightarrow\! K^-\pi^+$ decays.
Adding a  $p_T\!>\!400\MeVc$ pion yields $\sim\!0.5$M $D^{*+}$.
Adding another such pion leads to reference states $D^0_1(2420)$ or  $D^0_2(2460)$.
These are clearly seen in  Fig.~\ref{Fig:ThetacCDF},
even though partially overlapping due to their large natural widths.
Alternatively,  assigning a proton to the latter track produces
$\Theta^0_c$ candidates.

\begin{figure}[t]
\centerline{\psfig{file=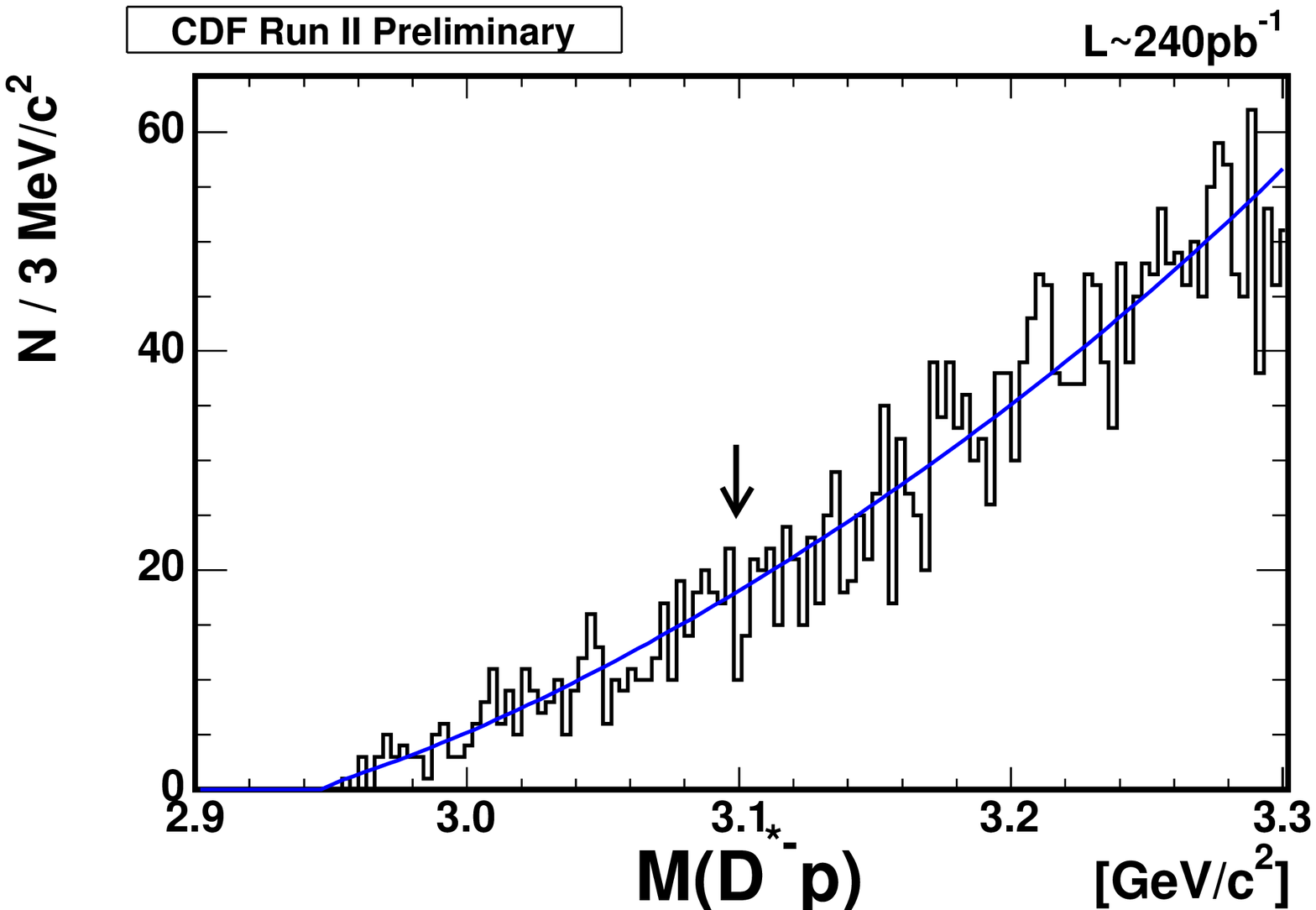,width=2.20in}            
            \psfig{file=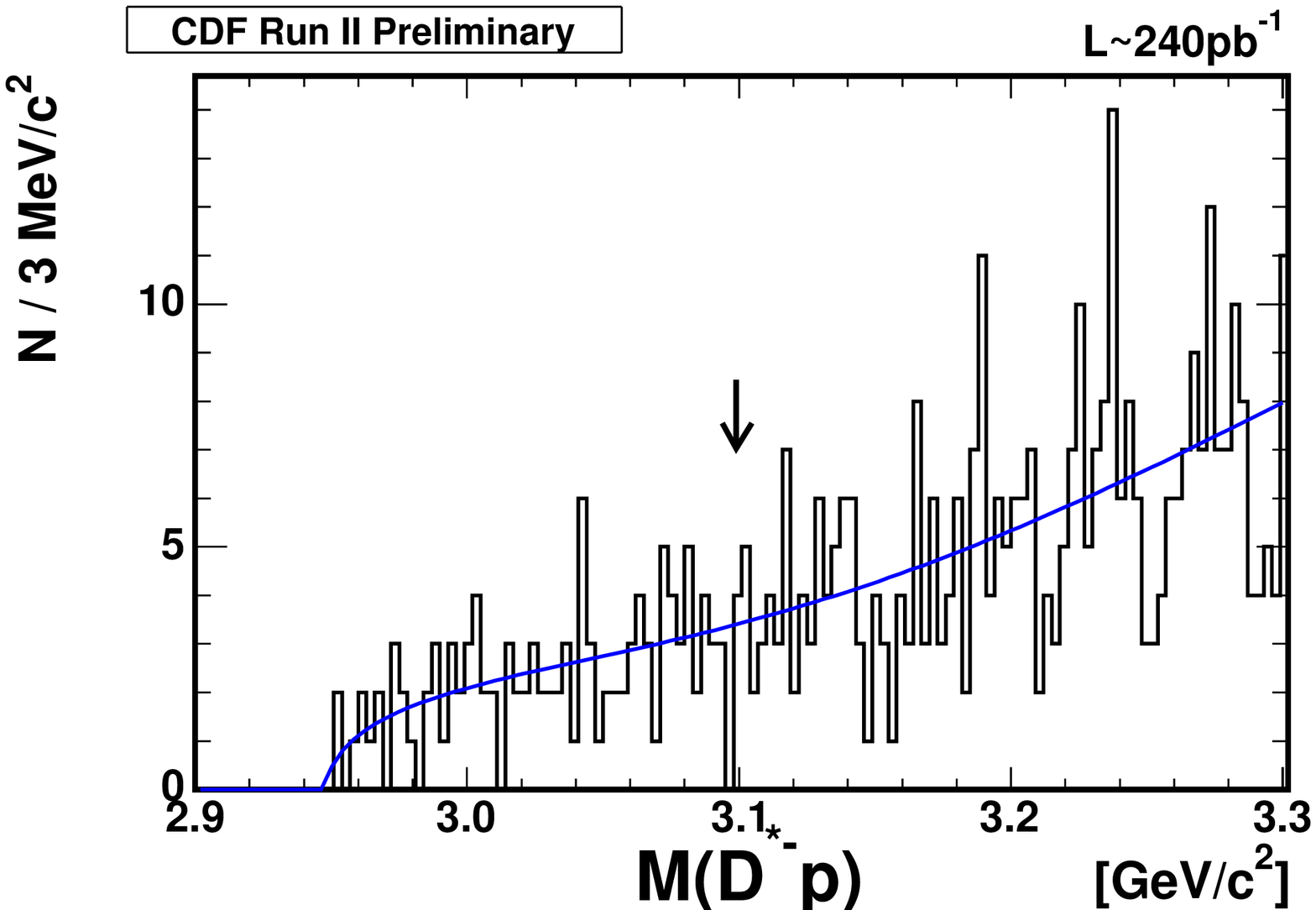,width=2.20in}      
           }
\vspace*{8pt}
\caption{ The $pD^{*-}$  mass spectra for prompt (left) and long-lived (right) 
           selections with  $p$-ID. 
\label{Fig:ThetacCDFwID}
}
\end{figure}

Since  $\Theta^0_c$'s might arise via long-lived $b$-decays,
or prompt production,   CDF distinguishes prompt
($|L_{xy}|\!<\!400\,\mu$m \& $|L_{xy}|/\sigma_{L_{xy}}\!<\! 3$)
and  long-lived   ($L_{xy}\!>\!400\,\mu$m \& $L_{xy}/\sigma_{L_{xy}}\!>\! 3$) samples.
No $D^{*-}p$  excess is seen at $\sim\!3099\MeVcc$ in 
either  case  (Fig.~\ref{Fig:ThetacCDF}).
Mass dependent 90\% CL  limits are  shown in Fig.~\ref{Fig:ThetacCDF} for the ``$b$-sample.''
In the signal region, $3100\!\pm\!18\!\MeVcc$, the maximum limit is 43 $\Theta^0_c$'s 
($\Gamma_\Theta\!=\!12\MeVcc$), or 71 for prompt.
Sensitivity is improved by particle ID.
Protons were identified using
a likelihood ratio ($e$, $\mu$, $\pi$, $K$, and $p$ hypotheses)
combining $dE/dx$ and TOF measurements, with the
cut optimized on 2.5k  $\Lambda_c\!\rightarrow\!pK^-\pi^+$ decays.
The new $pD^{*-}$ plots are in Fig.~\ref{Fig:ThetacCDFwID}.
The maximum  yields become 32 prompt and 15 long-lived $\Theta^0_c$'s,
although part
of this reduction is due to the efficiency ($\sim\!70$\%)
of the proton cut.

\begin{figure}[t]
\centerline{\psfig{file=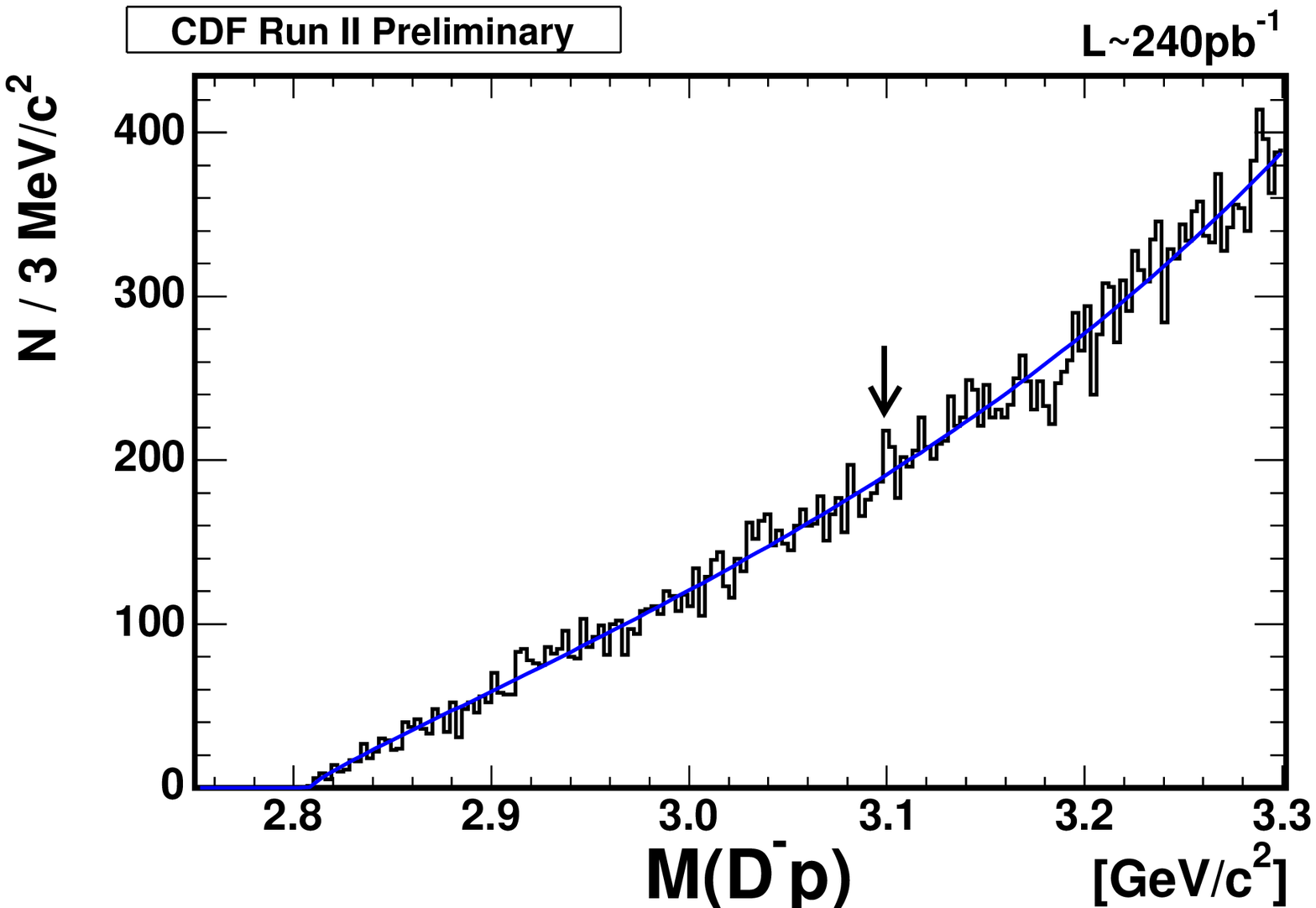,width=2.3in}            
            \psfig{file=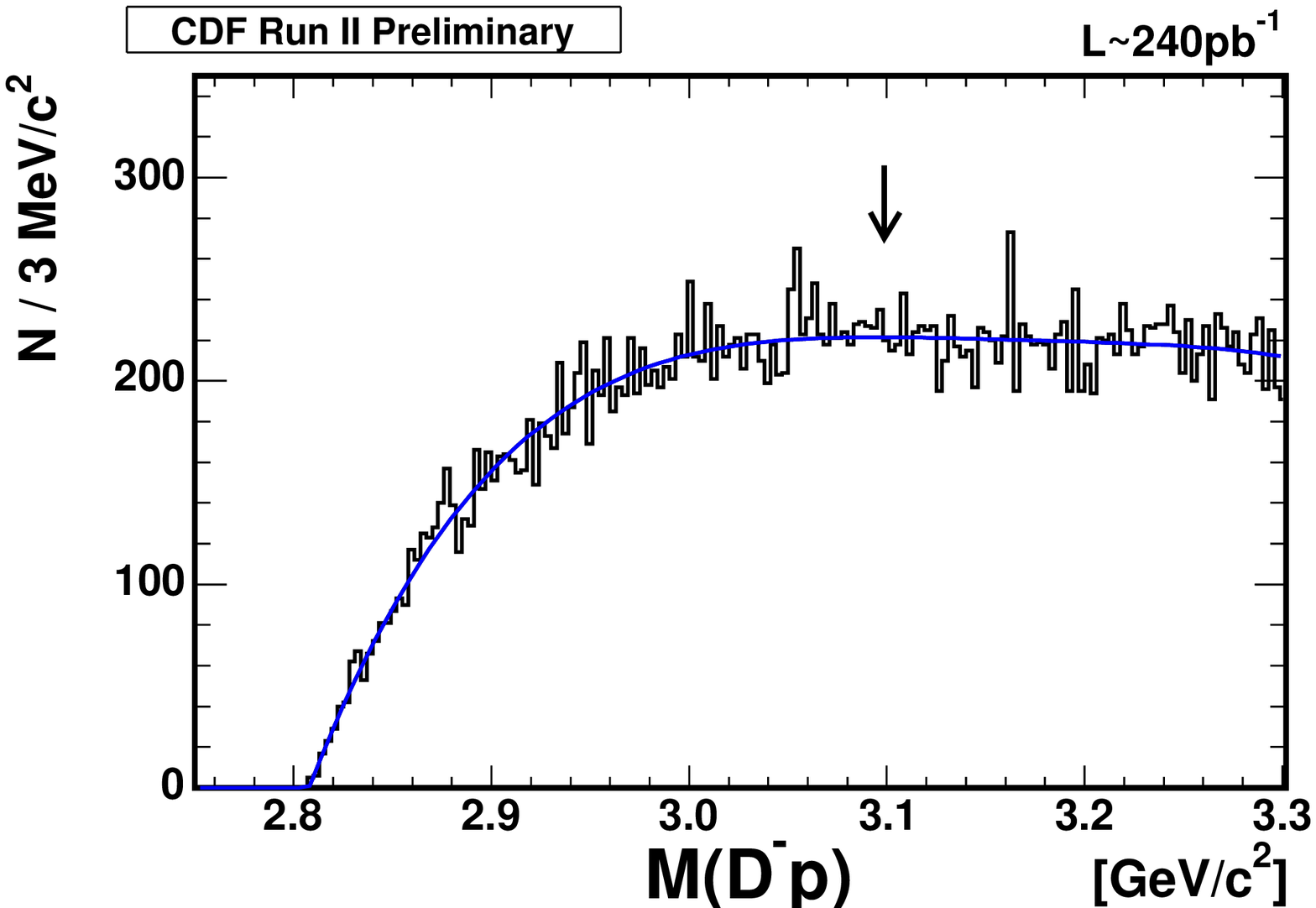,width=2.3in}      
           }
\vspace*{8pt}
\caption{The prompt (left) and long-lived (right)  $p{D}{^-}$ mass spectra (arrows mark H1 mass).
\label{Fig:ThetapDCDF}
}
\end{figure}

\begin{figure}[t]
\centerline{\psfig{file=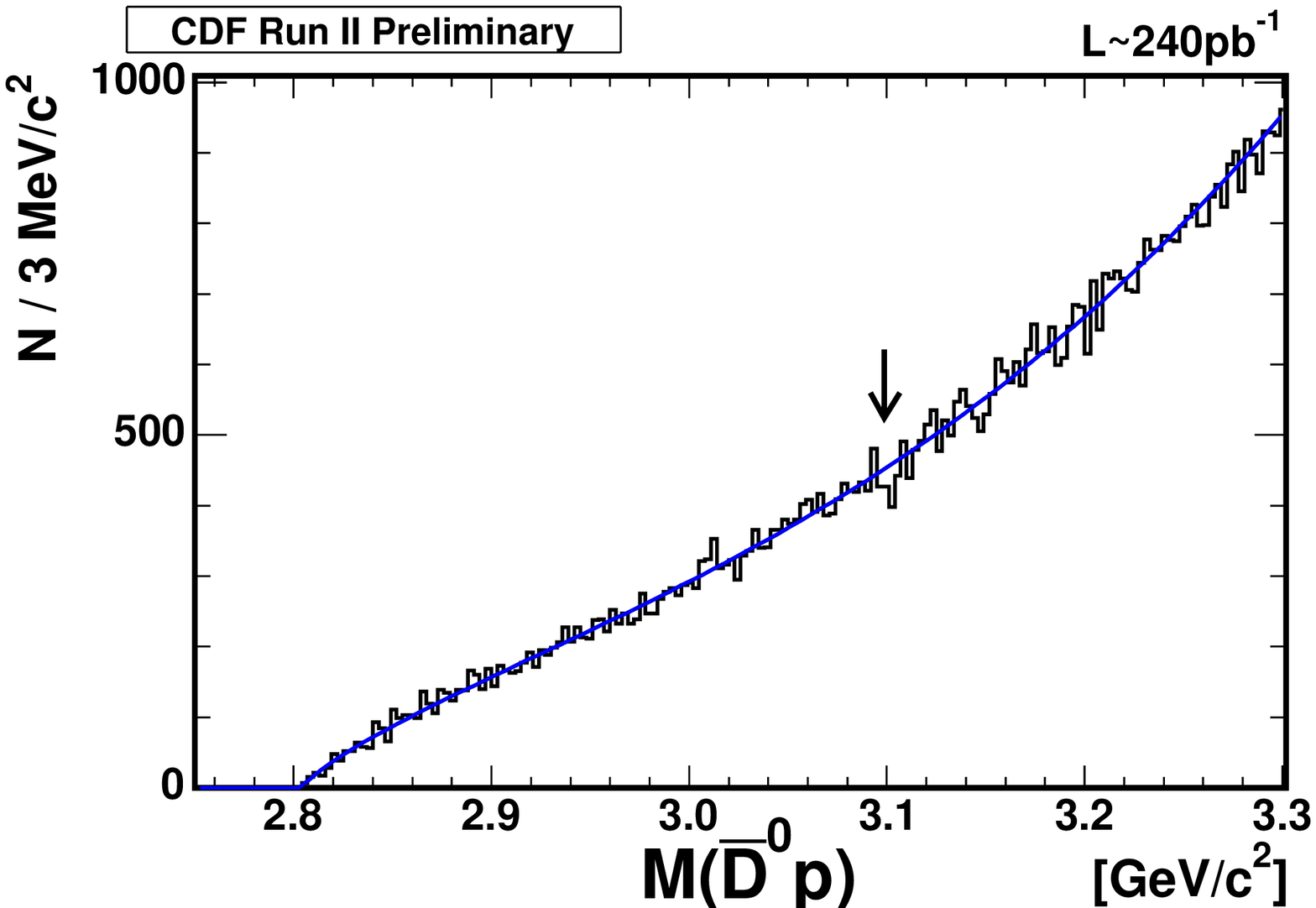,width=2.3in}
            \psfig{file=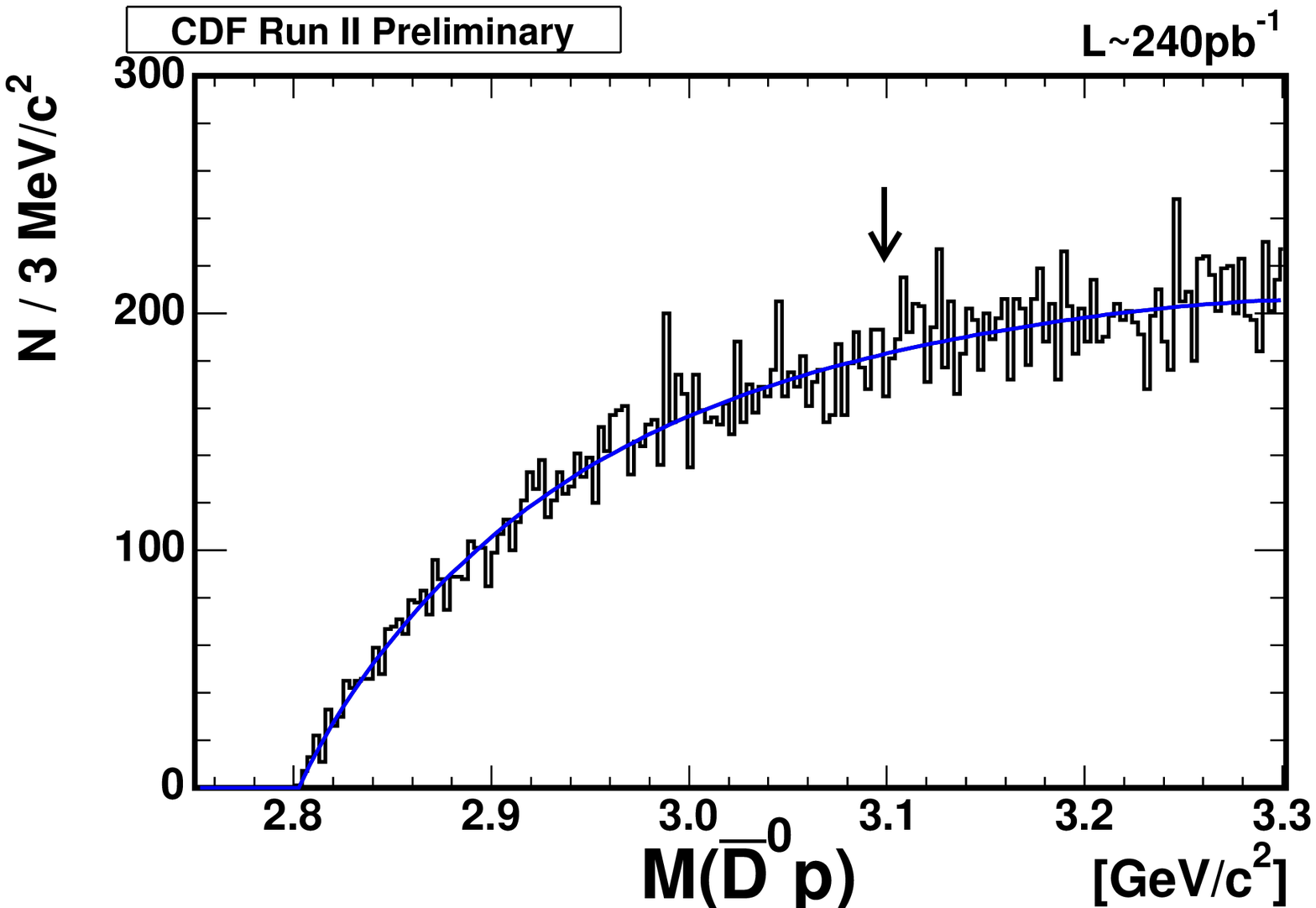,width=2.3in}
           }
\vspace*{8pt}
\caption{The prompt (left) and long-lived (right) $p\overline{D}{^0}$ mass spectra (arrows mark H1 mass).
\label{Fig:pDCDF}
}
\end{figure}

CDF extended their search\cite{CDFThetaCMore} to various analog channels:
$\Theta^0_c \!\rightarrow\! pD^-$, 
and  $\Theta^+_c  \!\rightarrow\! p\overline{D}{^0}$ ($uuud\bar{c}$),
and even  $p{D}{^0}$ ($uudc\bar{u}$).
Figure~\ref{Fig:ThetapDCDF} shows the results for $p{D}{^-}$  after
proton ID for prompt and long-lived samples.
The $p\overline{D}{^0}$ results are in Fig.~\ref{Fig:pDCDF}.
The  $p{D}{^0}$ plots are not shown here, but are similar
to Fig.~\ref{Fig:pDCDF}.
No signals are apparent, 
and the upper limits ($\Gamma_{\Theta}\!=\!12\MeVcc$) on candidates
may be summarized as:\\
\begin{minipage}[b]{1.0\textwidth}
\vspace*{5pt}
\begin{center}
\begin{tabular}{c @{}c r @{}r  r@{} r l @{}r}
  \underline{Mode} & ~\underline{Content}&  \multicolumn{4}{c}{\underline{Prmt \& L-L 90\% CL}}
                                                 & \multicolumn{2}{c}{\underline{Reference Mode \&  Yield}} \\
$pD^{*-}$          &$uudd\bar{c}$&~~&$<32$ & $<15$ &~~~&  $D^{*0}_1(2420) \rightarrow D^{*+}\pi^-$  & $3.7\pm0.9\,$k \\
                   &             &~~&      &       &~~~&  $D^{*0}_2(2460) \rightarrow D^{*+}\pi^-$  & $6.2\pm1.7\,$k  \\
$pD^-$             &$uudd\bar{c}$&~~&$<84$ & $<118$&~~~&  $D^{*0}_2(2460) \rightarrow D^{+}\pi^- $  & $31.7\pm1.3\,$k \\
$p\overline{D}{^0}$&$uuud\bar{c}$&~~&$<122$& $<214$&~~~&  $D^{*-}_2(2460) \rightarrow D^{0}\pi^- $  & $15.3\pm1.6$\,k \\
$p         {D}{^0}$&$uudc\bar{u}$&~~&$<245$& $<174$&~~~&  ~~~~~~~~~~~~``~~~~``  \\
\end{tabular}
\end{center}
\end{minipage}

CDF's  $\Theta^+_c(3100)$ limits are below  H1's report,
yet their precursor $D^{*-}$ sample dwarfs that of H1 by two orders of magnitude,
and all other null  searches\cite{ZEUSseminar,ALEPH}$^-$\cite{Belle5Q} 
by more than ten times. 
If the $\Theta^+_c$ exists, it is remarkably suppressed at the Tevatron!

\subsection{Bottom Pentaquarks  at CDF$\,$\protect\cite{CDFPsiP}}

\begin{figure}[t]
\centerline{\psfig{file=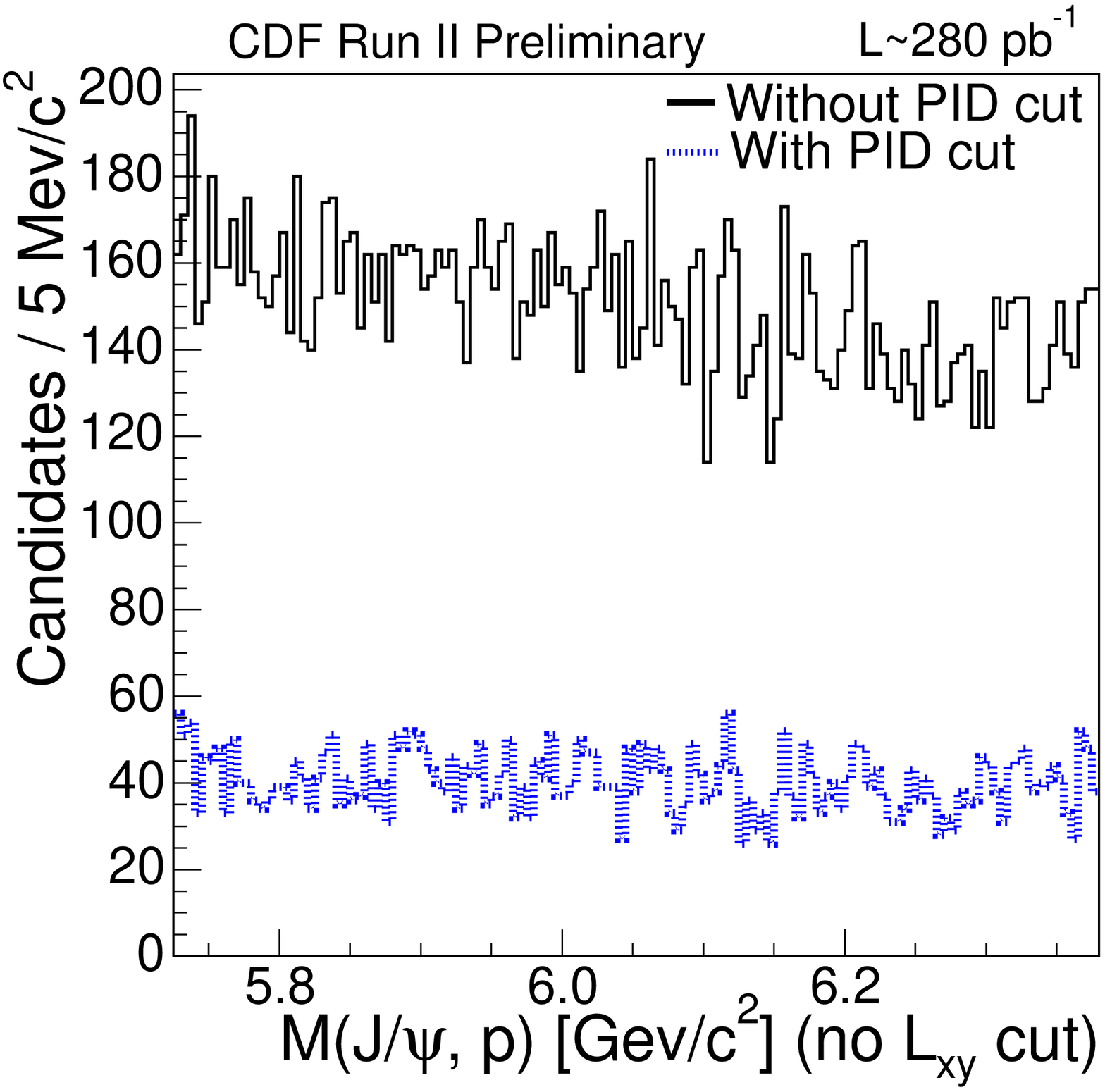,  width=1.65in}\psfig{file=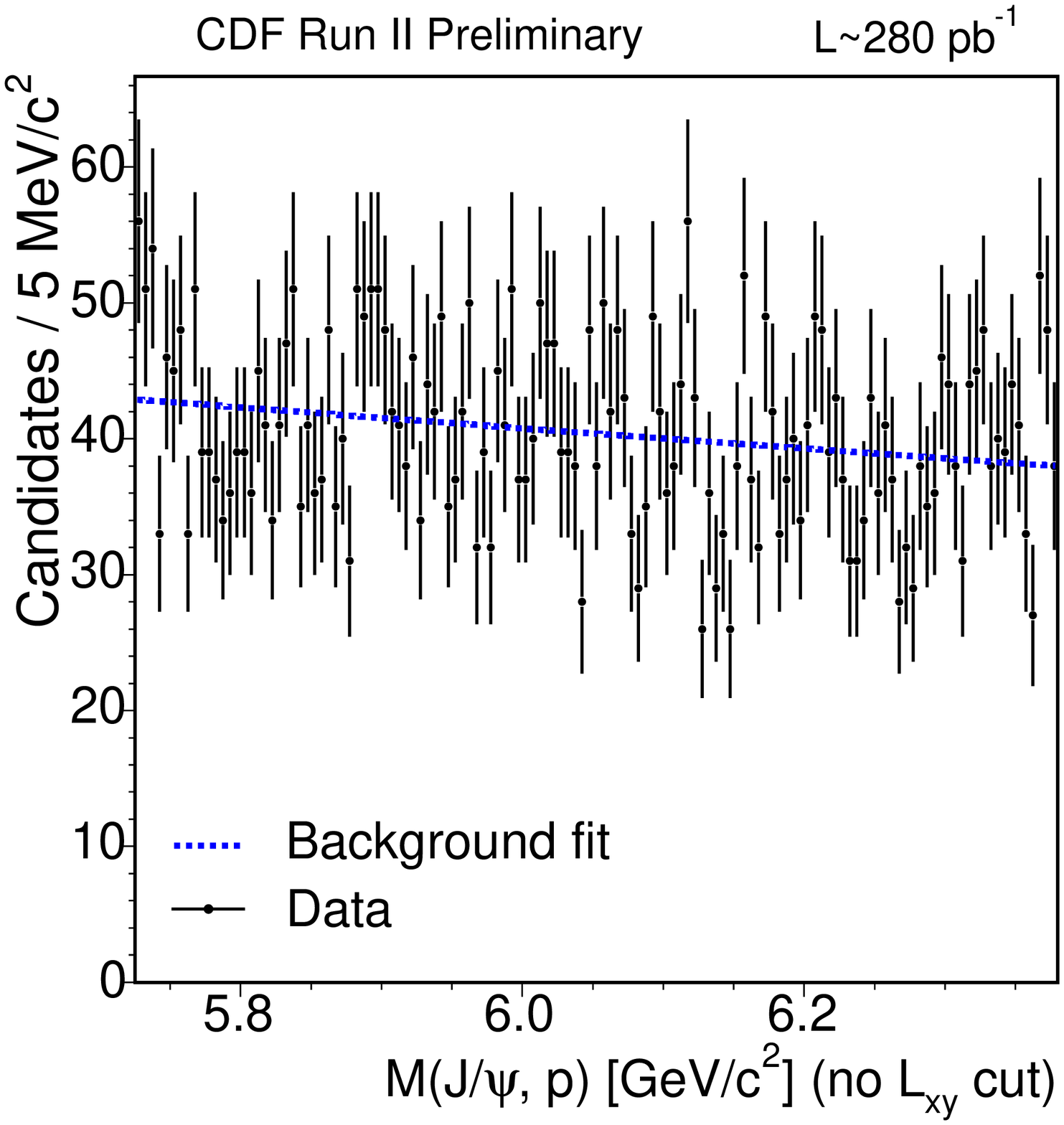,width=1.55in}
            \psfig{file=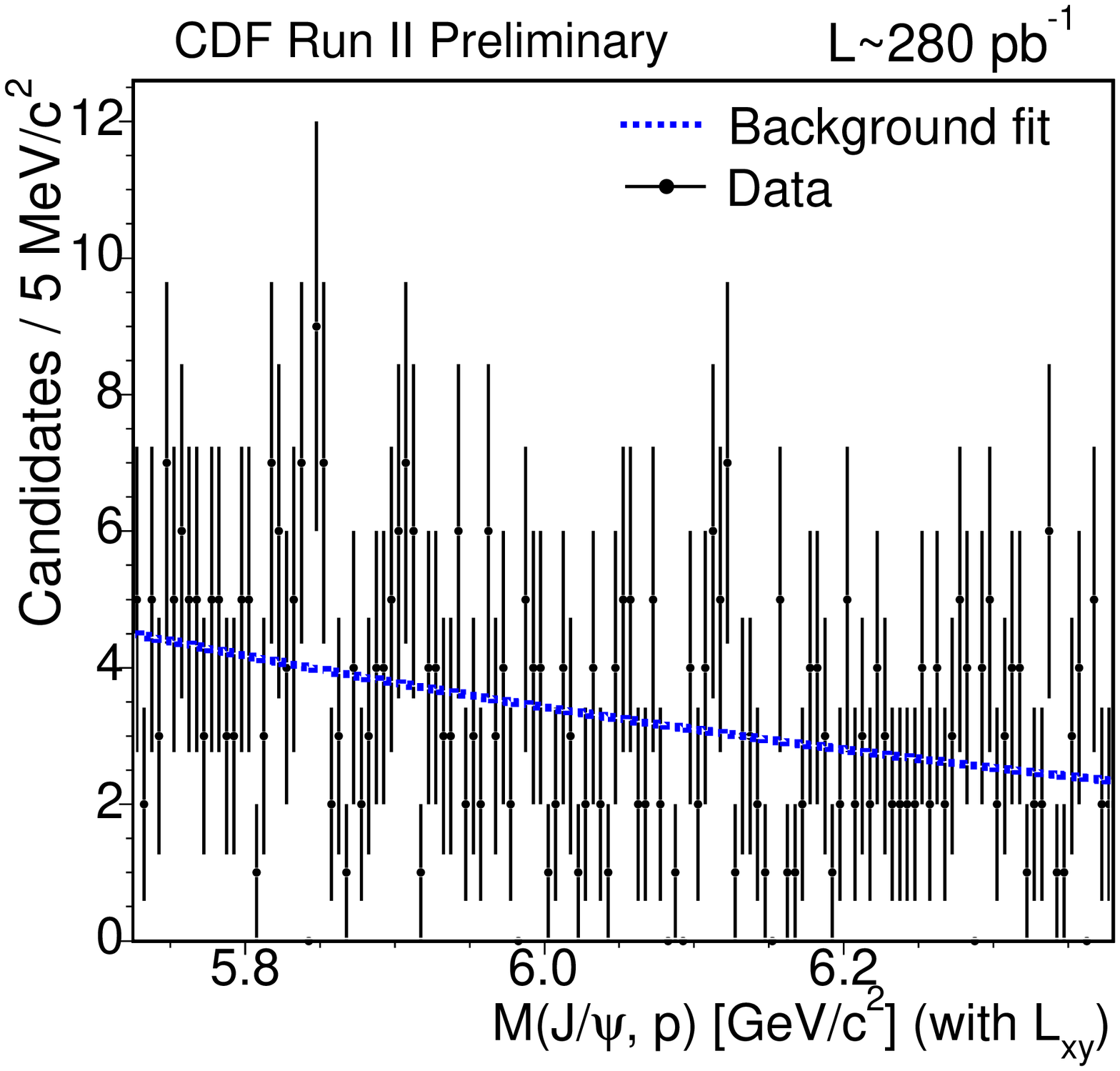,width=1.7in}
           }
\vspace*{8pt}
\caption{{\bf LEFT:}  The $pJ/\psi $ mass distribution without particle ID (top histogram)
                      and with proton ID cuts (bottom).
         {\bf CENTER:} The $pJ/\psi$ masses  with  proton ID (enlargement of lower
                       histogram in the left plot), with a linear background fit.
         {\bf RIGHT:}  The   $pJ/\psi$ masses  with  proton ID and $L_{xy} \!>\! 100\,\mu$m cut
                       for a long-lived pentaquark search.
\label{Fig:JPsiPCDF}
}
\end{figure}

The Tevatron offers potentially  exclusive access  to $b$-pentaquarks.
CDF has made one such 
search:  $R^+_s(uuds\bar{b})$,  predicted at \mbox{$\sim\!5920\MeVcc$},\cite{Q5:JPsiP}  
decaying {\it weakly} to $pJ/\psi$.
Candidates are made by combining $J/\psi$'s ($280\ipb$) with a charged track.
The reference mode is 2.4k of $B^+\!\rightarrow\! J/\psi K^+$.
Proton ID again uses the combined  likelihood.
The $ p J/\psi $ spectrum both before and
after the ID  is shown in Fig.~\ref{Fig:JPsiPCDF}.
With proton  ID the maximum 90\% CL over  $5800$-$6305\MeVcc$ is 76  $R^+_s$'s.
As a weak decay, $R^+_s$  could be long-lived:
for  $L_{xy}\!>\! 100\,\mu$m (Fig.~\ref{Fig:JPsiPCDF}) the limit is  21 $R^+_s$'s.

\subsection{Pentaquark Reprise}

All CDF searches lack any hint of pentaquarks,
 even though the size of precursor samples 
 exceeds the most comparable positive experiment.
 But in this, CDF is not unique.
A wide range of   experiments  now report null results (Table~\ref{Tab:Null5Q}).
 Many also have larger reference signals than do claimants. %---though often(?) less than CDF samples.
 {The $\Phi$ and $\Theta_c^{0}$} have a single sighting
 in contrast to a mounting number of non-observations.
 The $\Theta^+$ has about a dozen confirmations to its credit, 
 but they are now outnumbered by null searches.

\begin{table}[t] 
\tbl{Summary of experiments reporting  negative pentaquark searches
since LEPS reported  the $\Theta^+$.
 Entries are the citation number in this review.
 Instances where one of these experiments has also reported a signal are indicated by a ``$\sqrt{}$.''
 For the production modes
 ``$A$'' represents a nucleus, and ``$h$'' some set of hadron projectiles (e.g. $p$, $\pi$,\ldots).
\label{Tab:Null5Q}  }
{
\begin{tabular}{|@{}c@{}|c|@{}c@{}c@{}c@{}c@{}c@{}c@{}c@{}c@{}c@{}c@{} | @{}c@{}c@{}c@{}  | c@{}c@{}c@{}c@{}c@{}c@{}c@{}|}
\hline\hline
 \multicolumn{1}{|c|}{ } & \multicolumn{1}{|c|}{ }  & \multicolumn{20}{|c|}{Negative Pentaquark Search Exps.}\\
 \multicolumn{1}{|c|}{ } & \multicolumn{1}{|c|}{ }  & \multicolumn{10}{|c|}{Fixed Target}  
             & \multicolumn{3}{@{}c@{}}{$\,Low-$E $e^+e^-$} 
             & \multicolumn{7}{|c|}{High-$E$ Collider} \\ 
\multicolumn{1}{|c|}{\raisebox{2.5ex}{\parbox{1.5cm}{\centering Pentaquark \\  Channel\\~} }} %-8.2
&
&\rotatebox[origin=lB]{90}{CLAS}
&\rotatebox[origin=lB]{90}{HERMES}
&\rotatebox[origin=lB]{90}{SPHINX}
&\rotatebox[origin=lB]{90}{FOCUS}     %fermi
&\rotatebox[origin=lB]{90}{COMPASS}     %CERN
&\rotatebox[origin=lB]{90}{Hyper$CP$}         %fermi
&\rotatebox[origin=lB]{90}{SELEX}         %fermi
&\rotatebox[origin=lB]{90}{WA89}
&\rotatebox[origin=lB]{90}{E690}      %fermi
&\rotatebox[origin=lB]{90}{HERA-B}
&\rotatebox[origin=lB]{90}{BES} 
&\rotatebox[origin=lB]{90}{\sc BaBar}
&\rotatebox[origin=lB]{90}{Belle}
&\rotatebox[origin=lB]{90}{PHENIX}
&\rotatebox[origin=lB]{90}{STAR}
&\rotatebox[origin=lB]{90}{ALEPH}
&\rotatebox[origin=lB]{90}{DELPHI}
&\rotatebox[origin=lB]{90}{L3}
&\rotatebox[origin=lB]{90}{Zeus}
&\rotatebox[origin=lB]{90}{CDF~II}\\      %\cline{3-16}
&\rotatebox[origin=rB]{90}{\parbox{0.2cm}{\mbox{~1st Observation}}}
&{$\,\gamma p$}
&{$\,\,\gamma D$}
&{$\,\,pA$}
&{$\,\,\gamma A$} %HyperCP mixed beams (pi, p, hyperon) on ?
&{$\,\,\mu A$}
&{$\,\,hA$}
&{$\,\,hA$}
&{$\,\,\Sigma A$}
&{$\,\,pA$}              %Hera-B
&{$\,\,p A$}
%&{$\,\,\psi(^3S_1)$} 
&{$\,\,\psi(S)$} 
&\multicolumn{2}{@{}c@{}|}{$\Upsilon(4S)\,$}
&\multicolumn{2}{|@{}c@{}}{--$AA$--$\,$}
&\multicolumn{3}{@{}c@{}}{--$\,Z^0\,$--}
&{$\,ep\,$}
&{$\,\bar{p}p\,$}\\
\colrule
$\Theta^+ \rightarrow NK$     &$\!\!$LEPS\cite{LEPS5Q}$\!$        
& $\sqrt{}$  & $\sqrt{}$      & \refcite{SPHINX}  &  \refcite{FOCUSdpf}  & ---  & \refcite{HyperCP}   
&\refcite{SELEX5Q} &---  & \refcite{E690} 
&\refcite{HeraB} &\refcite{BESTheap} & \refcite{BaBarTheta} &  \refcite{Belle5Q}
& \refcite{Phenix}$\,$ &   \refcite{STARtheta}
& $\,$\refcite{ALEPH}  & $\,$\refcite{DELPHI}  & $\,$\refcite{AlephiDelphiL3Beach} 
&  $\sqrt{}$  &   \refcite{CDFThetaC}
\\
$\Theta^{++} \rightarrow pK^+ $     &$\!\!$---$\!\!$        
& \refcite{CLASThetapp} & \refcite{Hermes}  & ---  &---  &---  &---&--- &--- &---  
&---  &---  &  \refcite{BaBarThetapp}   &\refcite{Belle5Q}  &  ---  &  ---  &  ---
& \refcite{DELPHI} &---   & \refcite{ZEUS5Qlight} & ---
\\
$\,\,N_5/\Xi_5 \rightarrow  \Lambda K\,\,$     &  $\!\!$STAR\cite{STARN5}$\!\!$  
&---  &  ---  &  --- &  ---  &---  &---&--- &---  &---  &---&---   & \refcite{BaBarXi}
&--- &---  &$\sqrt{}$ 
& ---  & --- &--- &--- &---
\\  % aleph showed  exclusion at DIS04 in slovakia?
~~~~~~~~$\rightarrow \Sigma^0  K \,$     &--- 
&---  &  ---  &  --- &  ---  &---  &---&--- &---  &---  &---&---   & \refcite{BaBarStrangePentaPub}
&--- &---  &---
& ---  & --- &--- &--- &---
\\ 
$\Phi \!\rightarrow\! \Xi^-\pi^\pm$ &$\!\!$NA49\cite{NA49}$\!$        
&  --- &  \refcite{Hermes1860}  & --- &  \refcite{FOCUSdpf} &  \refcite{Compass} & ---  &---  
&\refcite{WA89}  & \refcite{E690} 
&\refcite{HeraB}  &---  &   \refcite{BaBarXi} & \refcite{Belle5Q1860}
& ---  & ---  &  \refcite{ALEPH}   & --- &--- &\refcite{ZEUS5Qlight}&  \refcite{CDFXi}
\\ 
$\Theta^{0}_c\rightarrow  pD^{*-}$& $\!$H1\cite{H1}         
& --- &  ---  & ---   & \refcite{FOCUSthetac} &---  & ---& --- &--- &---  &---  
&---  & --- & \refcite{Belle5Q}  &---  & ---  
&  \refcite{ALEPH}  
&---&--- &  \refcite{ZEUSseminar} & 
\refcite{CDFThetaC} 
\\
~~~~$\rightarrow  pD^{-}$& --- &  --- & ---  & ---   
&  \refcite{FOCUSthetac} &---  &---  &---  & ---  &--- &---  &---  
& ---& \refcite{Belle5Q} &  ---  & --- &  \refcite{ALEPH}  
& ---  & --- &---& \refcite{CDFThetaCMore}  
\\
$\Theta^{+}_c\rightarrow  p\overline{D}^{0}$& ---  &  ---     &---     &---   
& ---  & ---   &--- &---  &--- &--- &---  &---  & ---& \refcite{Belle5Q} 
& ---  &---  &  \refcite{ALEPH}  & ---   
& --- &---&\refcite{CDFThetaCMore}
\\
~~~~~$\rightarrow  p{D}^{0}$& ---        &  ---  &---  &---
& ---  & ---   &--- &---  &--- &--- &---  &---  & ---& ---  &---  & --- & ---  & --- 
& --- &---&\refcite{CDFThetaCMore}
\\
$ R^+_s \rightarrow  p J/\psi$& ---      &  ---    &---    &---  
& ---  & ---   &--- &---  &--- &--- &---  &---  & ---& ---  &---  & --- & ---  & --- 
& --- &---&\refcite{CDFPsiP}
\\
\hline\hline
\end{tabular}
}
\end{table}

The primary refuge for reconciling null  searches
with sightings lies in the possible peculiarities of production. 
 Most sightings are at low energies,
 often in exclusive reactions.
 Production at higher energies is predominantly
 through fragmentation,  or via $B$-decay, which are quite different from
 low-energy processes.
 Models of inclusive penta\-quark production are rudimentary,
 {but several have been proffered.}

 In one, the fragmentation probability,
 $f(\bar{c}\!\rightarrow\! \Theta^0_c)$, is estimated from that 
 of $D$ and $\Lambda^+_c$.\cite{Q5Frag}
 That author finds
 $f(\bar{c}\!\rightarrow\! \Theta^0_c) \!\simeq\! (2$-$7)\!\times\!10^{-3}$, 
 consistent with  H1's raw $D^{*-}$ and
 $\Theta^0_c$ rates.
 Translating to the Tevatron for $200\ipb$: 
 $8$-$28\,$M $\Theta^0_c$'s are produced!
 Alternatively, a ``coalescence'' model\cite{Coalescence}
 scales the joint $p$ and $D^{*-}$ production
 rates to a regime where the $p$ and $D^{*-}$ 
form a $\Theta^0_c$.
 Using H1's rate to set the absolute scale, there are 
$\sim\!50$M $\Theta^0_c$'s for $200\ipb$.
CDF efficiencies have not  been applied,
  but  it is surprising that   a signal should elude CDF
with H1-like $\Theta^0_c$-rates.

Another approach is a statistical (``microcanonical'') model for $pp$ 
 inter\-ac\-tions.\cite{LiuMicroCanonical} 
 This {\it does} favor low-energy $\Theta^+$ production due to the importance of
 $p\!+\!p \!\rightarrow\! \Theta^+ \! + \! \Sigma^+$.
But even so, the model predicts a fairly flat high-energy limit 
 of   $\sim\!1\%$ $\Theta^+$'s/event---a huge rate for CDF, even if  low-$p_T$ is favored.
 The prediction for the  $\Phi^{--}/\Xi^-$  ratio is 
 $\sim\! 2\%$ at the SPS---in line with NA49.
 But the ratio increases with energy by $\sim\!3\times$ at the Tevatron, 
 exacerbating the inconsistency posed by CDF's null result.

If the  key to  $\Theta^+$ and $\Phi$ production at low energies
is the incorporation of quarks from an  initial baryon, then it is difficult
to translate lessons from  low-energy experiments to 
the central rapidities studied by CDF.
One such model\cite{Bleicher} predicts  high  rates 
 ($\gtrsim \! 10^{-3}\,\Theta^+$/event for $pp\!\rightarrow\!\Theta^+\ldots$ )---but
 at high-rapidities/low-$p_T$'s---making {\it these} $\Theta^+$'s invisible to CDF.
Similarly, it has been argued\cite{HicksFermiAPS} that the apparent production discrepancies
may be due to the  kinematic and combinatoric advantages of low-energy, or particularly, 
exclusive reactions, where most claims arise.
This is based, in part, on an analysis which concludes
that $\Theta^+$ production in a range of processes
falls more rapidly with energy ($p_T$) than normal hyperons,\cite{Titov} 
undermining   high-energy searches.
But as these authors\cite{Titov} note:
 the processes considered, including a {\it target} fragmentation model,
 are kinematically linked to the initial baryons    
and are {\it not relevant}
to the {\it central} production of CDF.
While this particular suppression is not in play, 
what suppression lurks in the parton fragmentation  is another matter.

 One may hesitate relying  on these production models for pentaquarks, 
 particularly when  ``data points'' used to normalize some models are themselves uncertain. 
A simple  empirical foil to consider is  deuteron production
as a stand-in for pentaquarks.
 The ratio of anti-deuteron to anti-proton production scales 
 well across many high-energy processes
 (expected in coalescence models).
 For example, the ratio is very similar in $pp$ collisions at the ISR 
 and photo\-produc\-tion at HERA. 
 The $\bar{d}/\bar{p}$ ratio is $\sim\! 10^{-3}$ at $p_T/M \!=\! 0.2$,
 and falls by half at $p_T/M \!\sim\! 0.5$.\cite{AntiDeuteron}
If one takes  $\Phi/\Xi^-$ ratio as the appropriate analog to  $\bar{d}/\bar{p}$,
the NA49 ratio of $\sim\!3\%$ is at least an order of magnitude more plentiful
than implied by the deuteron analogy.
Similar scaling of  $\Theta^+$ reports 
gives ratios spanning several factors of ten. 
Scaling\cite{UA1Tan} CDF limits gives $\Theta^+/\Lambda^0 \!\lesssim\!0.02\%$---below
the deuteron-inspired rates---while the Zeus\cite{ZEUStheta} signal gives  $\Theta^+/\Lambda^0 \!\sim\! 0.1\%$.
The above comparisons cavalierly ignore detection efficiencies,
which maybe quite important as the  $\bar{d}/\bar{p}$-ratio  falls with $p_T$.
As noted by critics, this is an important  weakness of fragmentation dominated
experiments compared to the low-energy $\Theta^+$ sightings.
However, the suppression suggested by  $\bar{d}/\bar{p}$ is no where 
as extreme as sometimes claimed  for pentaquarks
({\it e.g.} $\Theta^+/\Lambda(\underline{\underline{1520}})\!<\!10^{-3}$)\cite{RateWarning}

The contrast between high-energy fragmentation {\it \`{a} la} CDF and
low-energy, especially exclusive, $\Theta^+$ production is sufficient
that little inference can be drawn from one to the other without
a robust theoretical link.
Low-energy $\Theta^+$ proponents can justifiably raise production arguments
to explain away high-energy null searches---but only at the risk of abandoning  their  high-energy compatriots: 
such as $\Theta^+$ by ZEUS.
Indeed, the quantity and quality of negative searches
present an impressive challenge, and  it seems  likely that 
at least {\it some}  claims will fall.
The strongest case rests with the  $\Theta^+$, where production advantages
may truly favor some observations.
Of critical importance are high-statistics studies from experiments claiming signals.
These have been advertised as imminent,\cite{HicksFermiAPS}
and the first preliminary result has just appeared from from CLAS: 
a search for $\gamma p\!\rightarrow\! \Theta^+ \overline{K}{^0}$ has
{\it failed} to observe a signal with 95\%CL limit of $\Theta^+/\Lambda(1520)\!<\! 0.2\%$!\cite{HighStat5Q} 
 If any  pentaquark claims are yet vindicated, 
 it will be interesting to learn why they are 
\mbox{so suppressed at the Tevatron.}

\section{``Anomalous'' $D^+_{sJ}$ States}

Pentaquarks were only the start of spectroscopic excitement in 2003.
{\sc BaBar}  announced 
a narrow state $\sim\!2317\MeVcc$ decaying to $D_s^+\pi^0$ in April.\cite{BabarDs2317}
Based on a hint from {\sc BaBar},\cite{BabarDs2317} CLEO quickly claimed
another  at $\sim\!2460\MeVcc$ in  $D_s^{*+}\pi^0$.\cite{CLEODs2460} 
The benign interpretation is that these are the missing
$0^{++}$ and $1^{++}$  $D^{**}_s$ states, which would
complete the $L\!=\! 1$ family along with  $D^+_{s1}(2536)$ ($1^{+-}$)
and  $D^+_{s2}(2573)$ ($2^{++}$).
But as such, these new states were much lighter and narrower ($<\! 10\MeV$)
than expected.
The  $D^{**}_s$ were thought  to follow  the  non-strange $D^{**}$'s: very broad 
$0^{++}$ and $1^{++}$ states which 
recent measurements put $\Gamma \!\sim\! 240$-$400\MeV$.\cite{D2StarBroadStates}
The $D^+_{sJ}(2317)$ did not look as  the $D^{+}_{s0}(0^+)$ should.
{\sc BaBar} suggested it might be a $q\bar{q}c\bar{s}$ state.\cite{DsJExotic}

\begin{figure}[t]
\centerline{\psfig{file=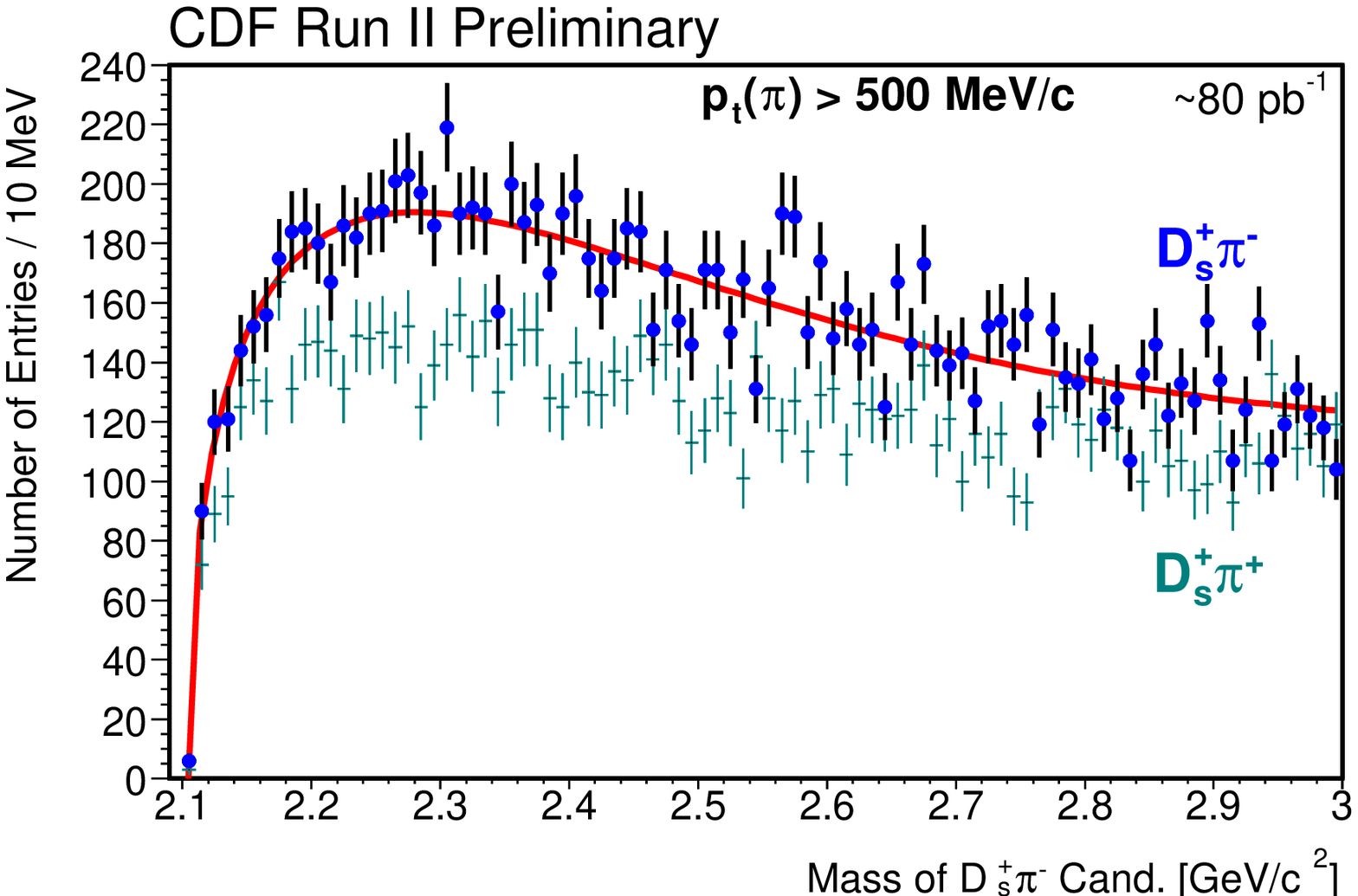,width=2.6in}\hspace*{-0.3cm}
            \psfig{file=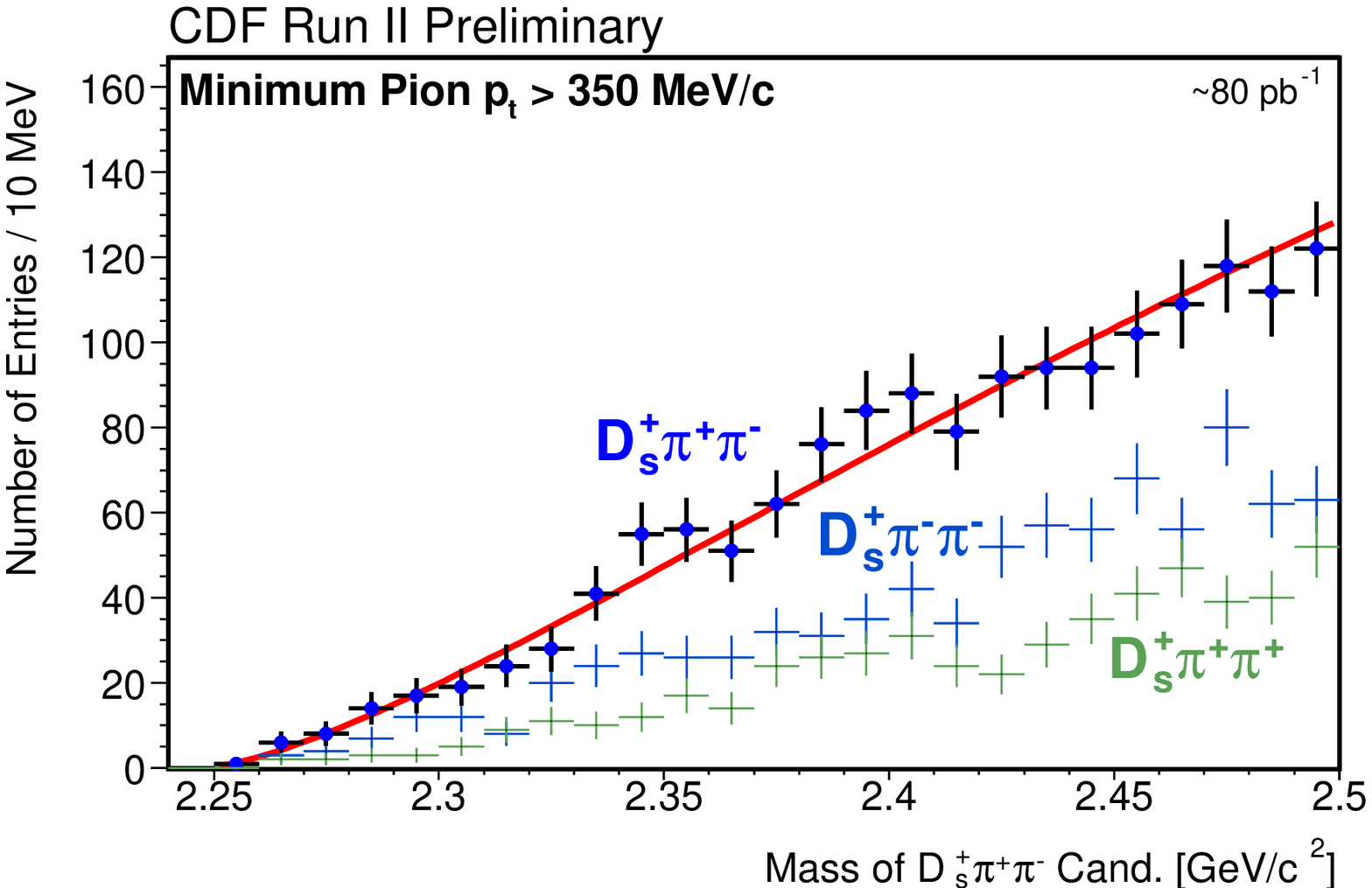,width=2.6in}}
\vspace*{8pt}
\caption{{\bf LEFT:}  Mass distributions of $D_s^+\pi^-$ and $D_s^\pm\pi^\pm$  
                      for pions with $p_T\!>\!500\MeV/c$.
         {\bf RIGHT:} Mass distributions of $D_s^+\pi^+\pi^-$  and  $D_s^+\pi^\pm\pi^\pm$   
                      for pions with $p_T\!>\!350\MeV/c$.
\label{Fig:DsJSearch}
}
\end{figure}

CDF is ill-suited for low-energy $\gamma$-detection,
and thus for  $D_s^{(*)+}\pi^0$.  
If, however, these new states were 4-quark systems,
or more generally had isospin  partners, 
there could be  $D_s^+\pi^-$ or   $D_s^+\pi^+\pi^-$ resonances. 
The latter decay is also allowed if the  $D^+_{sJ}(2460)$ 
is a  $1^{++}$, but forbidden for  $0^{++}$.
CDF searched for   these  using  80\ipb\ ($24.6$k  $D_s^+$'s),
resulting in the spectra of Fig.~\ref{Fig:DsJSearch}---no signals are seen.\cite{CDFDsPiPi}
To gauge the sensitivity, {\sc BaBar}'s 
$\sim\!1300$ $D^+_{sJ}(2317)$'s were based
on $\sim\!80k$  $D_s^+$'s, or $\sim\!3\times$ that of CDF.
While the origin of $D_s^+$'s can be 
different for the two experiments, 
CDF is in the ball-park to see a $D_s^+\pi^-$
analog given the large {\sc BaBar} signal.\cite{BaBarIsoAnalog}
For a $1^{++}$,  $D_s^+\pi^+\pi^-$ would be suppressed 
relative to $D_s^{*+}\pi^0$.
Belle later found a small signal  [$59.7\!\pm\! 11.5$ $D^+_{sJ}(2460)$'s]
and found the ratio of $D^+_{sJ}(2460)\!\rightarrow\! D_s^+\pi^+\pi^-$ 
 to $D_s^{*+}\pi^0$ to be $14\!\pm\! 4\!\pm\! 2\,$\%.\cite{Belle2460PiPi}
By na\"{\i}ve scaling,  \mbox{this  is below CDF sensitivity with 80\ipb.}

The new $D^+_{sJ}$'s excited spectroscopists,
but radical explanations now seem premature. 
Neither state is mysterious {\it once} their masses are understood.
Small widths arise naturally for the $D_{sJ}(2317)$  and  $D^*_{sJ}(2460)$
as $0^+$ and $1^+$ if they are below  the  $DK$ and $D^*K$  thresholds respectively.
As such, the preferred  decay  is excluded, and the isospin violating $D_s^{(*)}\pi^0$  is
the main hadronic mode.
It was soon noted\cite{CahnJackson} that  potential models are free to move $D^{**}_s$ masses
more than usually appreciated.
It was also argued,\cite{BardeenEichtenHill} light masses follow from chiral symmetry in QCD:
the ground state parity doublet,  $D_s^+$ and  $D_s^{*+}$  ($0^-$, $1^-$),
is paired with $0^+$ and $1^+$ excited states, and
chiral symmetry  breaking raises  the ($0^+$, $1^+$) doublet 
close to that of the  $D_{sJ}$'s.

Studies of decay modes and angular analyses support 
 $D^*_{s0}(2317)$ and $D'_{s1}(2460)$ assignments.\cite{DJstudies}
But there is not unanimity, and exotic proposals persist.\cite{ExoticDsjInterp,Maiani}
Lest the dust seem  settled, SELEX 
recently kicked up a new cloud with a narrow state
$D_{sJ}^+(2632)\!\rightarrow\! D^+_s\eta$, and a weaker $D^0K^+$ signal.\cite{SELEX2632}
New puzzles arise:\cite{Interp2632} 
Why so narrow?
Why is the $D^+_s\eta$ rate $\sim\!6\times$ larger than $D^0K^+$?
\mbox{The mystery is heightened by}   {\sc BaBar}'s failure to see
 $D_{sJ}^+(2632)\!\rightarrow\! D^0K^+$ while having a much larger
 $D_{s2}^+(2573)\!\rightarrow\! D^0K^+$ yield.\cite{VaBarNo2632}
  SELEX counters\cite{SELEX2632} that their production is distinctive by virtue 
  of their $\Sigma^-$ beam.
  CDF has a large  $D^+_{s2}(2573)\!\rightarrow\! D^0K^+$ sample---it
  will be interesting to see them search.
  But so far, the odds favor the $D_{sJ}$'s as just  $D_{s}^{**}$'s.

\section{The \protect\raisebox{-0.3ex}[1.0ex][1.0ex]{\Large $X$}-Files}

After a series of null results we close with a state
 CDF {\it has} confirmed, but whose nature is a mystery: the 
 $X(3872)$.
 It is a  tale we begin by recounting a bit of history.

\subsection{A Little Charmonium History}

 Our understanding of hadrons was revolutionized by studying
 $c\bar{c}$-states, starting
 with the  $J/\psi$ in  1974.\cite{TingRichter}
 Mapping $c\bar{c}$-states was largely done
 in the 70s in $e^+e^-$ annihilation.
 A limitation of  $e^+e^-$ is that only systems with 
photon quantum-numbers  are formed---{\it i.e.},
only $J/\psi$, $\psi(2S)$, $\psi(3770)$, \ldots are directly accessed.
  Almost all $c\bar{c}$-states below the  $\psi(2S)$
  ({\it i.e.} $\eta_c$ [$^1S_0$] and $\chi_c$ [$^3P_{0,1,2}$])
  were reached via  radiative $\psi(2S)$ decays.
Once these  were found,  $e^+e^-$ colliders were at a dead-end.
Heavier $1^{--}$  states, {\it e.g.}  $\psi(3770)$, are useless as 
they are above the $D\overline{D}$ threshold
and are broad, with tiny decay rates to lighter $c\bar{c}$-states.
The hunt shifted to other venues.

The $h_c$  ($^1\!P_1$) is the lone state {inaccessible\,}\cite{hcException} 
via $\gamma$-decays of the  $\psi(2S)$.
Searches for this state shifted to hadronic production, 
notably $\bar{p}p$ annihilation.
From the mid-1980s a few  $h_c$ claims surfaced.\cite{BaglinHc,E705} 
These were  consistent, but individually weak observations,
leading the PDG to classify the $h_c$ as ``needing confirmation.''

By the early 1990s all $c\bar{c}$-states below the $\psi(2S)$
were  ostensibly\cite{hcSaga} seen---only 
those above  $D\overline{D}$ remained. 
But such states  rapidly decay to open charm,
making them broad and difficult to find.
For example, the  $\psi(3770)$ ($^3\!D_1$)   is just
above $D\overline{D}$,
and yet $\Gamma \!\sim\!20\MeVcc$. Heavier  states grow ever  fatter.
The $^3D_2$ is an exception, its
spin-parity ($2^{--}$) prohibits $D\overline{D}$ decay.
The $^3\!D_2$ is  prime quarry  for 
charmonium hunters: a narrow state which
might be seen in the distinctive
$J/\psi\pi^+\pi^-$ mode.

In 1994 E705 ($300\GeVc$ $\pi/p$-Li)
published, along with a hint of the $h_c$,
a $2.8\sigma$ excess in  $J/\psi\pi^+\pi^-$
at $\sim\!3836\MeVcc$.\cite{E705} 
The  $^3D_2$ was the obvious interpretation, but
the $c\bar{c}q\bar{q}$ option\cite{DeRujula} was noted.
The  $58\!\pm\!21$ excess  was a large fraction of their  raw  $77\!\pm\!21$
 $\psi(2S)$     yield; but
no excess was seen by E672/E706\cite{E672E706} ($515\GeVc$ $\pi^-$-Be)---a higher 
statistics [$224\!\pm\!48$ $\psi(2S)$] result
with better resolution.
 A signal might also be  expected  in CDF Run~I data
 given  their much larger  $\psi(2S)$ sample [$\sim\!2k$] and superior resolution.
 Nothing was noticed there at $\sim\!3836\MeVcc$,\cite{CDF3836} \mbox{nor by  BES in} 
 $e^+e^-\!\rightarrow\! J/\psi\pi^+\pi^- \!+\!anything$.\cite{BES3836}  
 But it is unclear how the latter translates to  E705.

\subsection{Discovery of the $X(3872)$ \label{sec:XDiscov}}

In the early days of $b$-physics it was realized that
$b$-hadrons often decay to $c\bar{c}$ since a favored chain  is
$b \!\rightarrow\! c W^-$, $ W^- \!\!\rightarrow\! s\bar{c}$.\cite{bTOc}
Indeed, CLEO found $B \!\rightarrow\! J/\psi\! + \!${\it anything} to 
be $\sim\!\!1$\%.\cite{CLEObTOc}
In the early 1980's, this was viewed  as a tool for studying $b$-physics.
Decades later, some in Belle appreciated that
this could be ``inverted''  to  exploit  $B$'s 
{\it for studying charmonium}.     
The $c\bar{c}$ dead-end for $e^+e^-$  colliders
could  be evaded by using  feeddown from $B$'s instead of  $\psi(2S)$'s.
Belle demonstrated this by  observing 
$B^+ \!\rightarrow\!$ $ \psi(3770) K^+$,\cite{Belle3770}
and more significantly, used $B\!\rightarrow\! K  K_s K^\pm\pi^\pm$
to  rediscover  the   $\eta_c(2S)$.\cite {BelleEta2c}
Crystal Ball claimed\cite{CryStlBallEta2c}  the   $\eta_c(2S)$ 
at  $\sim\!3594\MeVcc$
over twenty years ago;
but Belle now found  it at $\sim\!3654\MeVcc$,
and was so confirmed.\cite{Eta2cConfirm}

\begin{figure}[t]
\centerline{
\psfig{file=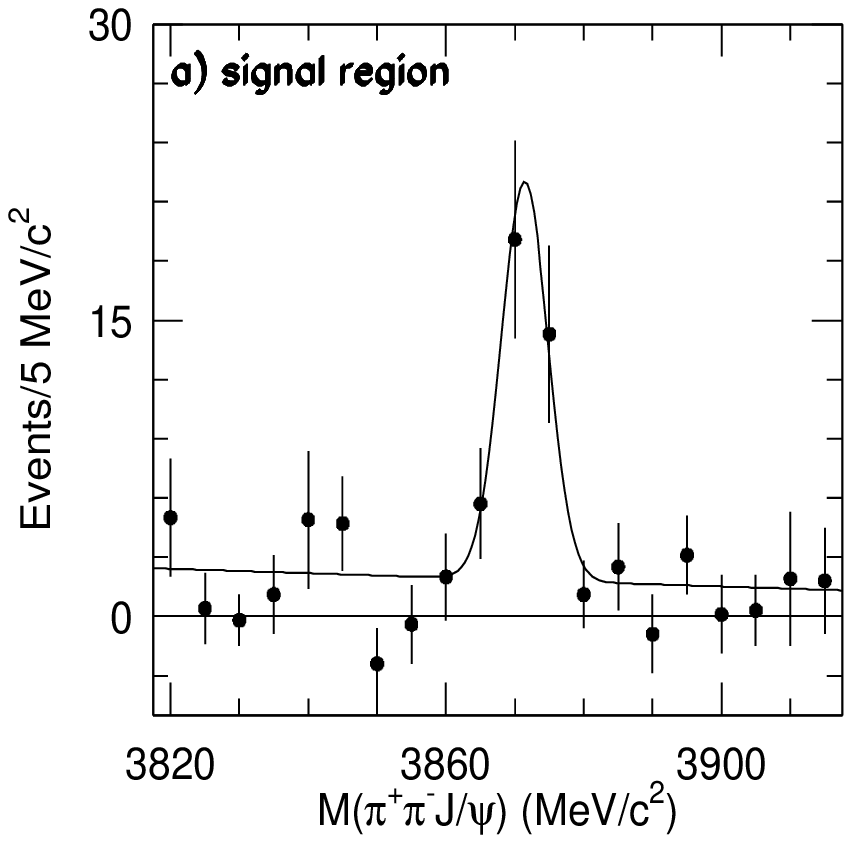,width=1.8in}
\psfig{file=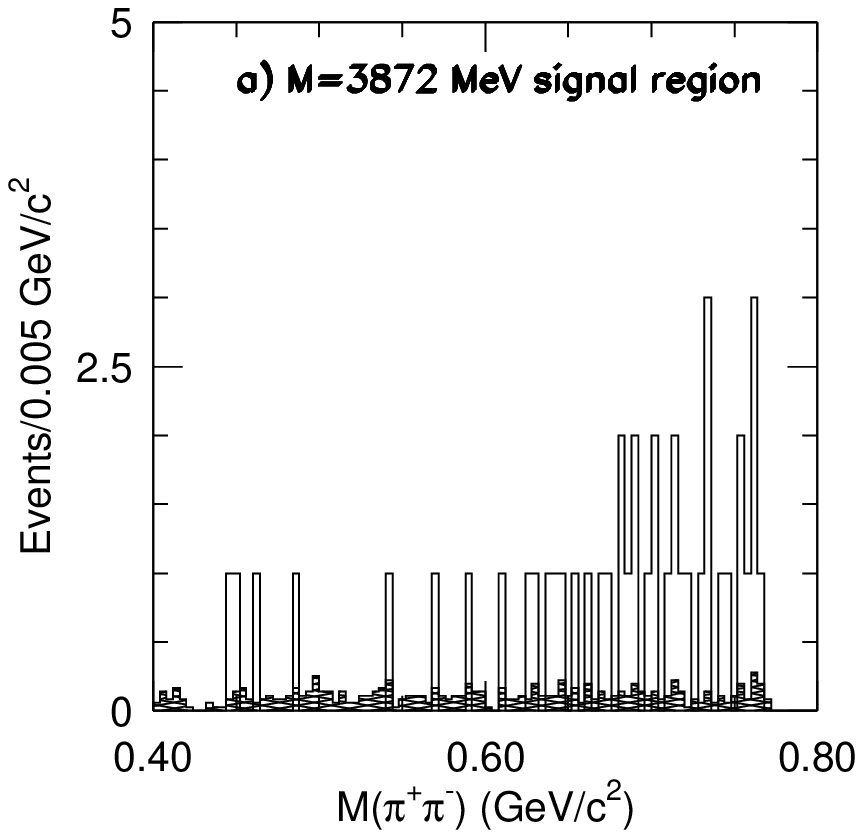,width=1.8in}
           }
\vspace*{8pt}
\caption{{\bf LEFT:} The  $J/\psi\pi^+\pi^-$ mass spectrum from
                     Belle\protect\cite{BelleX} showing the $X(3872)$.
         {\bf RIGHT:} The corresponding dipion masses.
                      The hatched histogram are sidebands normalized to signal area.
\label{Fig:XBelle}
}
\end{figure}

In Belle's $\eta_c(2S)$ studies a stray bump was spotted that turned out to be
a reflection of a new $J/\psi\pi^+\pi^-$ resonance
at  $3872.0   \pm \!0.6  \pm \!0.5\MeVcc$
(Fig.~\ref{Fig:XBelle}),\cite{BelleXPRL}
later dubbed  $X(3872)$.
The impulse was to take  this  as the long-sought $^3D_2$,  
but {\it that} was expected  at  $\sim\!3820\MeVcc$.\cite{XMassPredict}
It should also have a  prominent  $\chi_{c1} \gamma$  decay, 
 {which was  not seen}. 
Being  virtually at the  $D^0\overline{D}{^{*0}}$ mass,
Belle  speculated the $X(3872)$  could be  a $D^0\overline{D}{^{*0}}$  ``molecule.''\cite{DeRujula}  
 The exotic prospects\cite{Maiani,PreX4quark}$^-$\cite{SwansonMolecule}
provoked great interest, 
and it is questionable whether standard $c\bar{c}$\cite{ccbarOptions1,ccbarOptions2}  
\mbox{can accommodate this state.}

\subsection{The $X(3872)$ at CDF}
\subsubsection{Observation and Mass Measurement$\,$\protect\cite{XCDF}}

\begin{figure}[b]
\centerline{\psfig{file=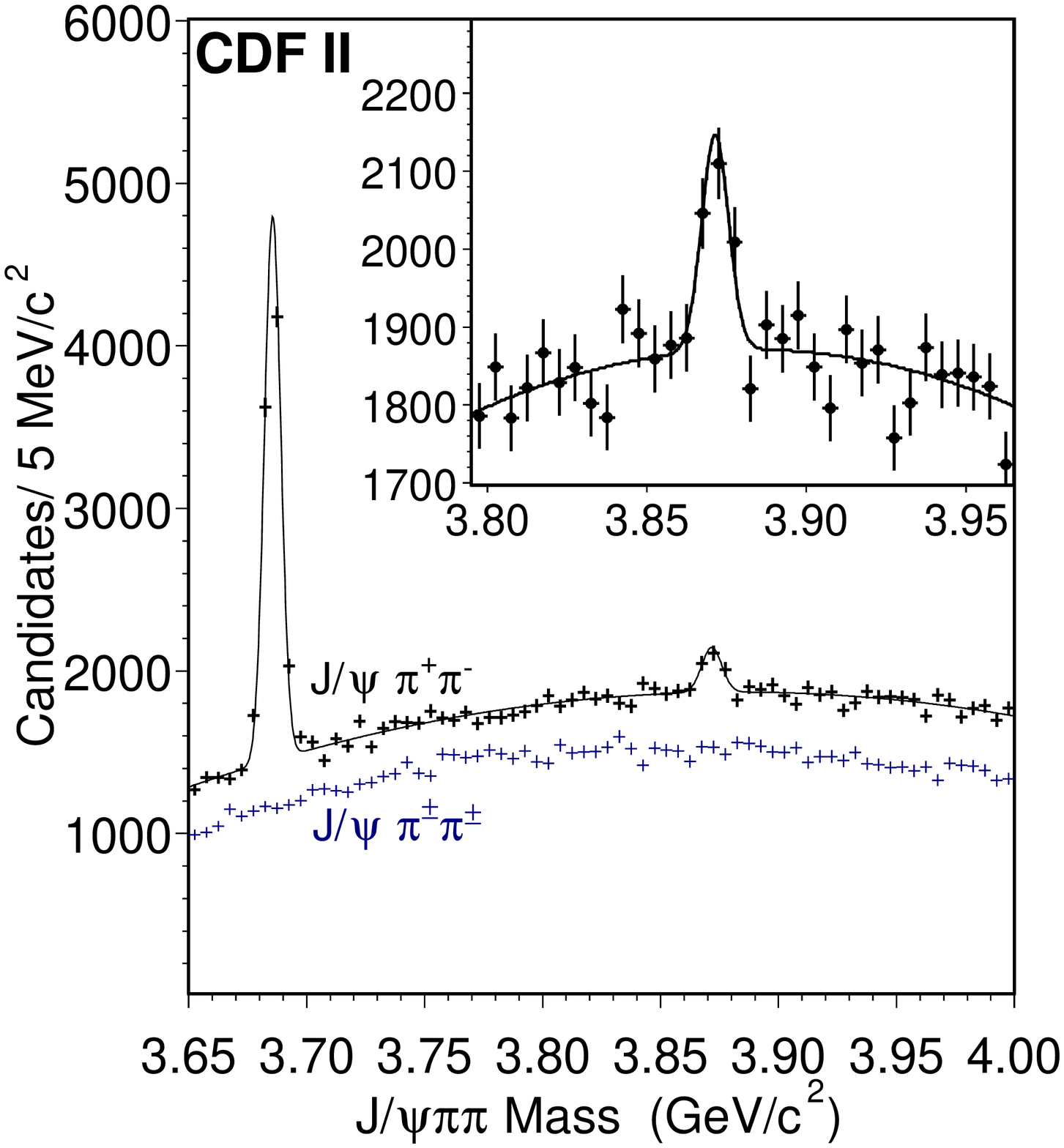,width=2.1in}~~\psfig{file=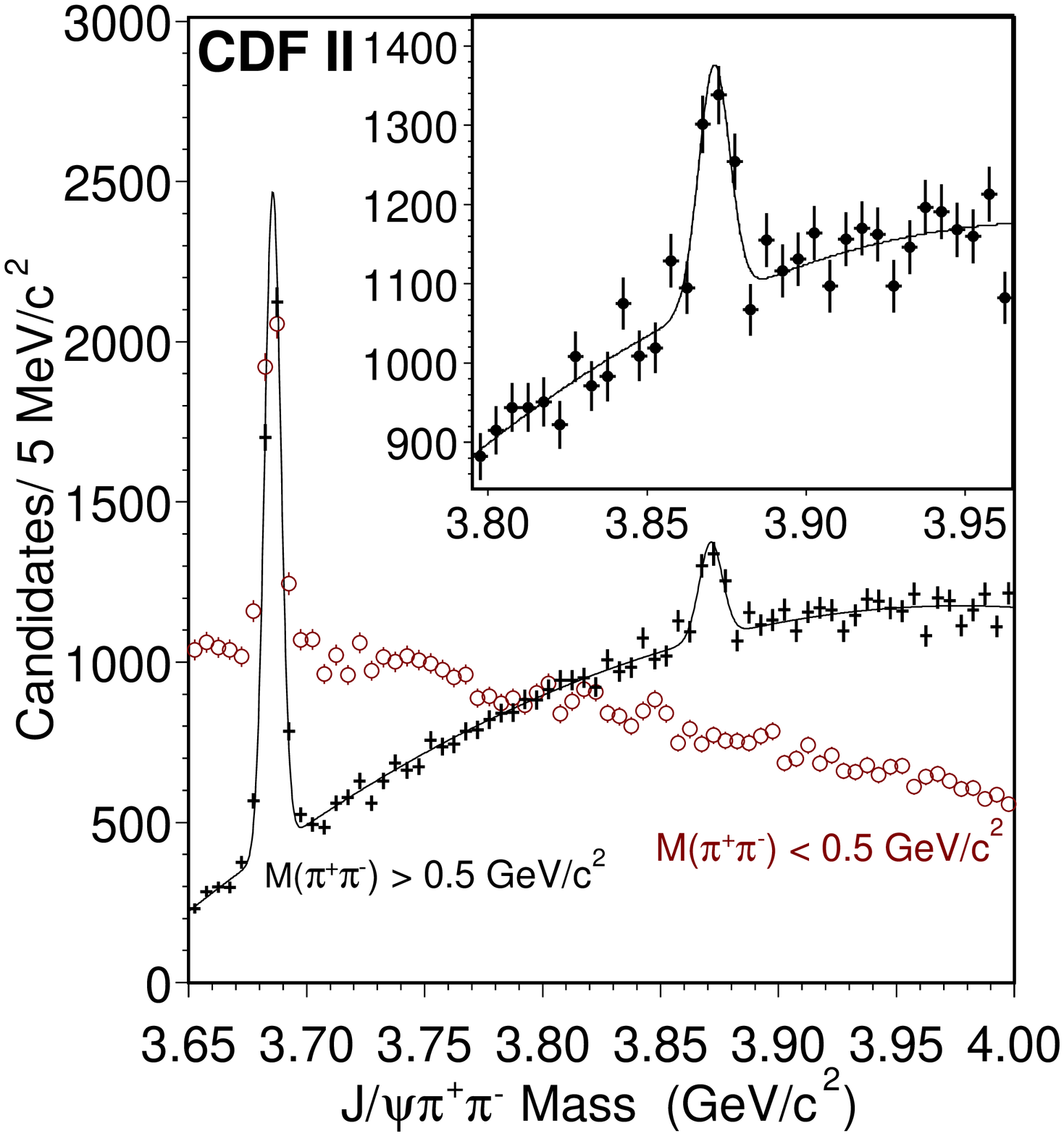,width=2.1in}
           }
\vspace*{8pt}
\caption{{\bf LEFT:} The $J/\psi\pi\pi$ mass distributions for same, and opposite,
          sign pions of the full selection. 
         {\bf RIGHT:} The $J/\psi\pi^+\pi^-$ mass for
           $M(\pi\pi)$ $<500$ and $>500\MeVcc$ subsamples.
         [{\it Figures reprinted with permission from D. Acosta  {\it et al.},  Phys. Rev. Lett. {\bf 93}, 072001 (2004).
         Copyright 2004 by the American Physical Society.}]
\label{Fig:XCDF}
}
\end{figure}

Belle  announced their discovery  of $B^+\!\rightarrow\!X(3872)K^+$ in  August 2003
at the Lepton-Photon Sym\-pos\-ium.\cite{BelleX} 
Coincidently, a continuation  of a Run~I search for the  $^3D_2$ 
was being prepared  in CDF.
Once Belle's preprint appeared, the search was expedited and 
$X \!\rightarrow\! J/\psi\pi^+\pi^-$ was sighted eight days later.
CDF publicly confirmed the $X(3872)$
at a Quarkonium Workshop held at Fermilab in September.\cite{XCDFQuarkOnia}

The CDF search began   with $220\ipb$
of $J/\psi\!\rightarrow\!\mu^+\mu^-$ triggers.
The challenge at the Tevatron is background,
and due to large particle multiplicities  per event
this can be fierce when  combining 
two charged particles to a $J/\psi$.
Because of fluctuations in multiplicity, some events have
many background candidates with little prospect of signal.
A loose preselection was made, and  events  with more
than 12 $J/\psi\pi\pi$ candidates with masses below $4.5\GeVcc$
were rejected.
The preselection was mainly based on track quality cuts
and fitting the  $J/\psi\pi\pi$ system to a common vertex.

The selection was tightened by demanding:
smaller $\mu^+\mu^-\pi^+\pi^-$-vertex fit $\chi^2$'s;
$M(\mu^+\mu^-)$ be within $60\MeVcc$ ($\sim\!4\sigma$)
of the $J/\psi$; $p_T(J/\psi)\!>\!4\GeVc$;
$p_T(\pi)\!>\!400\MeVc$; and  $\Delta R(\pi)\!<\!0.7$ 
for both pions, where $\Delta R(\pi)$ is relative to the  $J/\psi\pi\pi$ system.
The resulting  mass distributions are shown in Fig.~\ref{Fig:XCDF}.
A large  $\psi(2S)$ peak is seen, as well as a smaller bump 
at $\sim\!3872\MeVcc$.
No structure is apparent in  $J/\psi\pi^\pm\pi^\pm$. 
Gaussian fits to the peaks yield $5790\!\pm\!140$  $\psi(2S)$  
and $580\!\pm\!100$ $X(3872)$.

\begin{figure}[t]
\begin{minipage}[b]{0.53\textwidth}
\hspace*{-0.15cm}\mbox{\psfig{file=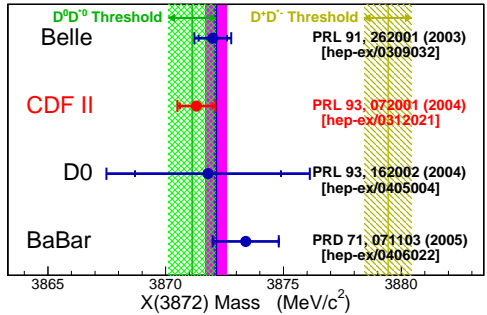,width=\textwidth}}
\end{minipage}\hfill
\begin{minipage}[b]{0.46\textwidth}
\caption{
  Summary of $X$-mass measurements for all observations\protect\cite{BelleXPRL,XCDF,XD0,BaBarX} 
     compared  to the  $D^0\overline{D}{^{*0}}$ and $D^+{D}{^{*-}}$ thresholds.
  Vertical lines indicate central values, and bands
  the range of uncertainty in measured masses---the dark solid band is for the  $X(3872)$. 
 \vspace*{18pt}
\label{Fig:XMassSum}
}
\end{minipage}\hfill
\end{figure}

Belle noted (Fig.~\ref{Fig:XBelle})  that the $X$ strongly 
favored high $M(\pi\pi)$.
CDF  confirmed this by splitting the sample into %those with
 $M(\pi\pi)$ above, and below, $ 500\MeVcc$ 
(Fig.~\ref{Fig:XCDF}).
No $X$-signal is discernible in the low-mass sample.
For high-$M(\pi\pi)$ the $X$-mass is
$3871.3\!\pm\!0.7\!\pm\!0.4\MeVcc$,
with a resolution dominated  $\sigma$  of $4.9\!\pm\!0.7\,(stat)\MeVcc$.
This mass is in good agreement with, and similar precision to 
Belle's (Fig.~\ref{Fig:XMassSum}).
The remarkable proximity of the $X$ to the  $D^0\overline{D}{^{*0}}$ threshold
fuels molecular speculations.

\subsubsection{$X(3872)$ Production at CDF$\,$\protect\cite{XCDFLife} \label{sec:XProd} }

Properties of $X$ production present an opportunity
to garner insights  into its nature.
Given Belle's discovery, $B$'s
are clearly  an important  source of the $X$,
but is this  how  CDF's signal arises?
If not, can direct $X$ production 
in $\bar{p}p$ collisions  shed light into its nature?
Specifically, does $X$ production in CDF differ from  charmonia?

Charmonia production has been extensively studied in $\bar{p}p$,\cite{CDFProdOctet}$^-$\cite{QuarkoniumYellowReport} 
and provided the experimental impetus\cite{CDFProdOctet} 
%for the so-called ``color octet'' model.\cite{ColorOctet}
for the so-called  "NRQCD factorization model."\cite{ColorOctet}
At the Tevatron, charmonia arise as a mixture of ``direct'' production
from fragmentation plus feeddown from higher-mass states.
An important source of feeddown is  $b$-hadrons:
they produce  $\sim\!10-20\%$ of $J/\psi$, $\chi_c$, and $\psi(2S)$.
The actual fractions depend upon species and $p_T$.
If the $X$ is not  simple $c\bar{c}$, it may have a very different
production rate, particularly if it is a fragile
molecule bound by only an\MeV\ or so.

A standard method\cite{CDFProdOctet}
to separate $b$ sources from ``prompt,''
{\it i.e.} either directly pro\-duced or from decays of short-lived particles,
is to measure a particle's apparent ``lifetime.''
Since the $X$ does not decay weakly,
its true  lifetime is far too short
for it to travel a discernible distance.
Any observed displacement, $ L_{xy}$ (Eq.~\ref{eq:Lxy}), is ascribed to
 ``$b\!\rightarrow\!X\ldots$'' decays.
In the $X$ selection   $p_T(J/\psi)$ is above $4\GeVc$, ensuring sufficient boost
such that $b$ decays can not mimic prompt production.
The  displacement is converted into 
``uncorrected proper-time'' by $ {ct}\! \equiv\! M\cdot\! L_{xy}/p_T$.
This is ``uncorrected'' because the mass and $p_T$ of the $J/\psi\pi^+\pi^-$
are only part of the  $b$-decay,
and so  $ct$ is not the true proper decay-time.
The $ct$ distribution will not give the correct $b$ lifetime,
but it still quantifies the {\it fraction}
of $b\!\rightarrow\!X\ldots$ decays.

D\O\ took a step in this direction when they
compared the fractions of signal that had  $ct\!>\!100\,\mu$m, 
and found $30.0\!\pm\!1.8\,(stat)\%$ for   $\psi(2S)$ and
$31.8\!\pm\!6.7\,(stat)\%$ for $X$.\cite{XD0}
By this measure the  states look identical, but 
the  prompt and  $b$   production sources are not 
actually disentangled, nor is the $ct$-resolution specified.
Parenthetically we note that
D\O\ considered other production features using this 
type of binary comparison.
In each case  the $X$ and $\psi(2S)$ were indistinguishable;
but lacking theoretical models one cannot assess the
significance of such null comparisons.

CDF's separation\cite{XCDFLife}   of prompt
and $b$ components begins with the same sample used
in the mass measurement.
Since precise vertexing is fundamental for measuring $L_{xy}$,
additional SVX and beamline criteria are applied.
The sample is reduced by $\sim\!15\%$, where
the main loss is from rejecting candidates
with $L_{xy}$ errors above $125\,\mu$m.
An unbinned likelihood fit is performed in mass and $ct$
to obtain the long-lived fraction.
The mass is modeled by a Gaussian for signal and a quadratic polynomial
for background.
In $ct$, the long-lived signal is an exponential smeared
by the resolution function (double Gaussian), 
and the prompt part is the resolution function.
Long-lived backgrounds
are also modeled by resolution smeared exponentials.

%%%%%%%keep all lifetime plots together!
\begin{figure}[t]
\centerline{\psfig{file=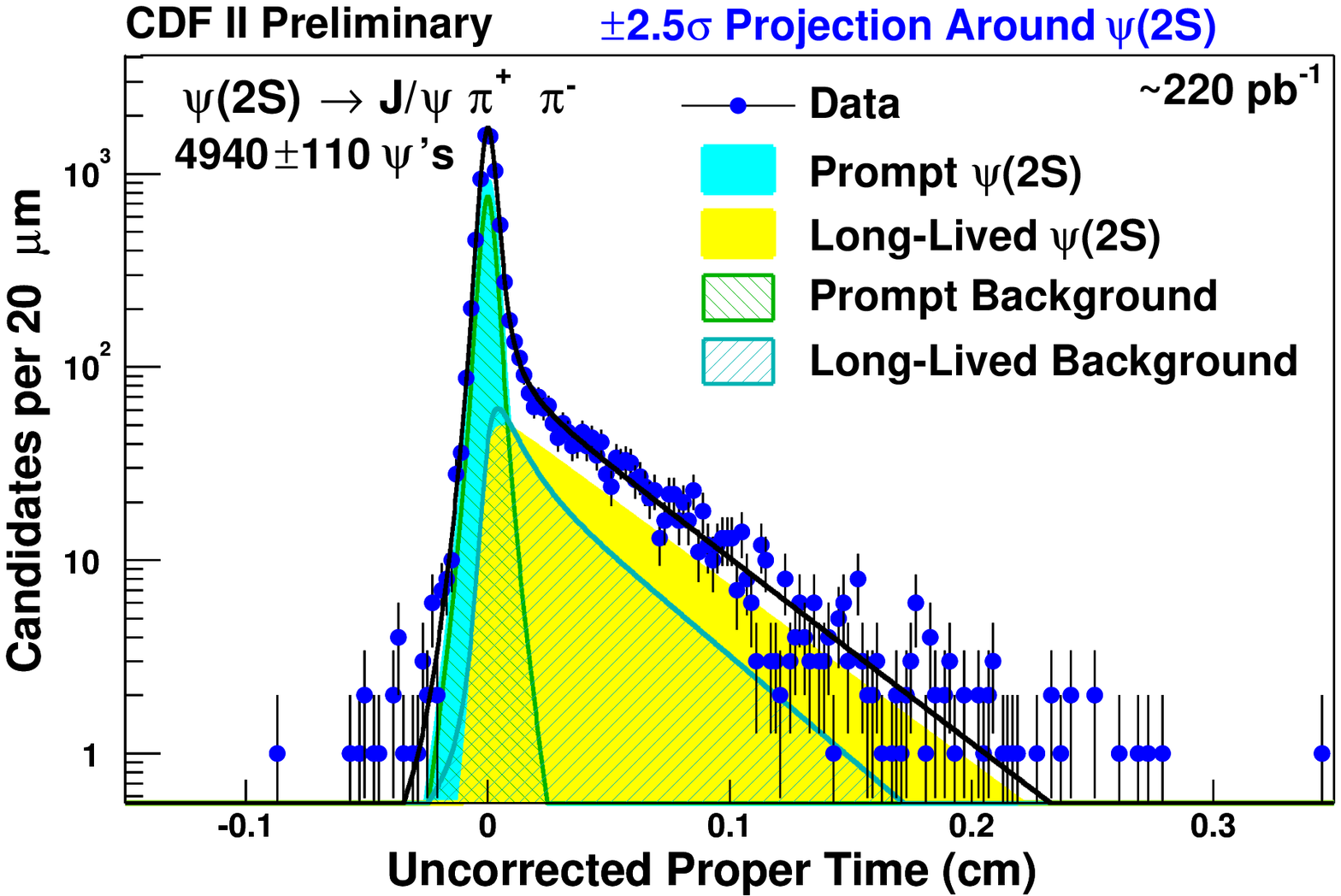,width=2.4in}\psfig{file=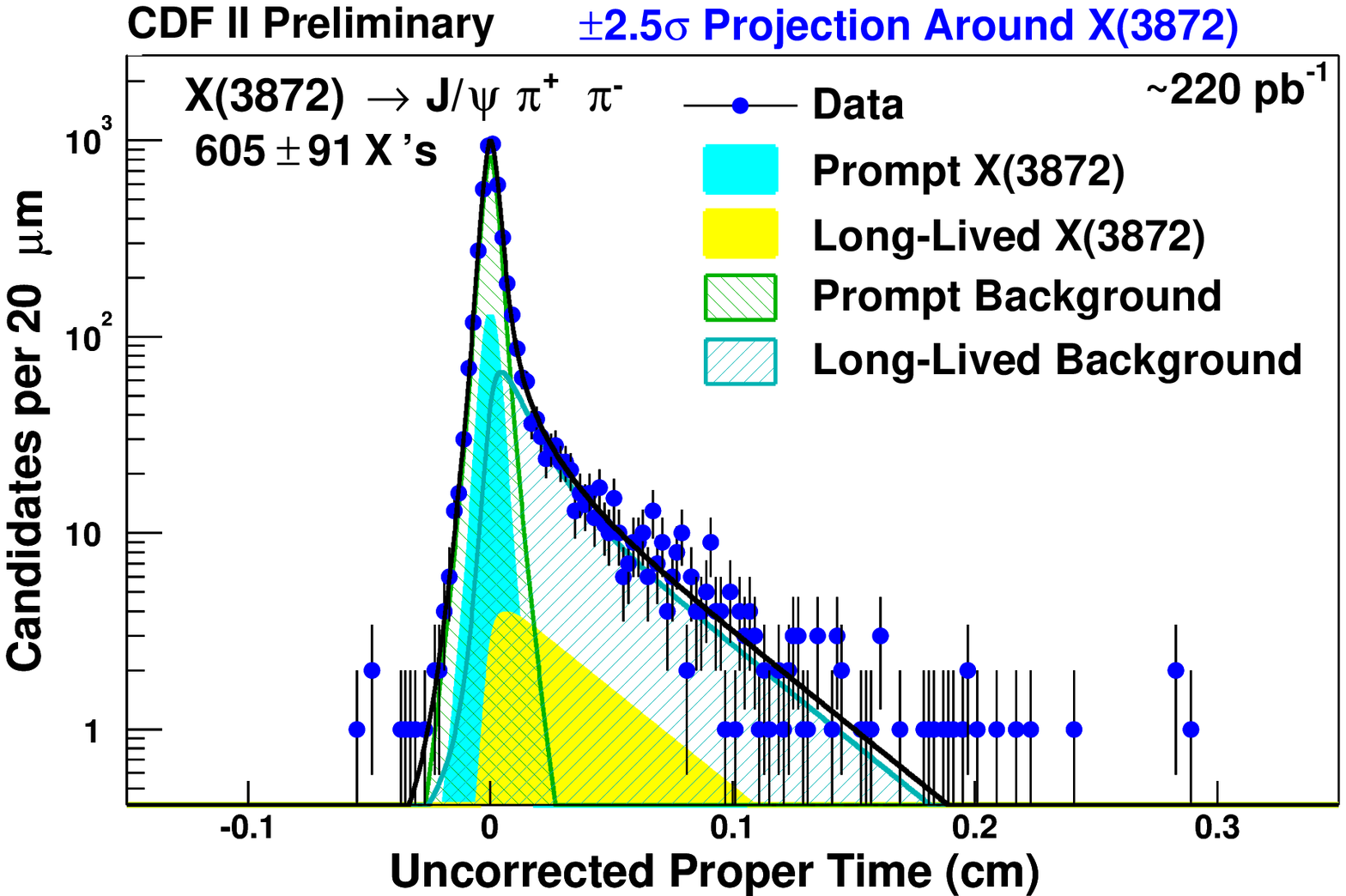,width=2.4in}
           }
\vspace*{8pt}
\caption{``Lifetime'' projections of likelihood fits onto data.
         {\bf LEFT:} The  $\psi(2S)$  distribution  with full PDF and its breakdown
         into signal (shaded) and background (hatched) classes.
         Signal and background are further separated into
         prompt and long-lived components.
         The projection is
         for candidates within $\pm2.5\sigma$ of the $\psi(2S)$ mass
         in order to be reflective of its signal-to-background ratio. 
         {\bf RIGHT:} Corresponding distribution for the  $X(3872)$.
\label{Fig:CDFLong-Lived}
}
\end{figure}

The fit results are portrayed in Fig.~\ref{Fig:CDFLong-Lived} 
by projecting the likelihood PDF onto  the 
$ct$ distribution of the data, which is well described.
In this sample  $28.3\!\pm\!1.0\!\pm\!0.7\,$\% of $\psi(2S)$'s 
are long-lived---similar to Run~I.\cite{CDFProdOctet} 
The $M(\pi\pi)\!>\!500\MeVcc$ sample is used for  the $X$ fit, 
but the signal is still deeply buried in background
in the $ct$ projection.
The long-lived $X$-fraction is  $16.1\!\pm\! 4.9 \!\pm\! 2.0\,\%$,
\mbox{which is smaller than the $\psi(2S)$,} 
but only by a bit more than $2\sigma$.
The  {\it absence} of $b\!\rightarrow\!X$-decays is excluded by  $3\sigma$ based on
Monte-Carlo  ``pseudo-experiments.''
It must  be stressed that these fractions
depend on the  sample selection, 
mainly $p_T$,\cite{CDFProdOctet} 
and are therefore {\it sample specific}.

CDF's long-lived fractions for $X$ and $\psi(2S)$ are quite similar, 
but factors that  might otherwise distinguish
$X$ production from $c\bar{c}$ may scale $\bar{p}p\!\!\rightarrow\!\!X$ 
and $b\!\!\rightarrow\!\!X$  rates together, {\it canceling} in the ratio.
Indeed, an analysis of inclusive $X$ production\cite{BraatenXPro}  in the
NRQCD  formalism\cite{NRQCD} lends credence to this view. 
Although posed in molecular terms,
the arguments are more general:
matrix elements for the~$X$~as~$1^{++}$ are argued
to scale  with  those of the $\chi_{c1}$, yielding universal
$X$-to-$\chi_{c1}$  scaling  in inclusive processes.
By setting the scale with  
a measured   $B\!\rightarrow\!X$ branching ratio, other
production ratios are predicted---like those below (Tables~\ref{Tab:XSecSumm} and~\ref{Tab:BrSumm}).
The predictions are crudely successful, but
they only   test internal consistency amongst the data,
as  an $X$ data-point must set the scale.
We take the larger lesson of this analysis to be a  case
for a  more {\it  general} insensitivity of inclusive production ratios, such as $B$
decay relative to $\bar{p}p\!\rightarrow\!X$.
Thus, the long-lived $X$ fraction measured by CDF
is probably not so telling.
A more incisive test is to consider the prompt and $b$ sources separate\-ly,
but we lack models for crisp predictions 
as well as knowledge of the  branching ratio 
${\cal B}_X \equiv {\cal B}_X[X\!\rightarrow\! J/\psi\pi^+\pi^-]$.
Still, we may  \mbox{forge ahead with some crude comparisons.}

Using CDF's $X(3872)$ and $\psi(2S)$ yields, $N_X$ and $N_{\psi}$ (Fig.~\ref{Fig:CDFLong-Lived}),
and long-lived fractions $f_{LL}$,
one can estimate the  production rate of $X$ 
relative to $\psi(2S)$,
{\it i.e.},
\begin{equation}
   \frac{\sigma(\bar{p}p\!\rightarrow\! X\ldots)}{\sigma(\bar{p}p\!\rightarrow\! \psi(2S)\ldots)} =
   \frac{ (1-f^X_{LL}) N_X       }
        { (1-f^\psi_{LL}) N_{\psi}  }\cdot
   \frac{{\cal B}_\psi[\psi(2S)\!\rightarrow\! J/\psi\pi^+\pi^-]}{{\cal B}_X[X \!\rightarrow\! J/\psi\pi^+\pi^-] }\cdot
   \frac{\epsilon_{\psi}}{\epsilon_X},
\label{eq:XSec}
 \end{equation}
where  ${\epsilon_X}/{\epsilon_{\psi}}$ is the (unreported) ratio 
of CDF efficiencies for $X$ and $\psi(2S)$.
Given the relatively soft kinematic cuts,  ${\epsilon_X}/{\epsilon_{\psi}}$
likely deviates from unity by
tens of percents rather than factors of two\cite{Sasha}---a modest uncertainty
for our purposes. 
The results are shown in Table~\ref{Tab:XSecSumm} along with CDF data
for $J/\psi$\cite{CDFProdOctet} 
and $\chi_c$,\cite{chicExp} where
the $b$-hadron feeddown was removed by a lifetime analysis,
as well as  that from $\psi(2S)$ and  $\chi_c$ to $J/\psi$.
These values are corrected for efficiency, unlike the crude 
estimate done here for the $X$---so that we must preserve
the ${\epsilon_X}/{\epsilon_{\psi}}$ factor.
The cross section ratios are  known to vary mildly with $p_T$,
making the values in Table~\ref{Tab:XSecSumm} depend on the  $p_T$ range.
This is a potentially important {\it caveat} for the $X$, as its $p_T$ behavior 
is (so-far) unknown.\cite{SashaPt}
With these qualifiers, we can compare the measured  production ratios.
It has been  estimated that production of some $D$-states can be nearly 
as large as the $\psi(2S)$.\cite{DProdChao}
The $X$ plausibly follows a $c\bar{c}$ pattern
{\it if} $\, 2\% \!\lesssim \!{\cal B}_X\!\lesssim \!10\%$.
A much larger ${\cal B}_X$  suppresses the cross section,
{perhaps indicating a non-$c\bar{c}$ character.}

\begin{table}[t] 
\tbl{Ratio of charmonium production cross sections 
relative to the $\psi(2S)$ derived from CDF measurements  at the Tevatron\protect\cite{CDFProdOctet,chicExp}
and PDG\protect\cite{PDG04} branching ratios.
The $X(3872)$ ratio is determined from the raw measurement of the CDF lifetime analysis,
and requires an efficiency correction, ${\epsilon_{\psi}}/{\epsilon_X}$.
\label{Tab:XSecSumm} }
{
\begin{tabular}{c c c c  c}
\hline\hline
State               & ~~~~~         & $p_T$ Range ($\!\!\GeVc$)  &~~  & $\sigma[c\bar{c}]/\sigma[\psi(2S)]$  \\ \hline
$J/\psi$            & ~~~~~         & $>5.5$ &  & $\sim\!5.0 \pm 1.0$          \\
$\chi_{c1}$         & ~~~~~         & $>5.5$ &  & $\sim\!4.3 \pm 1.1$          \\ 
$\psi(2S)$          & ~~~~~         &        &  &    1          \\ 
\hline
$X(3872)$           & ~~~~~         & $\int\!\epsilon$(CDF Analysis)$\cdot d p_T$ &   &$(0.045\pm0.008)/{\cal B}_X \cdot {\epsilon_{\psi}}/{\epsilon_X}$ \\
\hline\hline
\end{tabular}
}
\end{table}

Adapting Eqn.~\ref{eq:XSec} to
CDF's long-lived component, one can 
estimate the {\it inclusive} branching ratio
of  ``$B^+/B^0/B^0_s/b$-$baryon$'' mixture decaying to $X\!+\!anything$ 
relative to that for $\psi(2S)$.
Then, ${\cal B}(b\!\rightarrow\! X\ldots)$ may be obtained from multiplication
by the known  ${\cal B}(b\!\rightarrow\! \psi(2S)\ldots)$.
Table~\ref{Tab:BrSumm}  lists the result along with  known inclusive branching ratios for $c\bar{c}$ states,
as well as the  corresponding  {\it exclusive} ${\cal B}(B^+ \!\rightarrow\! [c\bar{c}] K^+)$.
${\cal B}(B^+ \!\rightarrow\! X K^+)$  is an average of  
$B$-factory measurements, up  to  the unknown   ${\cal B}_X$.
Both the {\it inclusive} and {\it exclusive} branching ratios tell a familiar story:
modest   ${\cal B}_X$ pushes  $b\!\rightarrow\!X$ branching ratios
into the $c\bar{c}$ realm, and large   ${\cal B}_X$   implies  suppression.
The last column shows the ratio of exclusive to inclusive
branching ratios: the $X$ is consistent---{\it independent}
of  ${\cal B}_X$--- with $c\bar{c}$,
albeit with very large errors.

%===================== Br Ratios:

\begin{table}[t] 
\tbl{Exclusive $B^+\!\rightarrow\! [c\bar{c}] K^+$ branching ratios are compared
to inclusive  branching ratios 
for ``$B^+/B^0/B^0_s/b$-$baryon$'' mixture decaying to  charmonium,
and to the $X(3872)$.
Charmonium values are from the PDG\protect\cite{PDG04} unless otherwise noted, 
the exclusive $X$ is a  Belle\protect\cite{BelleXPRL}  and {\sc BaBar}\protect\cite{BaBarX} 
 average (updated to PDG`04),
and the inclusive X is derived from CDF's lifetime analysis.
The $X$ values have residual unknowns: ${\cal B}_X(X\!\rightarrow\! J/\psi\pi^+\pi^-)$,
and CDF's $X$-to-$\psi(2S)$ efficiency ratio, ``${\epsilon_X}/{\epsilon_{\psi}}$.''
\label{Tab:BrSumm} }
{
\begin{tabular}{c @{}c c c c }
\hline\hline
\multicolumn{2}{c}{State} & ${\cal B}(B^+\!\rightarrow\! [c\bar{c}] K^+)$   $\times10^{-4}$
                 & ${\cal B}(b  \!\rightarrow\! [c\bar{c}]\ldots)$ $\times10^{-2}$ & Ratio \\ \hline
$\eta_c$     & $\,(^1S_0)$    &   $9.0    \pm 2.7$             &     ---             &  --- \\
$J/\psi$     & $\,(^3S_1)$    &  $10.0 \pm 0.4$~            &  $1.16\pm0.10$      &  $8.6\pm0.8$\,\% \\
$\chi_{c0}$  & $\,(^3P_0)$    &   $6.0 \pm 2.3$             &     ---             &  --- \\
$\chi_{c1}$  & $\,(^3P_1)$     &   $6.8 \pm 1.2$             &  $1.5\pm0.5$        &  $4.5\pm1.7$\,\%\\
$\psi(2S)$   & $\,(^3S_1)$    &   $6.8 \pm 0.4$             &  $0.48\pm0.24$      &  $14\pm7$\,\%\\
$\psi(3770)$ & $\,(^3D_1)$     &~~~~$4.8 \pm 1.3$\protect\cite{Belle3770}           &  ---      &  --- \\
\hline
$X(3872)$    & $\,(??)$   &   $(0.14\pm0.03)/{\cal B}_X$  &  $(0.011\pm0.006)/{\cal B}_X\cdot{\epsilon_{\psi}}/{\epsilon_X}$   
                          &  $(13\pm 8) \cdot {\epsilon_{\psi}}/{\epsilon_X}\,\%$ \\
\hline\hline
\end{tabular}
}
\end{table}

With modest ${\cal B}_X$, say $\sim\!2$-$10$\%, the $X$ falls into line with
the standard  $c\bar{c}$ in Tables~\ref{Tab:XSecSumm} and~\ref{Tab:BrSumm}. 
Alternatively, large  ${\cal B}_X$,
as in some exotic scenarios, could imply production and $b$-decay rates 
suppressed by up to an order of magnitude.
Thus the lesson to be learned  hinges upon the
size of  ${\cal B}_X(X\!\rightarrow\! J/\psi\pi^+\pi^-)$.
{\sc BaBar} has recently shown promising results indicating that they 
hope to soon measure ${\cal B}_X$.\cite{XBr}

\subsubsection{The Dipion Mass Spectrum$\,$\protect\cite{XCDFPiPi}  }
\label{sec:DiPionSec}

A feature of $X(3872)$ decay is its propensity for high-mass dipions 
(Figs.~\ref{Fig:XBelle} \& \ref{Fig:XCDF}).
Dipion spectra are often noted as window to the  $X$.
As is well known, $\psi(2S)\!\rightarrow\! J/\psi\pi^+\pi^-$ 
prefers high $M(\pi\pi)$.\cite{BESpsi2S}
High  masses are no surprise for the $X$ as  $c\bar{c}$ in a $^3S_1$---but this is untenable as it should
then be directly made in $e^+e^-$. 
Interest in   $\psi(2S)$ decay lead
to  general treatments of $\pi\pi$-transitions between quarkonia.
Dipion spectra  have been calculated using a QCD multipole expansion (ME)
of the color elec\-tric/mag\-net\-ic fields 
for  $^3S_1$,\cite{ME-TMYan} $^1P_1$,\cite{ME-Kuang} and $^3D_J$\cite{ME-TMYan}
$c\bar{c}$ going to  $^3S_1 \pi^+\pi^-$.
Other $J^{PC}$ states involve, at  lowest $L$, dipions
in a $1^{--}$, and for the masses of interest,  are dominated by the $\rho$-pole.
The  ME predicts that  $M(\pi\pi)$ favors low masses
for  $^1P_1$, and is  {\it relatively} {flat} for  $^3D_J$-states,
both at odds with Fig.~\ref{Fig:XBelle}.
The  $^3S_1$ and $\rho$ options do so peak. 
Normally  $[c\bar{c}]\!\rightarrow\!J/\psi \rho^0 $ is forbidden by isospin,
but  a state so close to the  $D^0\overline{D}{^{*0}}$
mass (Fig.~\ref{Fig:XMassSum}) can  violate isospin 
via  virtual coupling to  $D^0\overline{D}{^{*0}}$.

\begin{figure}[t]
\centerline{\psfig{file=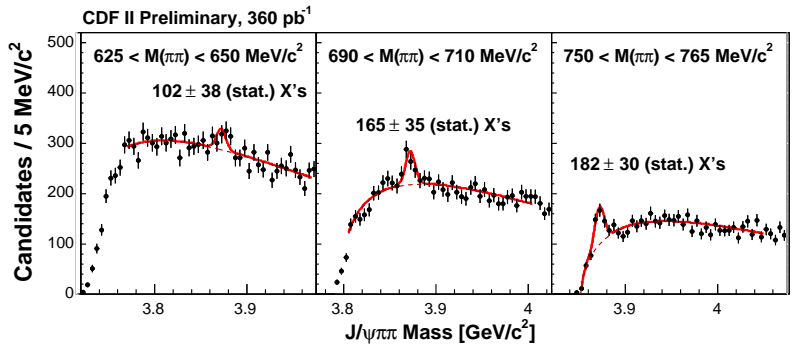,width=4.40in}
           }
\vspace*{8pt}
\caption{Three examples of $M(\pi\pi)$ ``slices'' around the $X$ of the
         $J/\psi\pi^+\pi^-$ mass distribution. % used to extract the $X$'s dipion spectrum.
\label{Fig:Slices}}
\end{figure}

Belle's original observation gave clear evidence for high  $\pi\pi$-masses, 
but only a rough shape.
CDF's large sample offers a sharper view.\cite{Sasha,XCDFPiPi,Kai}
An enlarged sample of  $\sim\!360\ipb$ is used.
The selection is as before, except fiducial cuts are applied to select 
a kinematic region of good
efficiency: {$p_T(X)\!>\!6\GeVcc$ and} $|\eta(X)|\!<\!0.6$. 
The sample is divided into slices of $M(\pi\pi)$,
and the  $J/\psi\pi^+\pi^-$  distribution is fit  to
obtain the signal yields for each slice (Fig.~\ref{Fig:Slices}).
The raw yields are corrected for detector and 
kinematic selection efficiencies using Monte Carlo simulation.
An  important ingredient is the simulation's $p_T$ spectrum.
This was varied so that the simulation matched 
the observed spectra for the $\psi(2S)$ and $X$.
In this way no assumption was made about the nature of $X$ production.
Within the limited precision, $p_T(X)$
is quite similar to that of the $\psi(2S)$.
The statistical error on the  $p_T(X)$ shape
is propagated into  a small systematic uncertainty on the $M(\pi\pi)$ efficiency 
corrections.

\begin{figure}[t]
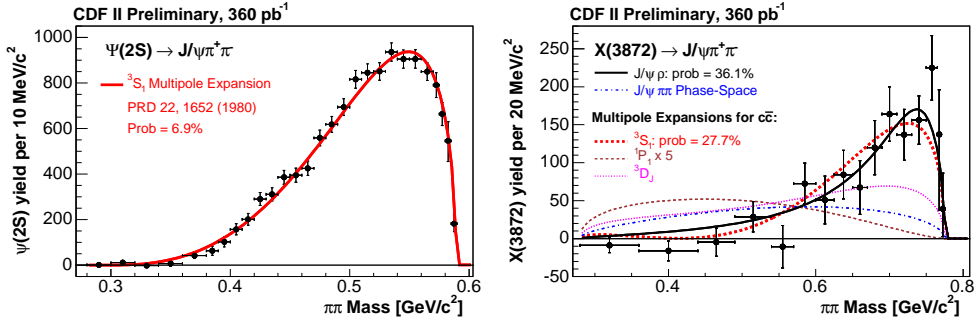

\centerline{\psfig{file=PsiTheoryFit-360-no12-ptX6-etaX06.eps-1,width=2.6in}\psfig{file=XTheoryFit-360-no12-ptX6-etaX06.eps-1,width=2.6in}
           }
\vspace*{8pt}
\caption{
         {\bf LEFT:}  Dipion spectrum for $\psi(2S)$ fit with a multipole
         expansion calculation.
         {\bf RIGHT:} Dipion spectrum for $X(3872)$ with fits 
         of multipole predictions for $^3S_1$, $^1P_1$, and $^3D_J$ charmonia,
         as well as a phase-space modulated Breit-Wigner (constant width) distribution
         for decay to $J/\psi \rho^0$, and three-body phase space.
         The $^1P_1$ fit is multiplied by 5 for better visibility.
\label{Fig:DiPionM}
}
\end{figure}

The efficiency corrected spectrum for the $\psi(2S)$ is shown
in Fig.~\ref{Fig:DiPionM}, along with a fit of  a multipole expansion model.\cite{ME-TMYan}
This model has been fit to higher statistics (23k) BES data,\cite{BESpsi2S}
and the CDF results agree with BES better than $1\sigma$.

Also  in  Fig.~\ref{Fig:DiPionM} is the corrected $X$ spectrum,
along with fits for $^3S_1$, $^1P_1$, and $^3D_J\!\rightarrow\!J/\psi\pi^+\pi^-$ ME's,
the $\rho$ (Breit-Wigner$\times$phase-space),
and simple phase space.
Only the  $^3S_1$ and  $\rho$ fits describe the data---the
two shapes are almost indistinguishable.
The  $^3S_1$  $c\bar{c}$  assignment for the $X$ being  untenable
 {\it  seemingly} forces the  $\rho$ option.

However, $\Upsilon$'s serve as a cautionary tale:
the basic ME {\it fails} to describe 
$\pi\pi$-masses for  $\Upsilon(3S)$ $\!\rightarrow\!\Upsilon(1S)\pi^+\pi^-$.\cite{ME-FailsCLEO} 
One hypothesis is that the $\Upsilon(3S)$ is  \mbox{so close  to the $B\overline{B}$}
threshold that coupling to $B\overline{B}$ distorts the  spectrum.\cite{ME-Moxhay}
This scenario has been challenged as inadequate,\cite{ME-AntiMoxhay}
but the mechanism itself is quite conventional.
Whatever the $X$ is, it is well situated to couple to $D^0\overline{D}{^{*0}}$,
potentially affecting $M(\pi\pi)$.

A definitive test for the $\rho$ is $X\!\rightarrow\! J/\psi\pi^0\pi^0$---forbidden 
for $\rho$'s, \mbox{but half the $\pi^+\pi^-$} rate for $I\!=\!0$  dipions.
But  $B$-factories are not  yet  sensitive.\cite{BelleOmega}
Belle has reported $X\!\rightarrow\!$ $J/\psi\pi^+\pi^-\pi^0$,
where the pions look like a virtual $\omega$.\cite{BelleOmega} 
The case would be complete with $J/\psi\,\omega$ decay:
the $\omega$  requires the dipions in $J/\psi\pi^+\pi^-$ 
to be odd $C$-parity, and thus a $\rho$.
Belle quotes an  $\omega$-to-$\rho$ branching ratio of 
 $1.0\!\pm\!0.5$\cite{BellePsiGammaSIS05},    
signaling large isospin breaking.
Very recently Belle reported  $J/\psi\gamma$ decay,\cite{BellePsiGammaSIS05} 
providing compelling support for the $\rho$. 
Confirmation may be desired,  but all this fits neatly  into a  picture 
where the $X$ has  $C\!=\!+$, 
and decays into $J/\psi\rho$  and $J/\psi\,\omega$ with isospin  badly broken.

Belle has pushed the $\rho$-analysis a step further by noting that
a Breit-Wigner is distorted by a centrifugal barrier 
if the $J/\psi$-$\rho$ angular momentum, $L_{\psi\rho}$, is non-zero.
A phase-space factor, the $J/\psi$ momentum in the $X$ rest-frame, $q^*_{\psi}$,
generalizes to $(q^*_{\psi})^{2L_{\psi\rho}+1}$.
Higher $L_{\psi\rho}$ softens the  $M(\pi\pi)$ fall-off at the upper limit ($q^*_{\psi}\!\rightarrow\!0$), 
and the  $\pi\pi$-peak shifts to lower masses.
The fit in Fig.~\ref{Fig:DiPionM} corresponds to $L_{\psi\rho}\!=\!0$, and 
CDF has not yet provided an $L\!=\!1$ fit.
But, as with Belle data,\cite{BelleNewMass} 
the agreement will clearly deteriorate---favoring 
an $S$-wave decay, and even parity for the  $X$.

\subsection{$X(3872)$ Reprise}

The identity of the $X(3872)$ is a pressing issue in spectroscopy.
The natural interpretation is a $c\bar{c}$ state.\cite{ccbarOptions1,ccbarOptions2}
In  an effort to sort out options,
an extensive search has been made for other decays---none are seen in:
$\chi_{c1}\gamma$,\cite{BelleXPRL}   
$\chi_{c2}\gamma$,\cite{BelleFailSearch}
$J/\psi\eta$,\cite{NoJPsiEta}
$D^+D^-$ and  $D^0\overline{D}{^0}$,\cite{Belle3770}
but, very recently, $J/\psi\gamma$\cite{BellePsiGammaSIS05} 
and  $D^0\overline{D}{^0}\pi^{0,}$\cite{BelleDDPi0} have been.
In the end, a case can be made against {\it all}  $c\bar{c}$ can\-di\-dates, 
as is summarized in Table~\ref{Tab:AntiCharmOnia}.
But the {\it caveat} is:
once one concedes that the $X$ is unusual---and sitting on $D^0\overline{D}{^{*0}}$ 
offers some grounds---then the usual $c\bar{c}$ expectations may be questioned.
But we go on to consider alternatives:
 1)~four-quark  states\cite{Maiani,XMolecule,SwansonMolecule}, 
2)~$c\bar{c}g$  hybrids,\cite{XHybrid}$^-$\cite{CloseYamamotoHybrid} 
3)~$c\bar{c}$-glue\-ball mixtures\cite{ccGlueMix}, or
4)~dynamic ``cusp''  from the   $D^0\overline{D}{^{*0}}$ threshold.\cite{XCusp}

\begin{table}[t] 
\tbl{Summary of arguments against $c\bar{c}$ assignments for the $X(3872)$.
This ignores mass predictions from potential models,
which also creates varying degrees of problems for $c\bar{c}$ states.\protect\cite{ccbarOptions1,ccbarOptions2}
The dipion  $J^{PC}$ is for lowest $L$.
``Unseen modes'' are expected to have been observed if the $X$ is that state.
\label{Tab:AntiCharmOnia} }
{
\begin{tabular}{c@{} c@{} c  c | c c c}
\hline\hline
\raisebox{-1.0ex}[0.0ex][0.0ex]{$n\,^{2s+1}L_J$}    &
\raisebox{-1.0ex}[0.0ex][0.0ex]{   $J^{PC}$       } &
\raisebox{-1.0ex}[0.0ex][0.0ex]{  State           } & $\pi\pi$ &Unseen& 
\raisebox{-1.0ex}[0.0ex][0.0ex]{  Other Objections  }     \\
                   &            &              &  $J^{PC}$ & Mode      &           \\ \hline  
 \raisebox{-1.0ex}[0.0ex][0.0ex]{$1\, ^1\!D_2$}   &\raisebox{-1.0ex}[0.0ex][0.0ex]{ $2^{-+}$}   
&\raisebox{-1.0ex}[0.0ex][0.0ex]{$\eta_{c2}$}   &\raisebox{-1.0ex}[0.0ex][0.0ex]{$1^{--} $} &      
& $J/\psi\pi^+\pi^-$ expected to be very small \\ 
                   &            &              &          &           & ($\eta_c\pi^+\pi^- \gg J/\psi\pi^+\pi^-$)\cite{ccbarOptions1}\\ 
                   &            &              &          &           & $M(\pi\pi)$ in $J/\psi\rho$ decay favors $S$-wave $\rightarrow$ Even Parity\\ 
$1\, ^3\!D_2$        & $2^{--}$   & $\psi_2 $    &$0^{++} $ & $\chi_{c1}\gamma$\protect\cite{BelleXPRL}  &  
      $J/\psi\rho$,\cite{XCDFPiPi,BelleNewMass}  $J/\psi\omega$,\cite{BelleOmega} \&  $J/\psi\gamma$\cite{BellePsiGammaSIS05} decays $\rightarrow C\!=\!+$\\ 
$1\, ^3\!D_3$        & $3^{--}$   & $\psi_3 $    &$0^{++} $ & $\chi_{c2}\gamma$\protect\cite{BelleFailSearch}  &  
      $J/\psi\rho$,\cite{XCDFPiPi,BelleNewMass}  $J/\psi\omega$,\cite{BelleOmega} \&  $J/\psi\gamma$\cite{BellePsiGammaSIS05} decays $\rightarrow C\!=\!+$\\
\hline  
$2\, ^1\!P_1$        & $1^{+-}$   & $h_c' $      &$0^{++} $ &           &  Wrong $\cos\theta_{J/\psi}$ distribution\protect\cite{BelleOmega}    \\
$2\, ^3\!P_0$        & $0^{++}$   & $\chi_{c0}'$ &$1^{--} $ & $D\overline{D}$\cite{Belle3770} &$D\overline{D}$ not suppressed $\rightarrow$ too broad\\
                     &            &              &          &    & Wrong $\ell$-$\pi$ angular dist. in $J/\psi\pi\pi$ decay\cite{BellePsiGammaSIS05}\\ 
                     &            &              &          &           &Not Seen in $\gamma\gamma$ Fusion\protect\cite{CLEOXgammagamma}\\
$2\, ^3\!P_1$        & $1^{++}$   & $\chi_{c1}'$ &$1^{--} $ & % $J/\psi\gamma$\protect\cite{BelleFailSearch}     &  \\
                                      & $Br(J/\psi\gamma)/Br(J/\psi\pi\pi)\!=\!0.14\!\pm\!0.05$\cite{BellePsiGammaSIS05}---too small\cite{BelleNewMass} \\
$2\, ^3\!P_2$        & $2^{++}$   & $\chi_{c2}'$ &$1^{--} $ & $D\overline{D}$\cite{Belle3770}   &$D\overline{D}$ not suppressed $\rightarrow$ too broad\\ 
                     &            &              &          &           &Not seen in $\gamma\gamma$ Fusion\protect\cite{CLEOXgammagamma}\\
\hline
 \raisebox{-1.0ex}[0.0ex][0.0ex]{$3\, ^1\!S_0$}  & \raisebox{-1.0ex}[0.0ex][0.0ex]{$0^{-+}$}   
&\raisebox{-1.0ex}[0.0ex][0.0ex]{$\eta_c'' $}  & \raisebox{-1.0ex}[0.0ex][0.0ex]{$1^{--}$} &           
&spin splitting ties mass to $\psi(4040)$ $\rightarrow$ too heavy \\
                   &            &              &   &     &$\Gamma(\eta_c,\,  \eta_c')\!\sim 20$ MeV $\rightarrow$ too broad\\
\hline\hline
\end{tabular}
}
\end{table}

%------------CUSP
In this last scenario the $X$ arises dynamically as a cusp due to the ``de-excitation'' 
of the $D^0\overline{D}{^{*0}}$ threshold.\cite{XCusp}
Very close to threshold the $S$-wave $D^0\overline{D}{^{*0}}$ de-excitation cross section
follows a $1/velocity$ dependence, which competes with the available phase space.
If the  $D^0\overline{D}{^{*0}}$ interaction is at all attractive, the $1/v$ factor can dominate
and produce a peak, but one which is not a true resonance.
A preferred decay is likely  $D^0\overline{D}{^0}\pi^0$ and/or  $D^0\overline{D}{^0}\gamma$, and
indeed Belle claims a quite large  $D^0\overline{D}{^0}\pi^0$ rate.\cite{BelleDDPi0}

% ------------GLUEBALL MIXTURE:

Another suggestion is that the $X$ 
is a vector glueball mixed with  $c\bar{c}$.\cite{ccGlueMix}
Although a $1^{--}$ state, it would be  highly suppressed in $e^+e^-$ 
since photons do not couple to gluons.
However, $X\!\rightarrow\! J/\psi \rho$, $J/\psi \omega$, and $J/\psi \gamma$
all refute this hypothesis.

The $X(3872)$ as a  $c\bar{c}g$ hybrid\cite{XHybrid}$^-$\cite{CloseYamamotoHybrid}  
is not very popular 
as the lightest states are estimated to be $\gtrsim\!4\GeVcc$,
albeit with a fair uncertainty. 
Numerous  states are expected, with exotic and non-exotic $J^{PC}$'s.
The $X$'s proximity to the $D^0\overline{D}{^{*0}}$ mass is
explained by assuming strong coupling to  $D^0\overline{D}{^{*0}}$. 
The main decays are 
normally  $[c\bar{c}g] \!\rightarrow\![c\bar{c}]  gg$
(including  $J/\psi \pi^+\pi^-$), and 
to light hadrons via $gg$ annihilation for $C\!=\!+$.
A  negative-$C$ hybrid is more likely to be narrow, but is {excluded} by $C\!=\!+$ decays like $J/\psi\rho$.
Mixing with $c\bar{c}$ or  {$D^0\overline{D}{^{*0}}$ opens up}
typical $c\bar{c}$ modes.\cite{CloseYamamotoHybrid} 
Branching ratios of $B\!\rightarrow 0^{+-}$ (exotic) hybrid, thought to be among
the lightest, is estimated
to be $\sim\!10\times$ lower than for normal $c\bar{c}$;\cite{ChiladizeHybridProd}
but other hybrids could have higher  rates.
Models of hybrid production at the Tevatron are less developed,  but since 
there are common  matrix elements,
presumably  hybrids are similarly suppressed in $\bar{p}p$.
But in the end, hybrid models must contend with the low $X$-mass and even $C$.

%--------molecule:

The idea of the $X(3872)$ as four-quark state spans  a range of extremes:
from bag-like models  in which all quarks play an equal role,
to scenarios where quarks act in  pairs. 
The latter can be a  deuteron-like ``molecule'' of two $q\bar{q}'$-pairs,
or $qq'$-$\bar{q}\bar{q}'$ diquarks.
Bag models often serve for light-quark exotics; but  
for the $X$, four-quark models gravitate to paired quarks given it
contains heavy quarks, and is so near the $D^0\overline{D}{^{*0}}$ mass. 
A diquark model envisages
a rich family of $[qc][\bar{q}\bar{c}]$ states:
various pairings with $u$ and $d$, and
two each of $0^{++}$ and $1^{+-}$, and one  $1^{++}$ and  $2^{++}$.\cite{Maiani}
The $X$  is proposed to be the  $1^{++}$.
In addition to charged $X^+$'s,
{\it two} neutral states are expected: $X^0_u\!=\![cu][\bar{c}\bar{u}]$ 
and $X^0_d\!=\![cd][\bar{c}\bar{d}]$.
These can mix with some angle, $\theta$, and the mass difference
between eigenstates is estimated to be:
$\Delta M_X \sim\!(7\!\pm\!2)/\cos(2\theta)\MeVcc$.
Since isospin is broken, both $X^0$ eigenstates
decay to $J/\psi\rho$ and $J/\psi\omega$.
From the fact that Belle reported a {\it single narrow} state the
authors argue that one $X^0$ dominates in $B^+\!\rightarrow\!XK^+$
decay, and the other in  $B^0\!\rightarrow\!X'K^0$.

CDF data bring constraints to this model.
While Belle supposedly produces only  one of the $X^0$'s, CDF's search is
inclusive:  $X^0_u$ and  $X^0_d$ are produced equally.
As is apparent from Fig.~\ref{Fig:XCDF},  no twin of the $X(3872)$
is visible, except for the possibility that CDF sees an unresolved
mixture of {\it both} $X^0_u$ and  $X^0_d$.
CDF fits their $X$ peak by a (resolution dominated)
Gaussian with $\sigma\!=\!4.9\!\pm\!0.7\,(stat)\MeVcc$.
From ``toy'' Monte Carlo studies I find it is difficult to accommodate
two peaks with $|\Delta M_X| \!\gtrsim \!8\MeVcc$.

A more restrictive condition comes from mass measurements.
As an equal mixture of unresolved $X$'s, CDF's mass  is the {\it average}
of $X^0_u$ and  $X^0_d$, and if  $B^+\!\rightarrow\!XK^+$
is a pure species: $|\Delta M_X| \!= \! 2|M_{Belle}-M_{CDF}|\!=\! 1.4\!\pm\!2.2\MeVcc$.
For a $1.64\sigma$ excursion ($95\%$ 1-sided CL),
the mass splitting must be less than $5\MeVcc$.
CDF data do not exclude a pair of $X^0$ states,
but they must have a small mass splitting, 
eroding the strength of isospin breaking, and some of the appeal of this model.
OR, the splitting is so large that new  modes open
up and  $J/\psi\pi^+\pi^-$ decays become invisible. 
{\sc BaBar} has recently reported a possible $B^0\!\rightarrow\!XK^0_s$
signal ($2.7\sigma$),\cite{XBr}
which if true, enables a direct measurement: $|\Delta M_X|\!= \! 2.7\!\pm1.3\!\MeVcc$.
By the same scaling used above, this translates into a 4.8\MeVcc\ limit,
similar to that inferred from CDF.

A molecule is the most popular exotic interpretation.
The proximity of the $X$ and  $D^0\overline{D}{^{*0}}$  masses
naturally incites such thinking.
A $J^{PC}$ of $1^{++}$, and possibly   $0^{-+}$, are thought 
the most promising  cases to be bound by  pion exchange.\cite{TornqvistMolec}
Generally,  $D^0\overline{D}{^{*0}}$,   $D^0\overline{D}{^{0}}\pi^0$,
and  $D^0\overline{D}{^{0}}\gamma$,
are expected to be major decay modes if energetically allowed.
Existence of a  $D^0\overline{D}{^{*0}}$ molecule suggests  
$D^+\overline{D}{^{*0}}$,  $D^+{D}{^{*-}}$,  $D^+_s{D}{^{*-}_s}$,\ldots\ analogs.
This simple scheme is  undermined by
a negative $X^+\!\rightarrow\!J/\psi\pi^+\pi^0$ search,\cite{BaBarXNeg}
which nominally\cite{IsoVecExcep}
excludes the $X$ as an isovector.
But in fact, binding by pion exchange is expected to be three times stronger for isosinglets
compared to isovectors; and the perturbation due to isospin breaking 
from the $D^0\!-\!D^+$ mass difference binds  $D^0\overline{D}{^{*0}}$ more
tightly while creating  repulsion for  $D^+\overline{D}{^{*0}}$ 
and   $D^+{D}{^{*-}}$ molecules.\cite{TornqvistMolec}
Thus, it is  in fact quite reasonable for there to be only
a {\it single}  $D\overline{D}{^{*}}$  molecule.

Swanson\cite{SwansonMolecule}  has built a  particularly detailed molecular model,
the crux of which is the near degeneracy of $D\overline{D}{^{*}}$, 
$J/\psi \rho$, and $J/\psi \,\omega$ masses.
The $X$ as $1^{++}$ will be a  mix of  these components.
In this model the latter two pairs are  {\it necessary} to achieve binding,
and no other $J^{PC}$ or charged states exist.
The $X$ is mostly $D^0\overline{D}{^{*0}}$ ($\gtrsim\! 80\%$), with
modest ($\sim\!10\%$) $D^+\overline{D}{^{*-}}$ and  $J/\psi \,\omega$ fractions,
and a tiny ($<\!1\%$)   $J/\psi \rho$. 
The  $J/\psi \rho$ is only a trace, 
but it has the largest branching ratio
because of the $\rho$'s large width.
Unlike many models,  $J/\psi \pi^+\pi^-\pi^0$  decay, through a virtual $\omega$,
is also large: $\sim\!60\%$ of  $J/\psi \rho$.
The next largest decay is  $D^0\overline{D}{^{0}}\pi^0$,
$\sim\!10\%$ of $J/\psi \rho$. 
The  $J/\psi\omega$ prediction prompted Belle to search for it,
and by measuring a $\omega$-to-$\rho$ branching ratio of $1.0\pm0.5$,\cite{BelleOmega,BellePsiGammaSIS05}
one can chalk-up a victory for this model. 
However,  Belle's preliminary report\cite{BelleDDPi0} of a $D^0\overline{D}{^{0}}\pi^0$ rate
more than $10\times$  that of $J/\psi \pi^+\pi^-$ is a  failure.

Na\"{\i}vely one  expects the formation of fragile states to be suppressed.
This is manifest in  ``low-energy universality.''\cite{BraatenXLowEUniv}
As an $S$-wave $D^0\overline{D}{^{*0}}$ system ($1^{+}$), 
the $X$ is so weakly bound that it is spatially large compared to its meson
constituents, and has an unnaturally
large $D^0\!-\!\overline{D}{^{*0}}$ ``scattering length.''
Important properties of the system are governed by this large 
scattering length rather than short-range details of its construction.
In particular, its cross section is $\propto\! \sqrt{E_B}$
for small binding energy $E_B$. 
One may imagine evading this suppression if
the $X$ is a {\it mixture} of $D\overline{D}{^{*}}$ 
and $c\bar{c}$ by coupling to the
$c\bar{c}$ wave-function to elevate production rates
to charmonium levels.
But by low-energy universality the  non-$D\overline{D}{^{*}}$ 
components of the wave-function also vanish  as $\sqrt{E_B}$, 
again enforcing $\sigma \!\propto\! \sqrt{E_B}$. 
\mbox{In fact, even if the $X$ arises {\it from}  $c\bar{c}$, say} 
$h_c'(2\,{^1\!P}_1)$ or $\chi_{c1}'(2\,{^3\!P}_1)$,
which is accidentally fine-tuned to the  $D\overline{D}{^{*}}$ mass,
the $c\bar{c}$ part is suppressed by $\sqrt{E_B}$,
and {\it again} $\sigma \!\propto\! \sqrt{E_B}$.
The same dependence is also present in branching ratios to the $X$.
One's prejudice for suppressed production is born-out
in this picture; and, as seen with NRQCD (Sec.~\ref{sec:XProd}), 
the suppression is similar
in {\it both} the production of, and in $B$ decays to,  the $X$.
Significant suppression can be accommodated by data (Table~\ref{Tab:XSecSumm})
{\it if} ${\cal B}_X$ is large---as in Swanson's model.

Low-energy universality has also been used to construct 
a model for $X$ formation  by
coalescence of  $D^0$ and $\overline{D}{^{*0}}$ 
in $B^+\!\rightarrow\! D^0\overline{D}{^{*0}} K^+$.\cite{BraatenXCoalesc}
It is estimated that
${\cal B}(B^+\!\rightarrow\! XK^+)\!\approx\! (2.7\!\times\! 10^{-5})
\Lambda^2_1/m^2_\pi \sqrt{E_B/0.5\!\MeV}$, where $\Lambda_1$ is a cutoff, 
and $E_B$ the binding energy.
The authors propose  $\Lambda_1\!\approx\!m_\pi$, 
and thus: {\it if} ${\cal B}_X$ is large, 
${\cal B}$ is close to the measured value  (Table~\ref{Tab:BrSumm}).
From this theoretical perspective we get the same message: decay rates
favor molecules {\it if} \mbox{$J/\psi\pi^+\pi^-$ is a very prominent mode.}

{\begin{center}
\vspace*{-0.5pt}
\raisebox{0.0ex}[-0.0ex][-0.0ex]{\LARGE $\sim$}\\
\vspace*{-0.5pt}
\end{center}}

After almost two years since its discovery the nature of the $X(3872)$
remains uncertain.
New pieces to the puzzle are available, and much is unfavorable to $c\bar{c}$ options.
A case has been made\cite{BelleNewMass} that the $X$ is most likely $1^{++}$---with the
$D^0\overline{D}{^{*0}}$ molecule an increasingly favored option.
But  as potentially the first unequivocally exotic hadron,
clear and compelling evidence must be required.

If one wants to cling to a $c\bar{c}$ assignment, 
$C$-parity eliminates all but two:  $1\,^1\!D_2$ and  $2\,^3\!P_1$.
The  $2\,^3\!P_1$ has the favored $J^{PC}$, but one must contend
with predictions that make it  $\sim\!100\MeVcc$ too heavy
and the  small $X\!\rightarrow\!J/\psi\gamma$ rate.

On the other hand, the  $1\,^1\!D_2$ prediction is only $\sim\!30\MeVcc$ below the $X$,
and it should be narrow because $D\overline{D}$ decay is forbidden.
CLEO's  $\gamma\gamma$-fusion search
was not sensitive enough to exclude it.\protect\cite{CLEOXgammagamma} 
An objection against the  $1\,^1\!D_2$ is that $\eta_c\pi^+\pi^-$ 
dominates its dipion transitions.
Barnes and Godfrey\cite{ccbarOptions1} estimate  $1\,^1\!D_2$ decay rates
but ignored the apparently significant  $D^0\overline{D}{^0}\pi^0$ decay.\cite{BelleDDPi0}
If we arbitrarily extend their model with a partial width $\Gamma(D^0\overline{D}{^0}\pi^0)\!=\! 1\MeV$,
then $\Gamma_{Tot}\!=\! 1.86\MeV$---a little less than Belle's 2.3\MeV\ limit on $\Gamma_X$.
The  $\eta_c\pi^+\pi^-$ fraction is then 11\%.
Belle's preliminary  $D^0\overline{D}{^0}\pi^0$ rate is $\sim\!15\times$ 
that of $J/\psi\pi^+\pi^-$,   but with $\sim\!50$\% error.\cite{BelleDDPi0}  
This rate limits  ${\cal B}_X(X\!\rightarrow\!J/\psi\pi^+\pi^-)\!\lesssim \!10\%$; 
but used  with  $\Gamma(D^0\overline{D}{^0}\pi^0)\!=\! 1\MeV$, 
we find  ${\cal B}_X\!\sim\! 3\%$.
This is, given the uncertainties, a ${\cal B}_X$ rate $\sim\!2$-$5\times$
below the  $\eta_c\pi^+\pi^-$   prediction,
thereby respecting  $\eta_c\pi^+\pi^-$ dominance.
Furthermore, estimates of $\pi\pi$ transitions usually 
 do not include resonant enhancements,
such as from the $\rho$. 
The  $1\,^1\!D_2$ can decay to   $J/\psi\rho$,
but not to $\eta_c\rho$. % ({\it i.e.}, $0^{-+}1^{--}$).
This could help boost   $J/\psi\pi^+\pi^-$ expectations, 
but only if one is willing to badly break isospin.

Isospin is a general  objection to $c\bar{c}$.
The $X(3872)$ is well positioned to break it by sitting on  $D^0\overline{D}{^{*0}}$.
Belle measures, with $\sim\!50\%$ errors, 
equal branching ratios  to $J/\psi\rho$ and $J/\psi\omega$.
However, these decays rely upon the width of the $\rho/\omega$ to populate the 
allowed phase space.
If one makes a simple estimation of the allowed $(phase~space)\times(Breit~Wigner)$,
the $\rho$ should have $\sim\!5\times$ the rate of the $\omega$.
Thus one can argue that  $J/\psi\rho$ may be suppressed by isospin, and, allowing for uncertainties,
by $\sim\!2$-$10\times$.
This is a far cry from the $\sim\!200\times$ one would expect from $\psi(2S)\!\rightarrow\! J/\psi \pi^0$
vs $J/\psi \pi^0\pi^0$ data.
This difference sets the scale of isospin breaking desired from  $D^0\overline{D}{^{*0}}$.

A final obstacle for the  $1\,^1\!D_2$ is the sharp fall-off of the $\pi\pi$-spectrum
seen by CDF (Fig.~\ref{Fig:DiPionM})  and Belle\cite{BelleNewMass}.
This favors $S$-wave decay, whereas the  $1\,^1\!D_2$ must go by $P$-wave.
The data are fairly striking in this respect.
A  loophole is the possibility of other effects intervening.
The  $S$-wave  argument is based on  the Breit-Wigner shape,
which ignores any more complicated {\it dynamics} in the decay.
In particular, the influence of  virtual $D^0\overline{D}{^{*0}}$ coupling
on $M(\pi\pi)$ is unknown---\mbox{recall the $\Upsilon(3S)$ tale.}

Admittedly the above arguments for $c\bar{c}$ rely as much on ignorance as they do 
on our knowledge. 
But we should  not  be swept away by the appealing prospects of an exotic $X$. 
Are the loopholes for  $c\bar{c}$ more contrived than  an exotic $X$ would be momentous?
There is even some  hints  {\it against} molecules. 
Belle's large  $D^0\overline{D}{^0}\pi^0$ rate  bounds ${\cal B}_X$ to be rather small,
thereby making $X$ production {\it very} charm\-on\-ium-like:
plug ${\cal B}_X\!=\!5\%$ into Tables~\ref{Tab:XSecSumm} \&  \ref{Tab:BrSumm}!
This begs the question of how a  $D^0\overline{D}{^{*0}}$ molecule
bound by only $\sim\!1\MeV$ can escape significant  suppression.
We may be on the verge of isolating the first unambiguous exotic hadron, 
or maybe not quite yet.

\section{Summary}

If 2003 was `the year of observation' for pentaquarks,
2004 may well be `the year of non-observation.'
CDF has searched in very large samples and found no 
evidence~for $\Theta^+(1540)$, $\Phi(1860)$, or $\Theta^0_c(3100)$. 
Whether this means  that one or more of these states are spurious,
or only that pentaquark production is highly
suppressed at the Tev\-atron, is unclear.
Both cases are interesting.
But the bulk of world data casts a
dark shadow over pentaquark prospects---if they 
are to revive, high-statistics signals will be pivotal.
Such  analyses are expected soon  from low-energy photo-production experiments
that have claimed the $\Theta^+$---early reports\cite{HighStat5Q}  are discouraging.

Irrespective of the fate of pentaquarks, 2003 also saw important,
and uncontroversial, discoveries of $D^+_{sJ}$ states and the $X(3872)$.
The  $D^+_{sJ}$'s look increasingly like $L\!=\!1$ $c\bar{s}$
states, albeit in conflict with prior potential models.
This is still exciting, if only to specialists.
The recent SELEX claim of $D^+_{sJ}(2632)$ kicks up new dust,
both because of its unusual properties and the  null
searches at $B$-factories.
It will be interesting whether CDF
can see $D^+_{sJ}(2632)\!\rightarrow\! D^0K^+$ in their large charm sample.

The $X(3872)$ remains an exciting  exotic candidate.
A case has been built against all  charmonium options,
and a  $D^0\overline{D}{^{*0}}$ molecule is increasingly popular.
The case against $c\bar{c}$ is, however, partially  predicated on conventional expectations, and 
the exceptional qualities of the $X$ creates
enough latitude to keep the   $c\bar{c}$ door open a crack.
Production data seem to point towards charmonium,
but a reliable measurement of ${\cal B}_X(X\!\rightarrow\! J/\psi\pi^+\pi^-)$
is needed.
More is to be learned from existing data, 
and samples are growing at the Tevatron and the $B$-factories.

Are we in the midst of a revolution in spectroscopy? 
Or only actors in the latest episode of a forty-year snark hunt?
We are hopefully on the cusp of learning which.

\section*{Acknowledgments}

I would like to thank 
K.-T. Chao, 
E.~Eichten,
Y.-P.~Kuang,
D. Litvintsev,
S. Olsen,
Ch.~Paus,
C.~Quigg,
A.~Rakitin,
K.~Sumorok,
and
K.~Yi
for stimulating discussions and helpful comments,
and my colleagues at CDF for an enjoyable atmosphere
and their very hard work to produce the results discussed here.
However, all opinions expressed---and errors committed---are the sole responsibility
of the author.


\begin{thebibliography}{0}
\bibitem{SashaPt}
Preliminary evidence from CDF's subsequent dipion mass analysis
indicate that the raw $p_T$ spectra for the $\psi(2S)$ and $X(3872)$
are, within statistics, indistinguishable.\cite{Sasha}
\bibitem{IsoVecExcep}
The authors of Ref.~\refcite{Maiani} argue that the exclusion limits on an $X^+$ state
are not sufficiently stringent to exclude their particular model.
\end{thebibliography}
\end{document}